\documentclass{jfm_arXiv}
\usepackage{graphicx}
\usepackage{epstopdf, epsfig}
\usepackage{units}
\usepackage{natbib}
\usepackage[percent]{overpic}
\usepackage{subfigure}
\usepackage{color}
\usepackage{tikz}
\usetikzlibrary{calc} 

\shorttitle{The linear instability of the stratified plane Couette flow.}
\shortauthor{G. Facchini, B. Favier, P. Le Gal, M. Wang,
 M. Le Bars}

\title{The linear instability of the stratified\\ plane Couette flow}

\author{Giulio Facchini\corresp{\email{facchini@irphe.univ-mrs.fr}}\aff{1}, Benjamin Favier\aff{1}, 
Patrice Le Gal\aff{1}, Meng Wang\aff{2}, Michael Le Bars\aff{1}}

\affiliation{\aff{1}Aix-Marseille Univ, CNRS, Centrale Marseille, Institut de Recherche sur les Ph\'enom\`enes Hors \'Equilibre, 49 rue F. Joliot Curie, 13013 Marseille, France
\aff{2}Department of Mechanical Engineering, University of California,
Berkeley, CA 94709, USA}

\begin{document}
\newcommand{\cor}[1]{\textcolor{red}{\textit{#1}}}

\maketitle

\begin{abstract}
We present the stability analysis of a plane Couette flow which is stably stratified in the vertical direction orthogonally to the horizontal shear.
Interest in such a flow comes from geophysical and astrophysical applications where background shear and vertical stable stratification commonly coexist.
We perform the linear stability analysis of the flow in a domain which is periodic in the stream-wise and vertical directions and confined in the cross-stream direction.
The stability diagram is constructed as a function of the 
Reynolds number $Re$ and the Froude number $Fr$, which compares the importance of shear and stratification. 
We find that the flow becomes unstable when shear and stratification are of the same order (i.e. $Fr\sim1$) and above a moderate value of the Reynolds number $Re\gtrsim 700$.
The instability results from a resonance mechanism
already known in the context of channel flows, for
instance the unstratified plane Couette flow in 
the shallow water approximation. 
The result is confirmed by fully non linear direct numerical simulations and to the best of our knowledge, constitutes
the first evidence of linear instability in a vertically
stratified plane Couette flow. 
We also report the study of a laboratory flow generated by a transparent belt entrained by two vertical cylinders and immersed in a tank filled with salty water linearly stratified in density.
We observe the emergence of a robust spatio-temporal pattern 
close to the threshold values of $Fr$ and $Re$
indicated by linear analysis, and explore the
accessible part of the stability diagram.
With the support of numerical simulations we 
conclude that the observed pattern is a signature 
of the same instability predicted by the linear theory,
although slightly modified due to streamwise confinement.

\end{abstract}
\begin{keywords}
Authors should not enter keywords on the manuscript, as these must be chosen by the author during the online submission process and will then be added during the typesetting process 
(see http://journals.cambridge.org/data/\linebreak[3]relatedlink/jfm-\linebreak[3]keywords.pdf for the full list)
\end{keywords}

\section{Introduction}\label{sec:intro}
Shear and density stratification are ubiquitous 
features of flows on
Earth and can strongly affect the dynamic of different fluids like air in the atmosphere or water in the ocean. 
More generally the interest for the stability of parallel flows
dates back to the second half of the nineteenth century \citep{Helmoltz1868,Kelvin1871} and the first crucial statement 
came with \cite{Rayleigh1879} who gave his name to the famous inflexion point theorem proving a necessary criterion for an inviscid homogeneous parallel flow to be unstable.
Contemporarily the first laboratory experiments performed by  \cite{Reynolds1883} showed that also inflexion-free flows can run unstable at sufficiently high $Re$, thus highlighting the need for a viscous analysis.
Still more than a century ago \cite{Orr1907} provided a viscous equivalent of the Rayleigh principle. Nonetheless, as reviewed  by  
\cite{Bayly1988}, providing a solution of the Orr-Sommerfeld equation at large $Re$ number turns out to be 
exceedingly difficult
and has drawn since then, the attention of many studies \citep{Heisenberg1924,Schlichting1933,Lin1955}.
Interestingly even for the simplest profile of parallel flow,
a conclusive answer as been lacking for almost a century 
as reported by \cite{Davey1973}: 
'It has been conjectured for many years that plane Couette flow is stable to infinitesimal disturbances although this has never been proved [...] We
obtain new evidence that the conjecture is, in all probability, correct'.
Since then the stability analysis of the plane Couette
(PC hereafter) 
flow continues to be of deep interest in studying the transition to turbulence via non-linear mechanisms \citep{Barkley2005}  
but its linear stability is nowadays no more questioned
\citep{Romanov1973}.
In the present work we show that by adding a vertical linear
(stable) density stratification, the PC flow becomes unstable,
at strikingly moderate $Re$ numbers, typically $Re\gtrsim 700$.
The observed instability relies on the
same resonance mechanism showed by \cite{Satomura1981a} for
shallow water waves, here extended to the case
of internal gravity waves.
An interesting feature of this finding is that density stratification is generally thought to be stabilising as it inhibits vertical motion.
Nonetheless our counter-intuitive result does not come as  
a prime novelty.
In the close context of rotating-stratified (and sheared) flows 
\cite{Molemaker2001} and \cite{Yavneh2001} questioned the
other Rayleigh celebrated criterion \citep{Rayleigh1916} 
and showed that Rayleigh-stable Taylor Couette flows may become unstable
when adding linear density stratification.
The Strato-Rotational-Instability as successively named by \cite{Dubrulle2005b} was observed in the laboratory a few years later \citep{LeBars2007LeGal} and is still the subject of experiments \citep{Ibanez2016}.
The stability analysis of parallel flows where shear coexists with stratification has also a long tradition.
The most famous shear instability,
i.e. the Kelvin-Helmoltz instability
was found indeed in the context of a two  
layers fluid endowed with different velocity and density
\citep{Helmoltz1868,Kelvin1871}. This work was extended
to the three density layers configuration, 
with constant shear in the middle one, 
by \cite{Taylor1931} and \cite{Holmboe1962}, who 
identified two different 
instability mechanisms, and later by \cite{Caulfield1994}
who isolated a third possibility.
\cite{Miles1961} and \cite{Howard1961} gave the
stability criterion of the Kelvin-Helmoltz instability,
for the case of continuous linear stratification.
Since then, most of the studies have focused on the configuration where density gradient and shear are parallel. 
On the contrary only a few 
\citep[e.g.][]{Deloncle2007,Candelier2011,ArratiaThesis2016} 
recently considered
the case of non alignment as reviewed by \cite{ChenThesis2016},  who also showed \citep{Chen2016} that a free-inflexion boundary layer profile is linearly unstable when linear stratification is added.
Restricting ourselves to the case of a vertically stratified and horizontally sheared plane Couette flow, \cite{Bakas2009b}
already considered the problem of the linear stability
while investigating the interaction between gravity waves and
potential vorticity perturbations, but curiously
did not explore the linearly unstable region.
We repeat their linear stability analysis using
a pseudo-spectral method (i.e. with the same approach as \cite{Chen2016}) and find that exponentially growing modes appear at moderate $Re$ number $Re\sim700$ and $Fr\simeq1$, for non vanishing vertical and horizontal wave number $k_x/k_z\sim0.2$.
Results are confirmed by fully non-linear 
direct numerical simulations (DNS).
We also analyse the laboratory flow produced by a shearing device immersed in a rectangular tank filled with salty water 
linearly stratified in density.
We verify that a fairly parallel PC flow can be generated and observe that beyond a moderate $Re$ number $Re\gtrsim1000$ and for $Fr$ number close to $1$, a robust velocity pattern appears in the vertical mid-plane parallel to the shear, that is where no motion is expected for a
stable PC flow.
In particular perturbations grow in an exponential manner and
looking at how their saturation amplitude varies in the $(Re,Fr)$ space, we find that an abrupt transition is present close to the marginal stability limit predicted by linear stability analysis.
The quantitative agreement of the observed spatio-temporal pattern
with the linear theory is only partial,
which we claim to be a consequence of the finite streamwise 
size of our device.
This hypothesis is largely discussed and supported by the results of additional DNS confirming that the
finite size of the domain weakly affects the base flow, but 
does modify the shape of the perturbation pattern.
We conclude that the observed instability indeed corresponds
to the linear instability of the vertically stratified PC flow modified by finite size effects and that a redesigned
experiment may reproduce more faithfully the spatio-temporal pattern predicted by the linear theory.

The paper is organized as follows.
In section \ref{sec:Theory} we define the observed flow with its governing equations and describe the linear stability approach.
In section \ref{sec:lin_results} we report the results of linear analysis and in section \ref{sec:DNS} those of direct numerical simulations.
The experiments are described in section \ref{sec:experiments} and the experimental results compared with the linear theory and direct numerical simulations
in section \ref{sec:discussion}.
In section \ref{sec:conclusions} we summarise our study and briefly 
discuss possible applications and future development
of the present work.

\section{Theoretical frame}\label{sec:Theory}
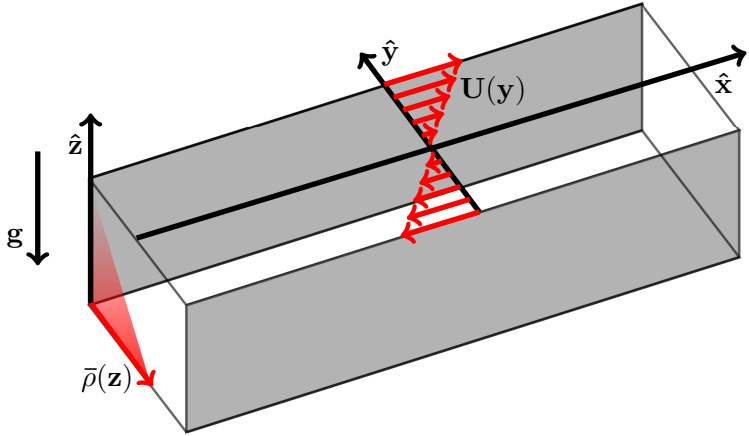
\begin{figure}
\centering
\begin{minipage}{0.8\linewidth}
\def\s{.8}			
\newlength{\SX}			
\setlength{\SX}{\s em}
\newlength{\SY}			
\setlength{\SY}{0.2425 em}	
\newlength{\NX}			
\setlength{\NX}{.60em}	
\newlength{\NY}			
\setlength{\NY}{\s em}

\begin{tikzpicture}
\coordinate (CS) at ($(current page.south) - \s*(0,4)$);
\coordinate (CN) at ($(current page.north) + \s*(0,4)$);
\coordinate (C) at ($(CN)!0.5!(CS)$);
\coordinate (CE) at ($(current page.east) + \s*(5,0)$);
\coordinate (CW) at ($(current page.west) - \s*(5,0)$);
\coordinate (CNE) at ($(CE)!0.5!(CN)$);
\coordinate (CNW) at ($(CW)!0.5!(CN)$);
\coordinate (CSE) at ($(CS)!0.5!(CE)$);
\coordinate (CSW) at ($(CS)!0.5!(CW)$);


\node[] at ($(C) + (0,0)$) {\includegraphics[width=\s\linewidth]{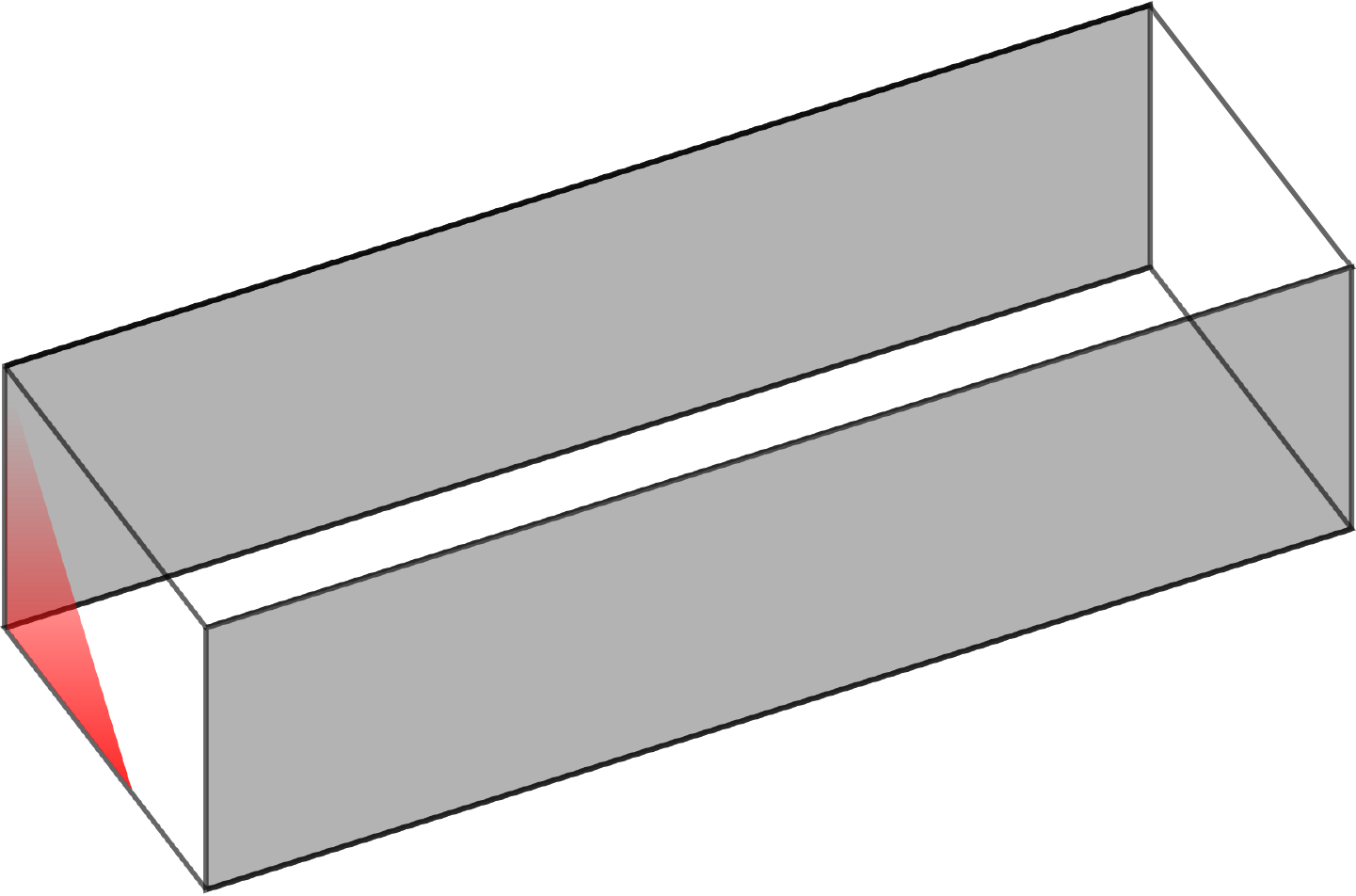}};
\coordinate (S) at ($(CNW) + \s*(2.,0.2)-0*15*(\SX,\SY)$); 
\draw[line width=2.,black,->] 
($(S) + 1*6.45*(\NX,-\NY)$) -- ($(S) - 0.25*6.45*(\NX,-\NY)$);
\node[] at ($(S) - 0.25*6.45*(\NX,-\NY) + (0.4,0)$ ) 
{\large $\mathbf{\hat{y}}$};
\node[] at ($(S)+16*(\SX,\SY)+0.5*6.45*(\NX,-\NY)+(-0.3,-0.4)$) 
{\large $\mathbf{\hat{x}}$};
\draw[line width=2.,black,->] ($(S)-15*(\SX,\SY)+0.5*6.45*(\NX,-\NY)$) -- ($(S)+16*(\SX,\SY)+0.5*6.45*(\NX,-\NY)$); 
\coordinate (N) at ($(CSW) + \s*(-2.85,.6)$); 
\draw[line width=2.,black,->] ($(N)$) -- ($(N)+2.5*(0,1)$); 
\draw[line width=2.,black,->] ($(N)+ (-.7,2)$) -- 
($(N)+(-.7,.5)$); 
\node[] at ($(N)+2.5*(0,1)+(-0.2,-0.4)$) 
{\large $\mathbf{\hat{z}}$};
\node[] at ($(N)+(-.7,.5)+(-0.3,0.35)$) 
{\large $\mathbf{g}$};
\node[] at ($(S)+4*(\SX,\SY)+(0.4,-0.4)$) 
{\large $\mathbf{U(y)}$}; 
\draw[line width=2.,red,->] (S) -- ($(S)+4*(\SX,\SY)$);
\draw[line width=2.,red,->] ($(S) +0.1*6.45*(\NX,-\NY)$) -- ($(S) +0.1*6.45*(\NX,-\NY)+ 3.2*(\SX,\SY)$); 
\draw[line width=2.,red,->] ($(S) +0.2*6.45*(\NX,-\NY)$) -- ($(S) +0.2*6.45*(\NX,-\NY)+ 2.4*(\SX,\SY)$); 
\draw[line width=2.,red,->] ($(S) +0.3*6.45*(\NX,-\NY)$) -- ($(S) +0.3*6.45*(\NX,-\NY)+ 1.6*(\SX,\SY)$);
\draw[line width=2.,red,->] ($(S) +0.4*6.45*(\NX,-\NY)$) -- ($(S) +0.4*6.45*(\NX,-\NY)+ .8*(\SX,\SY)$);  
\draw[line width=2.,red,->] ($(S) +0.9*6.45*(\NX,-\NY)$) -- ($(S) +0.9*6.45*(\NX,-\NY)- 3.2*(\SX,\SY)$); 
\draw[line width=2.,red,->] ($(S) +0.8*6.45*(\NX,-\NY)$) -- ($(S) +0.8*6.45*(\NX,-\NY)- 2.4*(\SX,\SY)$); 
\draw[line width=2.,red,->] ($(S) +0.7*6.45*(\NX,-\NY)$) -- ($(S) +0.7*6.45*(\NX,-\NY)- 1.6*(\SX,\SY)$);
\draw[line width=2.,red,->] ($(S) +0.6*6.45*(\NX,-\NY)$) -- ($(S) +0.6*6.45*(\NX,-\NY)- .8*(\SX,\SY)$);

\draw[line width=2.,red,->] ($(S) +6.45*(\NX,-\NY)$) -- ($(S) +6.45*(\NX,-\NY) -4*(\SX,\SY)$); 

\node[] at ($(N)+4.2*(\NX,-\NY)+(-0.6,0.1)$) 
{\large $\mathbf{\bar{\rho}(z)}$};
\draw[line width=2.,red,->] ($(N)$) -- 
($(N)+4.2*(\NX,-\NY)$); 

\end{tikzpicture}
\end{minipage}
\caption{
\label{fig:shear_strat_sketch}
Sketch of the analysed flow in a Cartesian reference
$\boldsymbol{\hat{x}}$, $\boldsymbol{\hat{y}}$,
$\boldsymbol{\hat{z}}$.
The base flow is aligned with the stream-wise
direction $\boldsymbol{\hat{x}}$, the 
constant shear is aligned with the cross-stream direction 
$\boldsymbol{\hat{y}}$ while density stratification
and gravity are aligned with the
vertical direction $\boldsymbol{\hat{z}}$.
We highlight in grey no slip lateral boundaries.
Open periodic boundaries are not coloured.
}
\end{figure}

We consider the plane Couette flow generated by two parallel walls moving at opposite velocity for a fluid which is stably stratified in density as sketched in figure \ref{fig:shear_strat_sketch}.
We denote $\boldsymbol{\hat{x}}$ the stream-wise direction, $\boldsymbol{\hat{y}}$ the cross-stream direction (i.e. the direction of the shear) and $\boldsymbol{\hat{z}}$ the vertical direction (i.e. the direction of the stratification).
The vector $\boldsymbol{g}$ denotes gravity while red
arrows sketch the shape of the constant shear profile $U(y)$
and red shading mimic vertical stratification  $\bar{\rho}(z)$. 
In the Boussinesq approximation we obtain the following system of equations:
 \begin{eqnarray}\label{eq.NS_full}
\frac{\partial\boldsymbol{u}}{\partial t}+ (\boldsymbol{u}\bcdot\bnabla)\boldsymbol{u} & = &
 -\frac{\bnabla p'}{\rho_0} -\frac{\rho'}{\rho_0}g\boldsymbol{\hat{z}} +\nu\nabla^2\boldsymbol{u},\\
 \bnabla \bcdot \boldsymbol{u} =0,\label{eq:NSgeneral2}\\
\frac{\partial\rho'}{\partial t}+(\boldsymbol{u\cdot\bnabla})\rho' 
-\frac{{N}^2}{g}\rho_0w \boldsymbol{\hat{z}} & = & 
k\nabla^2\rho',\label{eq:NSgeneral1}
\end{eqnarray}
where we decompose the pressure and density fields $p$ and $\rho$
in a perturbation $p'$ and $\rho'$ and a stationary part
$\bar{p}=p_0+\rho_0gz-N^2z^2\rho_0/2$ and $\bar{\rho}=\rho_0(1-N^2z/g)$,
with $p_0$ and $\rho_0$ two constant
reference values. 
We indicate with ${N}=\sqrt{-\partial_z\bar{\rho}(g/\rho_0)}$ 
the background Brunt-V\"ais\"al\"a frequency, 
while $\nu$ and $k$ denote viscosity and salt diffusivity.
\subsection{Linear stability analysis}
We perform the linear stability analysis of the equations (\ref{eq.NS_full})-(\ref{eq:NSgeneral1}) in a Cartesian
box of dimensions $(L_x,L_y,L_z)$ centered in $x=y=z=0$. 
To this aim we introduce the buoyancy $b=\rho'/\rho_0 g$ and decompose
the velocity perturbation $\boldsymbol{u}$ in a perturbation
$\boldsymbol{u}'$ 
and a base solution $\boldsymbol{U}=-U_0y\boldsymbol{\hat{x}}$.
Boundary conditions are periodic in the stream-wise and vertical directions and no-slip, i.e. $\boldsymbol{u}'=0$ at the rigid walls $y=\pm L_y/2$.
Buoyancy perturbations $b$ are also set to $0$ at the 
walls.
The system is made non dimensional using the length $L_0=L_y/2$,
the density $\rho_0$ and the velocity $U_0=\sigma L_0$ 
where $\sigma$ is the shear rate. 
This choice is coherent with 
\cite{Chen2016}
and gives the same set of dimensionless numbers which are the Reynolds number $Re=L_0U_0/\nu$, the Froude number 
$Fr=U_0 L_0/N =\sigma/N$ and the Schmidt number $Sc=\nu/k$. 
We then look for solutions of the non dimensional perturbations 
$\tilde{\boldsymbol{u}},\tilde{p},\tilde{b}$ in the form of normal modes 
\begin{equation}\label{eq.normal_modes}
\tilde{\boldsymbol{u}},\tilde{p},\tilde{b}=(\boldsymbol{u}(y),p(y),b(y))
e^{ik_xx+ik_zz-i\omega t},
\end{equation}
where we use again symbols $\boldsymbol{u}$, $p$ and $b$ 
to simplify notations.
Substituting in equations (\ref{eq.NS_full})-(\ref{eq:NSgeneral1}) and retaining only the first order terms we obtain:
\begin{eqnarray}\label{eq.eigen_all}	
-i\omega u  &=& ik_xuy + v -ik_xp +\frac{1}{Re}\Delta_y u,
\label{equ}\\ 
-i\omega v  &=& ik_xvy -\frac{dp}{dy} +\frac{1}{Re}\Delta_y v,    \label{eqv}\\ 
-i\omega w  &=& ik_xwy -\frac{b}{Fr^2} -ik_zp +\frac{1}{Re}\Delta_y w, \label{eqw}\\
 0 &=& ik_xu + ik_zw + \frac{dv}{dy},
 \label{eqi}\\ 
-i\omega b  &=& ik_xby +w+\frac{1}{ReSc}\Delta_y b,
   \label{eqb}
\end{eqnarray}
where we denote with $\Delta_y$ the Laplacian operator
$\Delta_y=d^2/dy^2-k_x^2-k_z^2$.
The system of equations above is solved using a pseudo-spectral approach similarly to \cite{Chen2016}, the only difference is that discretisation is made on the Gauss-Lobatto collocation points of the Chebychev polynomials (i.e. instead of Laguerre) because this choice is well adapted to a two-side bounded domain. 
The generalized eigenvalue problem $\boldsymbol{Af}=\omega\boldsymbol{Bf}$ for $\boldsymbol{f}=[u,v,w,b,p]$ is solved with the QZ algorithm.  
In parallel we also consider the inviscid approach 
which consists in neglecting both viscous dissipation and
salt diffusion, thus reducing the system 
(\ref{eq.eigen_all})-(\ref{eqb})
to one equation for the pressure:
\begin{equation}\label{eq:pressure_spectral}
\frac{\partial ^2 p}{\partial y^2} -\frac{2k_x}{\omega_*}\frac{\partial p}{\partial y} +\left( k_z^2\frac{\omega_*^2}{1-Fr^2\omega_*^2}-k_x^2\right)p=0,
\end{equation}
where $\omega_*=\omega+k_xy$.
The equation above is analogous 
to that provided by \cite{Kushner1998} who previously
studied the stability of a vertically stratified PC
flow in the presence of rotation $f$.
In the limit of no rotation ($f\rightarrow 0$) we verify that the two equations are the same but 
contrarily to \cite{Kushner1998} we could not find a 
meaningful limit in which our equation (\ref{eq:pressure_spectral}) becomes autonomous in $y$.
As a consequence we cannot provide a compact form for the dispersion relation.
Nonetheless, looking at equation (\ref{eq:pressure_spectral}) is still extremely instructive.
First, one observes that the second term in equation (\ref{eq:pressure_spectral}) possibly diverges at $y=0$ when considering stationary modes which are marginally stable 
(e.g. $\omega=0$). 
This corresponds to the existence of
a barotropic critical layer, which happens to be regularised because, from the symmetry  
of the base flow, we expect $\partial p/\partial y$ to be null in 
$y=0$ for a stationary mode.
Similarly the third term of (\ref{eq:pressure_spectral}) becomes critical in $y^*=\pm1/k_xFr$ when $\omega=0$.
These are baroclinic critical layers, i.e. 
the locations where the Doppler shifted frequency $\omega_*$
of internal waves matches the Brunt V\"ais\"al\"a frequency N.
In different contexts critical layers can be excited and
have been observed in experiments \citep{Boulanger2008} 
and numerical simulations \citep{Marcus2013Suyang}.
However in our configuration the most unstable mode is
always observed to be stationary and at wave numbers 
$k_{max}<1/Fr$,
which implies
that the corresponding critical layers $y*_{max}$
are always situated outside the numerical domain, $|y^*_{max}|>1$.

\section{Linear Stability results}\label{sec:lin_results}
We have already mentioned that for unstratified fluids (i.e. $Fr=\infty$) the PC (unperturbed) profile is linearly stable for any value of the Reynolds number $Re$, thus we expect the flow to be potentially unstable only at finite values of the Froude number.
The values of the Schmidt number for common salty water 
(i.e. in our experiments) is $Sc\sim700$ thus we 
preliminarily consider the limit $Sc=\infty$ and discuss 
the quality of this approximation at the end of this section.

\begin{figure}
\subfigure
{
\begin{overpic}[height=.19\linewidth]{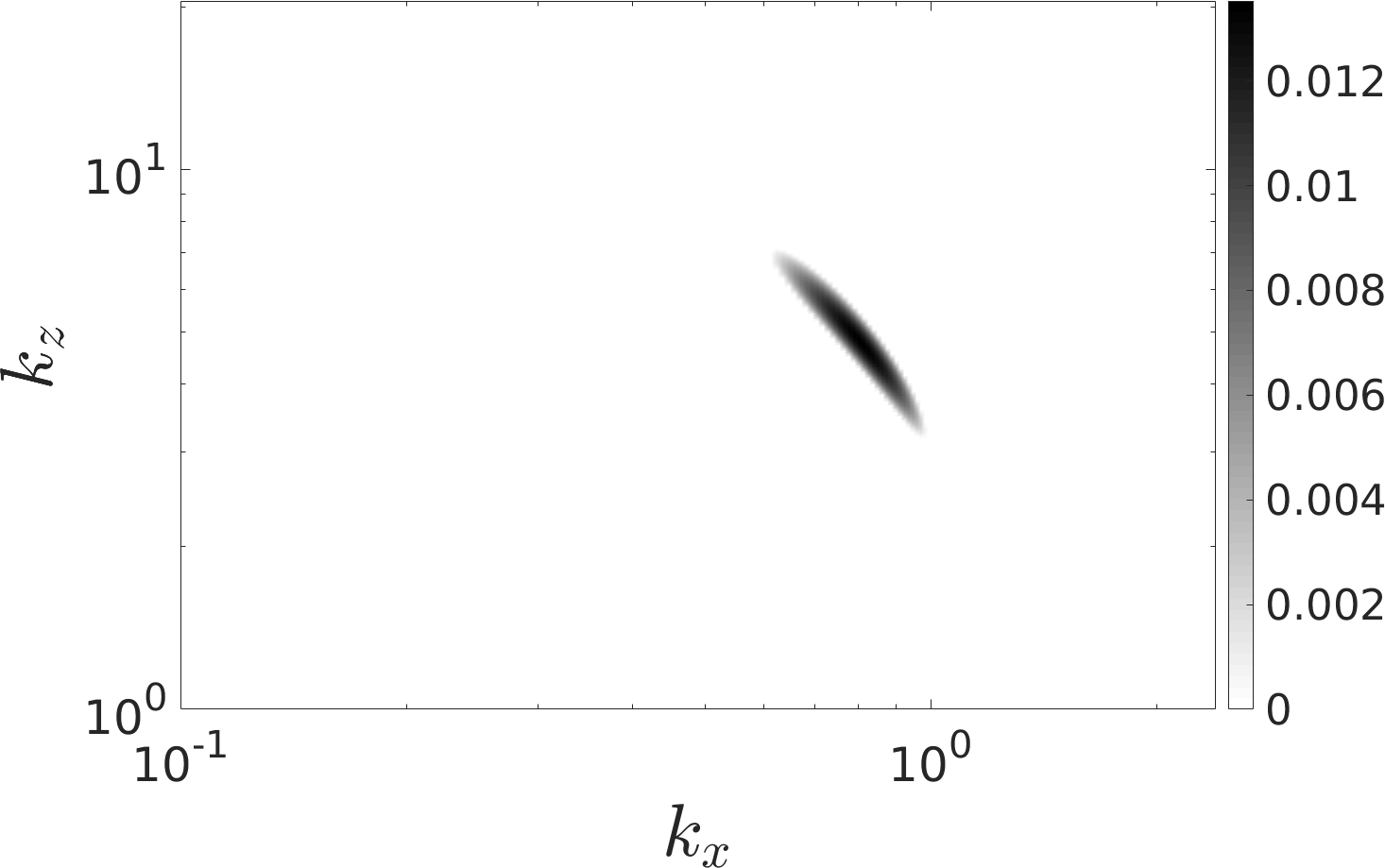}
\end{overpic}
}
\hfill
\subfigure
{
\begin{overpic}[height=.19\linewidth]{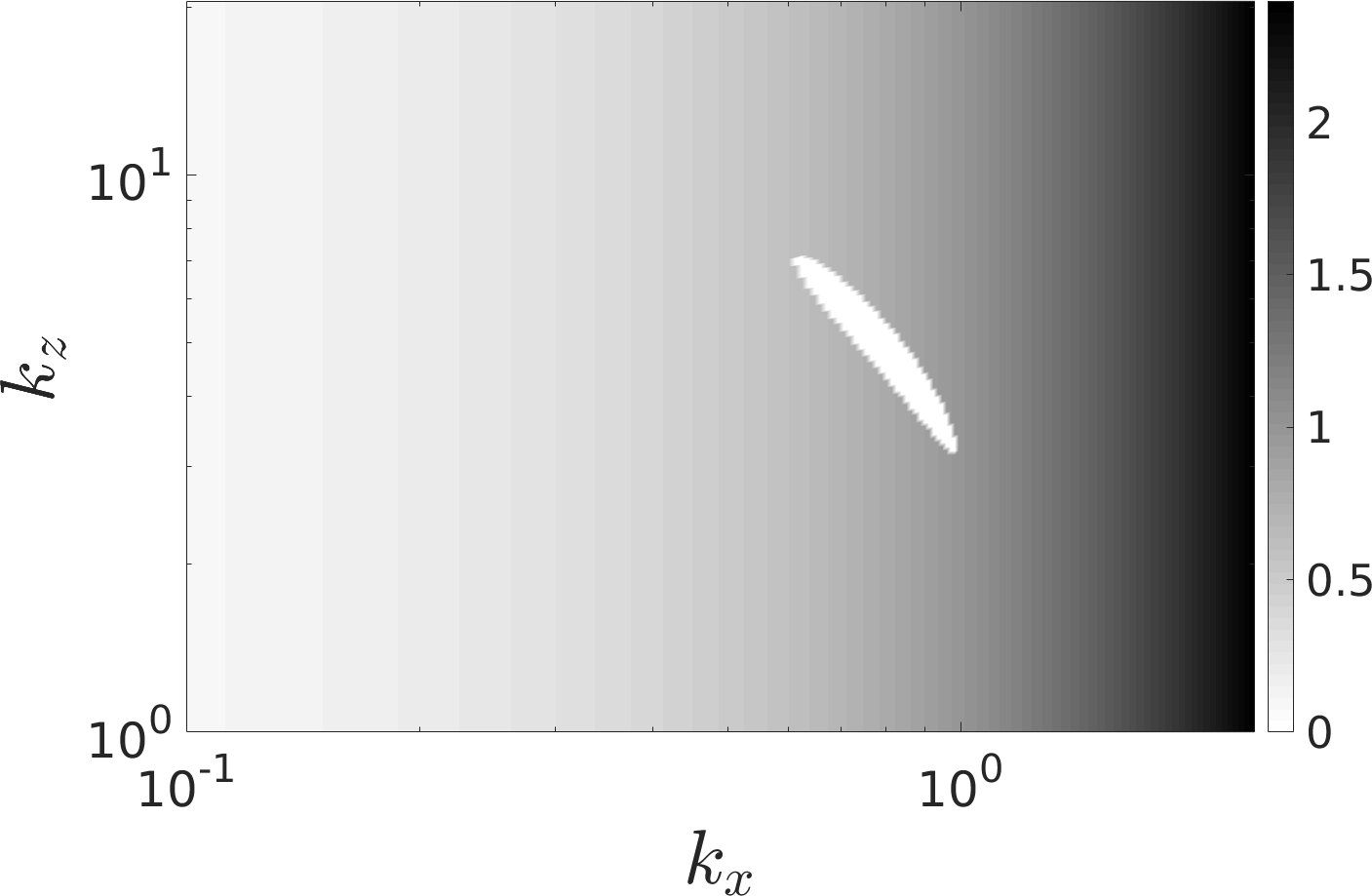}
\end{overpic}
}
\hfill
\subfigure
{
\begin{overpic}[height=.19\linewidth]{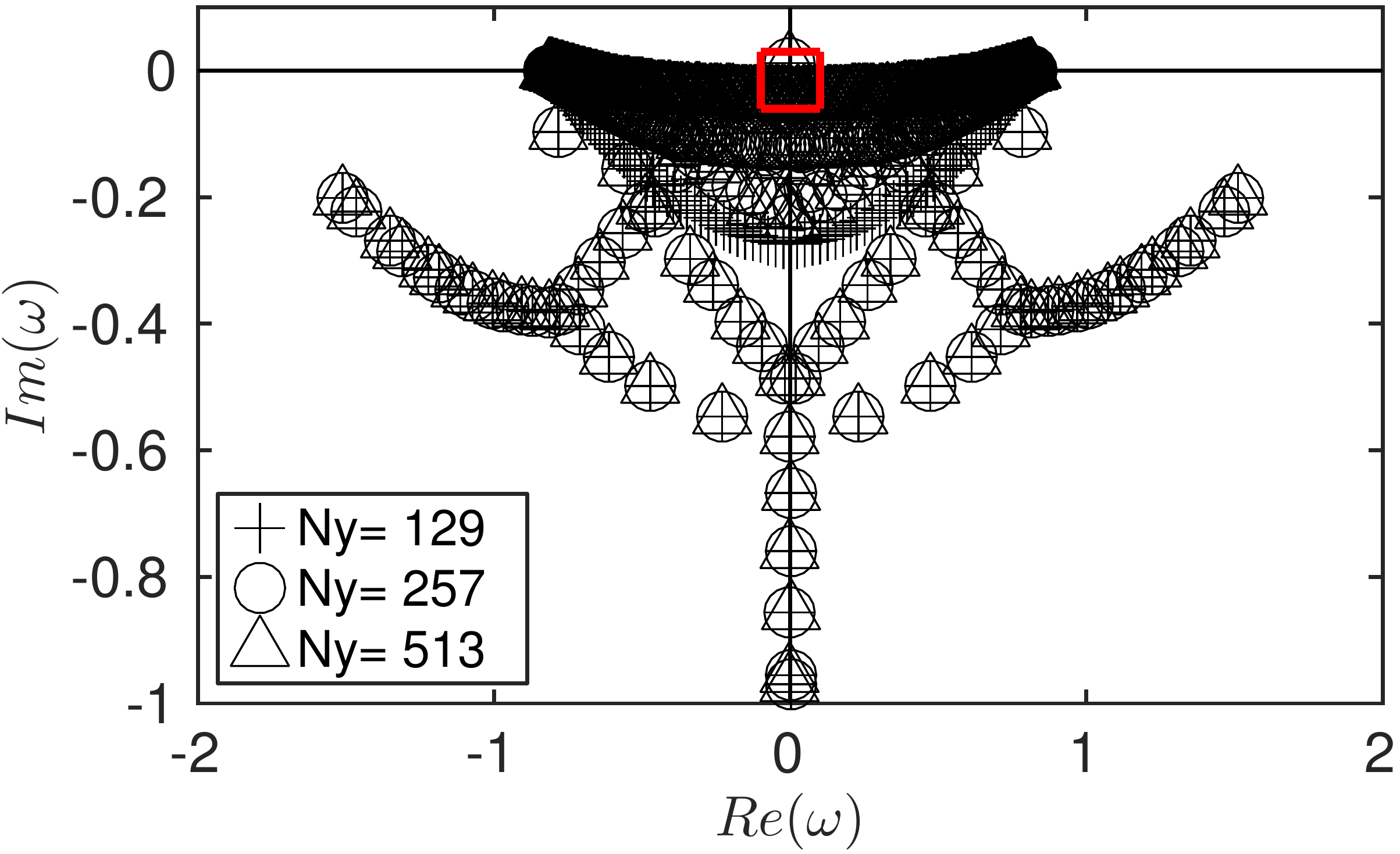}\rlap{\hspace{8.1em} \raisebox{1.6em}{\includegraphics[height=.75cm]{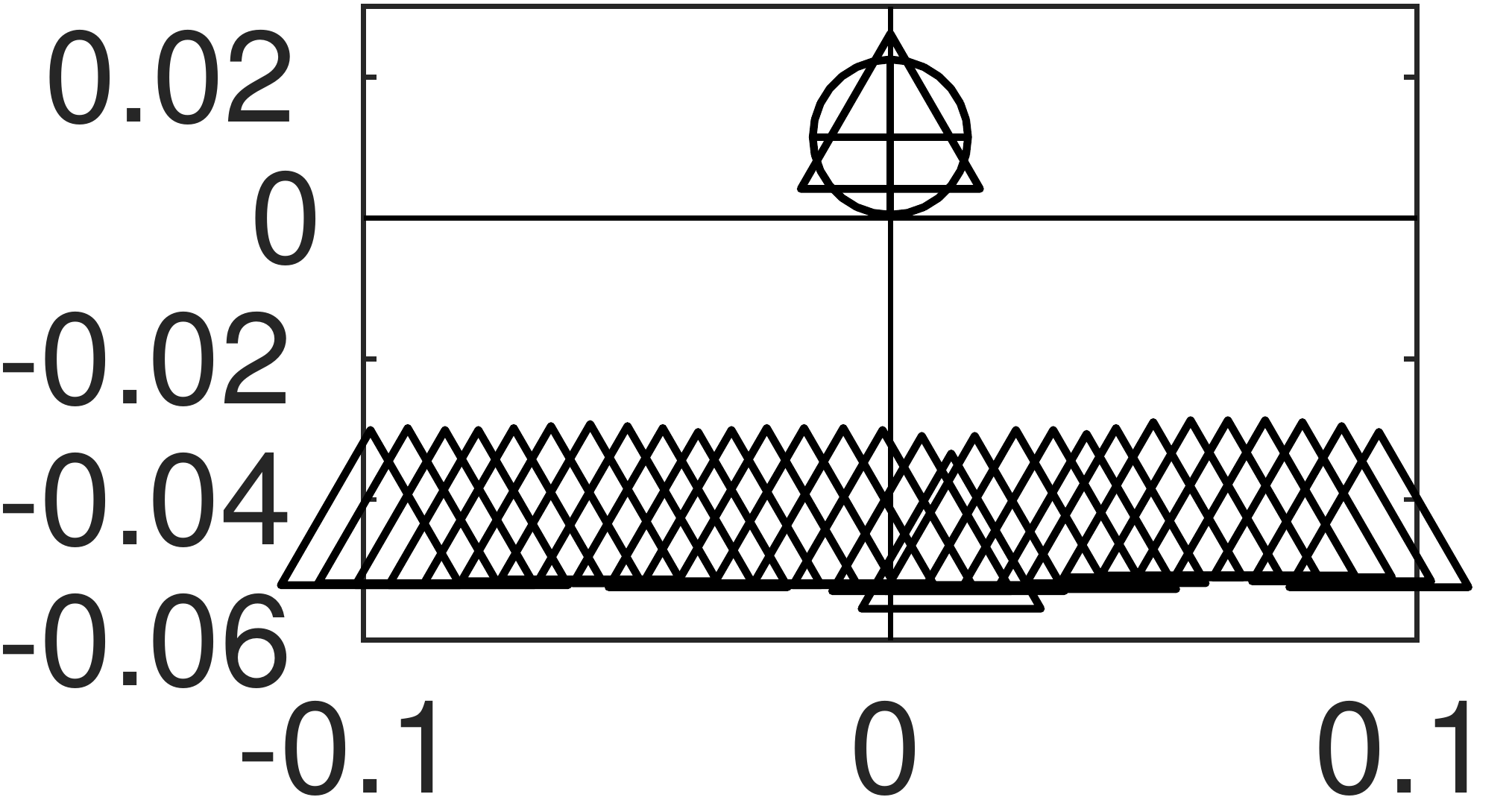}}}
\end{overpic}
}
\caption{\label{fig:evalFR1Re1000}
Left: growth rate $Im(\omega)$ of the most unstable mode 
in the space $(k_x,k_z)$ at $Fr=1$ and $Re=1000$.
Center: oscillation frequency $Re(\omega)$ 
of the most unstable mode 
in the space $(k_x,k_z)$ at $Fr=1$ and $Re=1000$.
Notes that unstable modes ($Im(\omega)>0$) are stationary ($Re(\omega)=0$).
Right: full spectrum at the most unstable
mode $k_x=0.815$ and $k_z=4.937$, $Fr=1$, and $Re=1000$.
Crosses refer to $N_y=129$ collocation points, circles to $N_y=257$ and triangles to $N_y=513$.
The inset at the bottom right coincides with the area delimited
by the red rectangle.
}
\end{figure}

As a first result we report that one stationary growing mode (i.e. $Im(\omega)>0,Re(\omega)=0$) 
appears at $Fr\lesssim1$, wave numbers $k_x\sim 0.8$, $k_z \sim 5$ and remarkably moderate Reynolds number $Re\simeq 700$.
In figure \ref{fig:evalFR1Re1000} we report the value of the imaginary part and the real part of the most unstable eigenmode for 
$Re=1000$ and $Fr=1$ as a function of $k_x$ and $k_z$. 
Looking at the imaginary part (left) one sees that the flow is unstable over a narrow elongated region centered in $k_x\sim0.8$, $k_z\sim5$ and stable elsewhere.
Correspondingly the real part (center) is zero whenever the flow is unstable and non zero elsewhere.
In figure \ref{fig:evalFR1Re1000} (right) we also report the values of the temporal frequency $\omega$ for all the eigenvalues and a various number of collocation points of $N_y=129$, $257$ and $513$ for the most unstable wave numbers ($k_x=0.815$, $k_z=4.937$).
Physical eigenvalues correspond to the points where
three different symbols are superposed, all other points 
corresponding to spurious numerical modes. 
The inset close to the origin of the diagram shows that a unique eigenvalue is present for which $Im(\omega_c)>0$. 
The value of $\omega_c$ is stable to the variation of $N_y$ the number of collocation points, which indicates that the mode we observe is a physical one.
We remark that this feature makes the systematic analysis of the 
space ($Re$, $Fr$) particularly simple, for example differently from \cite{Chen2016} the problem of non physical eigenvalues with positive $Im(\omega)$ is not present.
 
\begin{figure}
\centering
\subfigure
{
\begin{overpic}[width=.47\linewidth]{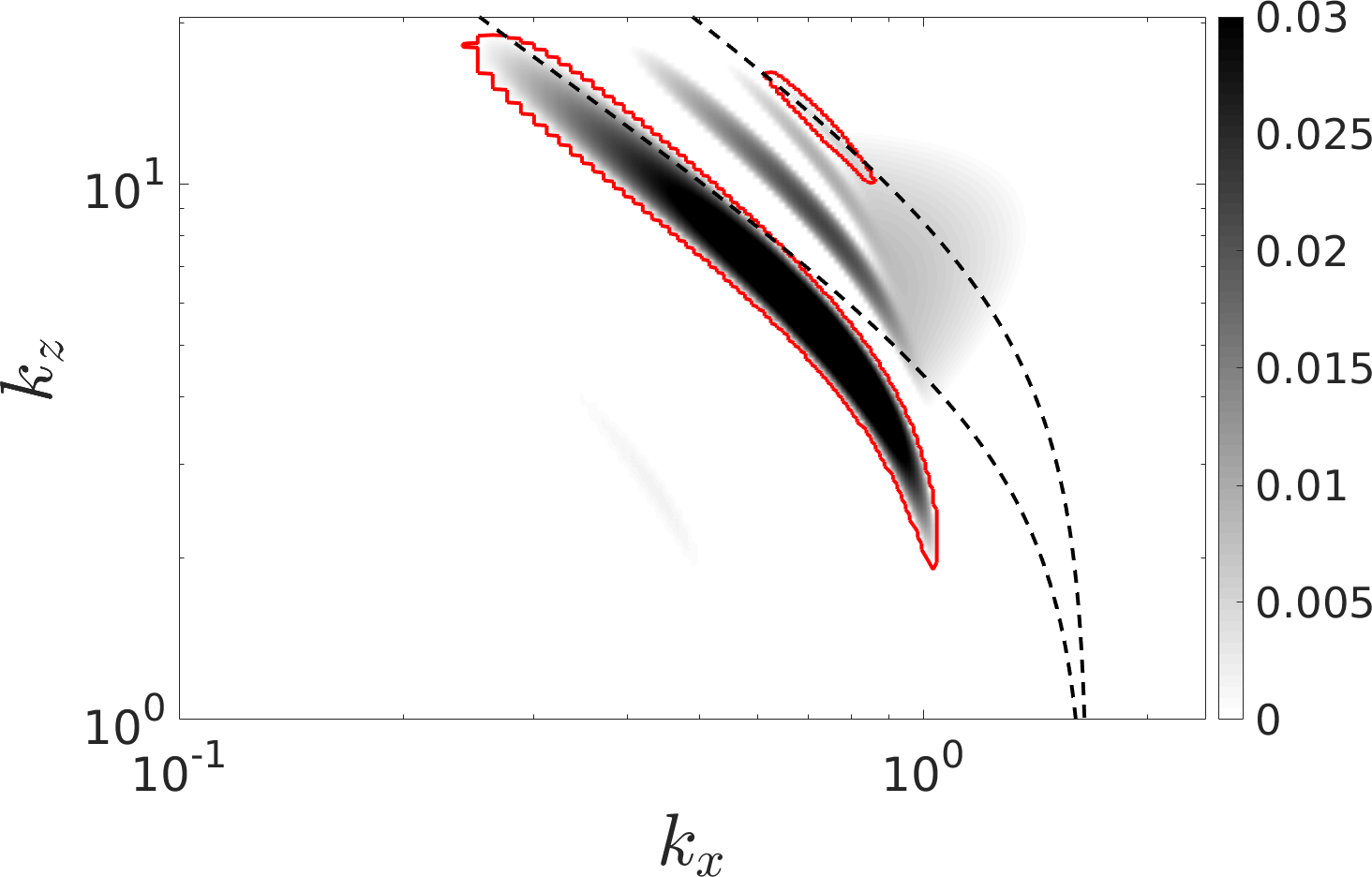}
\rlap{\hspace{2.3em}\raisebox{2.6em}{
\rotatebox{0}{\begin{overpic}[width=.2\linewidth]{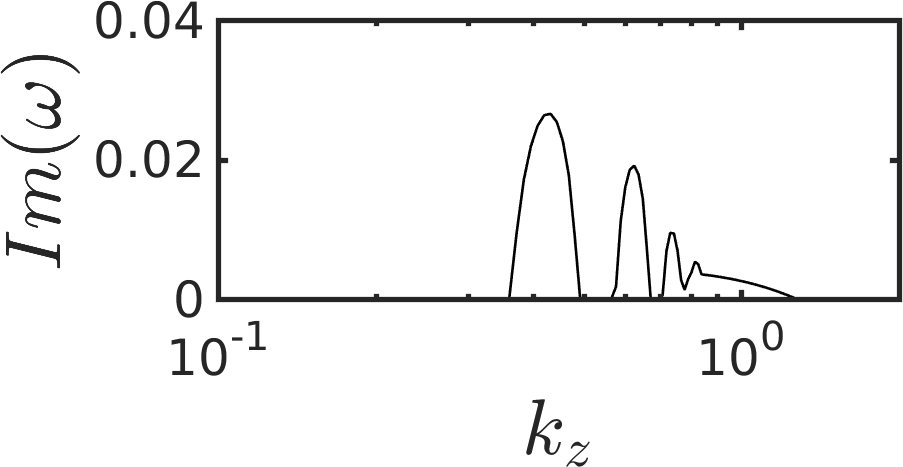}
\put (57,41) {\tiny \textcolor{red}{\textbf{a}}}
\put (67,34) {\tiny \textcolor{red}{\textbf{b}}}
\put (72,26) {\tiny \textcolor{red}{\textbf{c}}}
\put (75,23) {\tiny \textcolor{red}{\textbf{d}}}
\end{overpic}}}}
\end{overpic}
}
\hfill
{
\begin{overpic}[width=.47\linewidth]{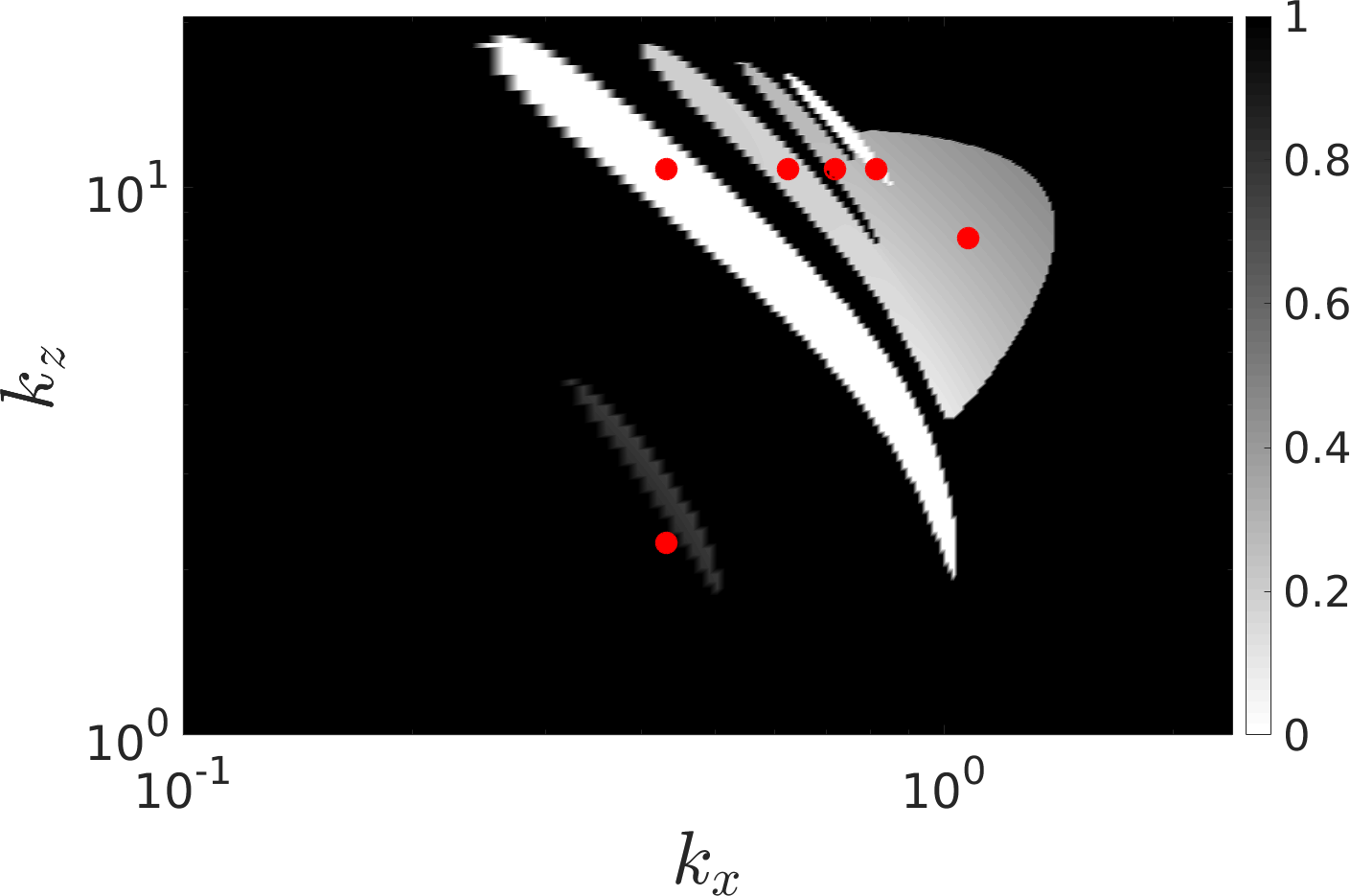}
\put (50,54) {\small \textcolor{red}{\textbf{a}}}
\put (57.,55) {\small \textcolor{red}{\textbf{b}}}
\put (60.,55) {\small \textcolor{red}{\textbf{c}}}
\put (62.5,55) {\small \textcolor{red}{\textbf{d}}}
\put (72.5,48) {\small \textcolor{red}{\textbf{e}}}
\put (49,27) {\small \textcolor{red}{\textbf{f}}}
\end{overpic}
}
\subfigure
{
\begin{overpic}[width=.47\linewidth]{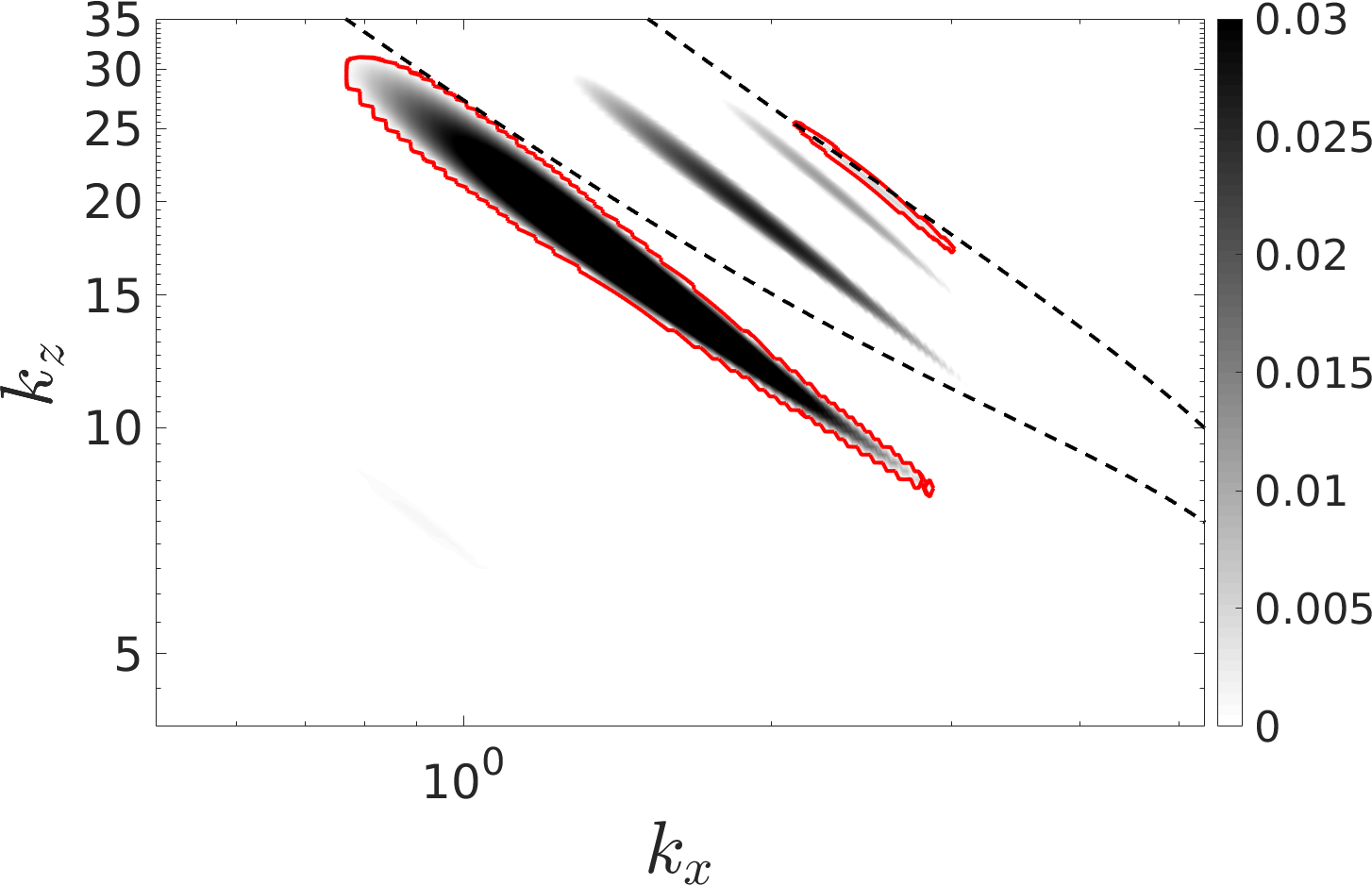}
\end{overpic}
}
\hfill
\subfigure
{
\begin{overpic}[width=.47\linewidth]{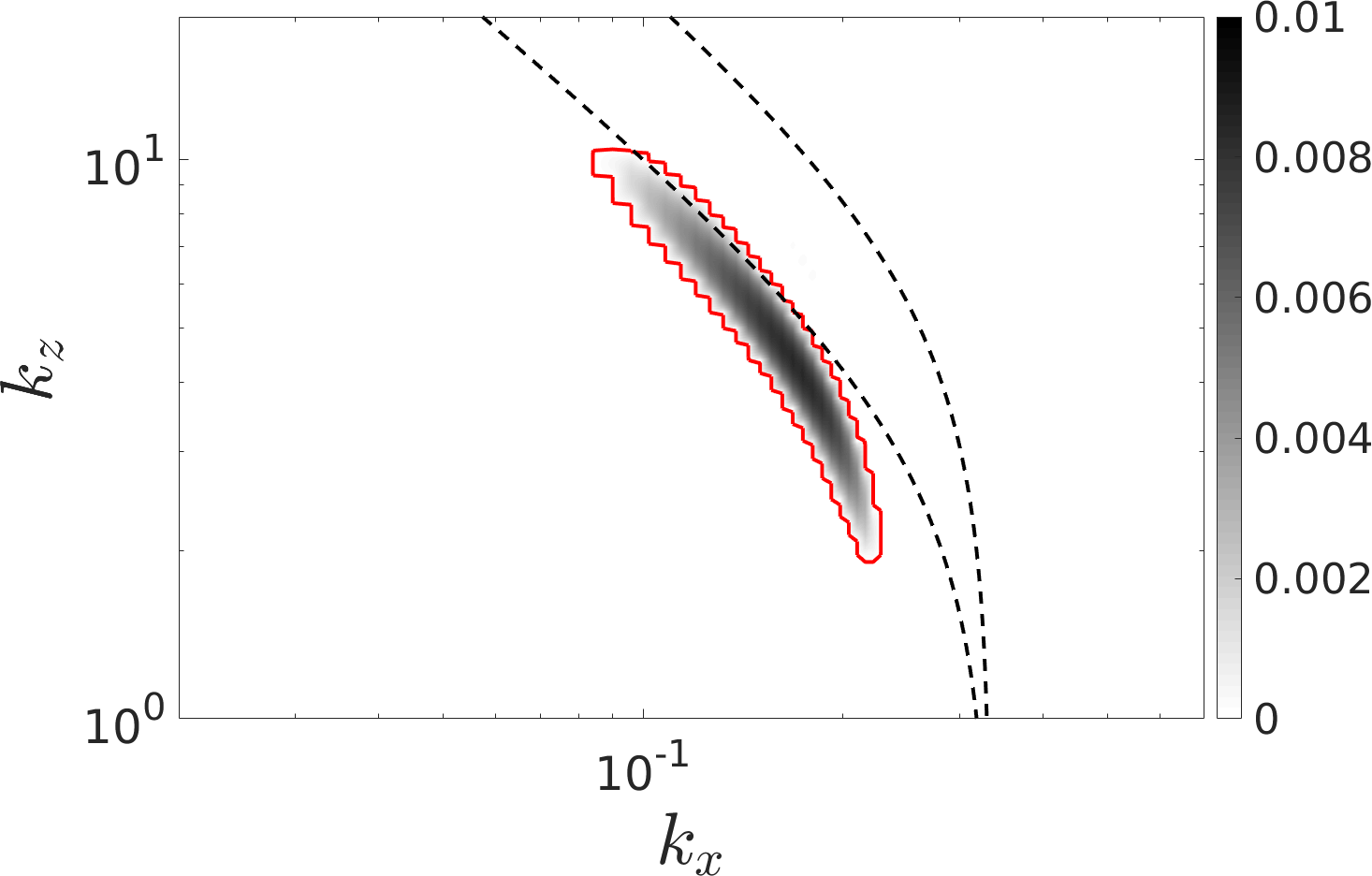}
\end{overpic}
}
\caption{ \label{fig:evalFR02_1_5Re1000}
Top: growth rate $Im(\omega)$ (left)
and real part $Re(C)$ of the velocity $C=\omega/k_x$ (right) 
of the most unstable modes in the space $(k_x,k_z)$ at $Re=10000$ and $Fr=1$.
Red dots and letters label modes of different shapes. 
Bottom: Growth rate $Im(\omega)$ of the most unstable modes
at $Re=10000$ and $Fr=0.2$ (left) and $Fr=5$ (right).
The red dashed contours distinguish
stationary branches from oscillating ones.
The black lines refer to the theoretical predictions we discuss
in section \ref{sec:inst_mechanism}.
Note that horizontal and vertical axes have different scales
depending on the $Fr$ number.
}
\end{figure}
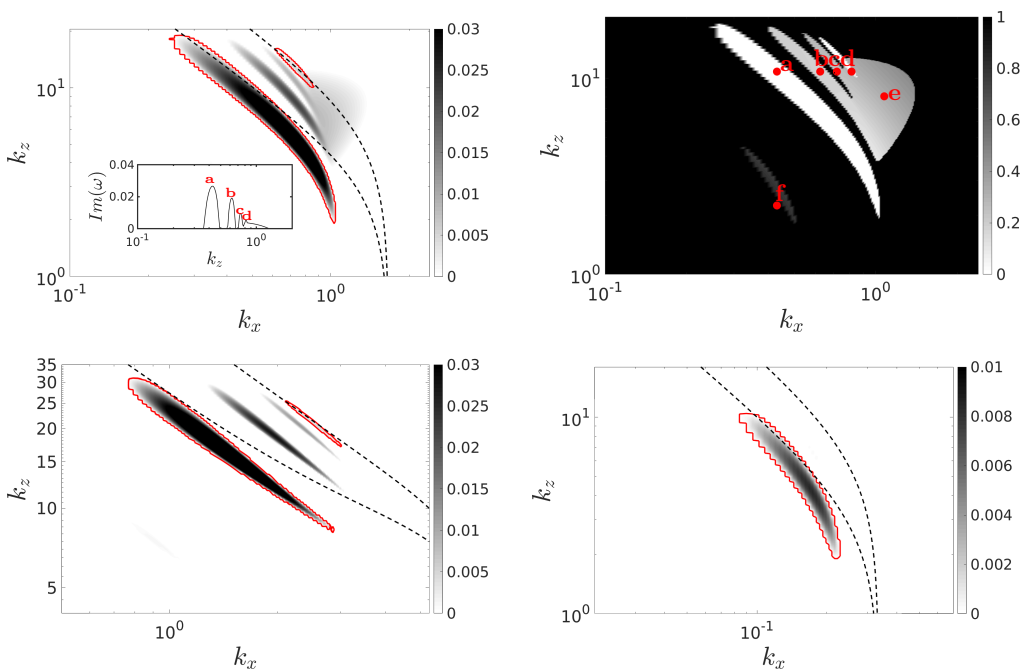

When increasing $Re$ the unstable region in the ($k_x$, $k_z$)
space increases and unstable modes exist over a 
larger range of Froude number.
On the top of  figure \ref{fig:evalFR02_1_5Re1000} we report the value of $Im(\omega)$ and $Re(C)$ for the most
unstable mode at $Re=10000$ and $Fr=1$,
where we define $C$ as $C=\omega/k_x$. 
In the same figure we report similar graphs for $Im(\omega)$
at $Re=10000$, $Fr=5$ (left) and $Fr=0.2$ (right).
Note that at these values of $Fr$ number there is
no unstable mode at $Re=1000$.

One sees that at $Fr=1$ (left panel) the diagram is now richer: besides the original unstable branch constituted by stationary modes (\textit{a}), new unstable branches appear at larger $k_x$ which correspond to oscillatory 
(\textit{b}, \textit{c}, \textit{e})
and stationary (\textit{d}) modes, as visible from
the value of $Re(\omega)/k_x)$ (right).  
Note that within a same branch the value of $Re(\omega)/k_x$ varies very little, while the value 
of $Im(\omega)$ shows a maximum and smoothly decreases to zero at
the branch boundaries.
The quantity $Re(\omega)/k_x$ then characterises each 
different branch. 
A new oscillating branch (\textit{f}) 
also appears at smaller $k_x$ 
but it 
is still very weak and poorly visible at this $Re$ number.
On the bottom of figure \ref{fig:evalFR02_1_5Re1000} we 
report the value of $Im(\omega)$ in the space ($k_x$, $k_z$)
at $Fr$ number smaller and larger than one.
When the Froude number is diminished to $Fr=0.2$ (left)
we recover almost the same scenario, even if different branches look now more spaced one from another and appear at larger $k_x$ and $k_z$ similarly to other kinds of shear flows
\citep{Deloncle2007,Park2013}.
On the contrary when the Froude number is increased to $Fr=5$
(right) the unstable region is drastically reduced, 
as well as the growth rate, which is dropped by an order of magnitude.
Also the most unstable mode moves toward lower value of $k_x$ while $k_z$ only slightly changes.
As a general remark we observe that the unstable branches,
i.e. the continuous regions defined by $Im(\omega)>0$,
show an elongated shape.
Precisely, unstable regions appear extended
when moving along the curve $k_xk_z=const$ while they are
quite narrow in the orthogonal direction.
We stress that this result is independent of the Froude number 
which suggests a self-similar behaviour as already observed by
\cite{Deloncle2007}.
Also unstable modes always appear at $k_z,k_x\neq0$, i.e. the flow is linearly unstable only to three-dimensional perturbations, which is different from the studies of \cite{Deloncle2007} and \cite{Lucas2017b}, 
performed on different vertically stratified and horizontally sheared flow (the hyperbolic tangent shear profile and Kolmogorov flow respectively).

\subsection{Stability Diagram}\label{sec:stab_diagram}
\begin{figure}
\begin{overpic}[width=1.\linewidth]{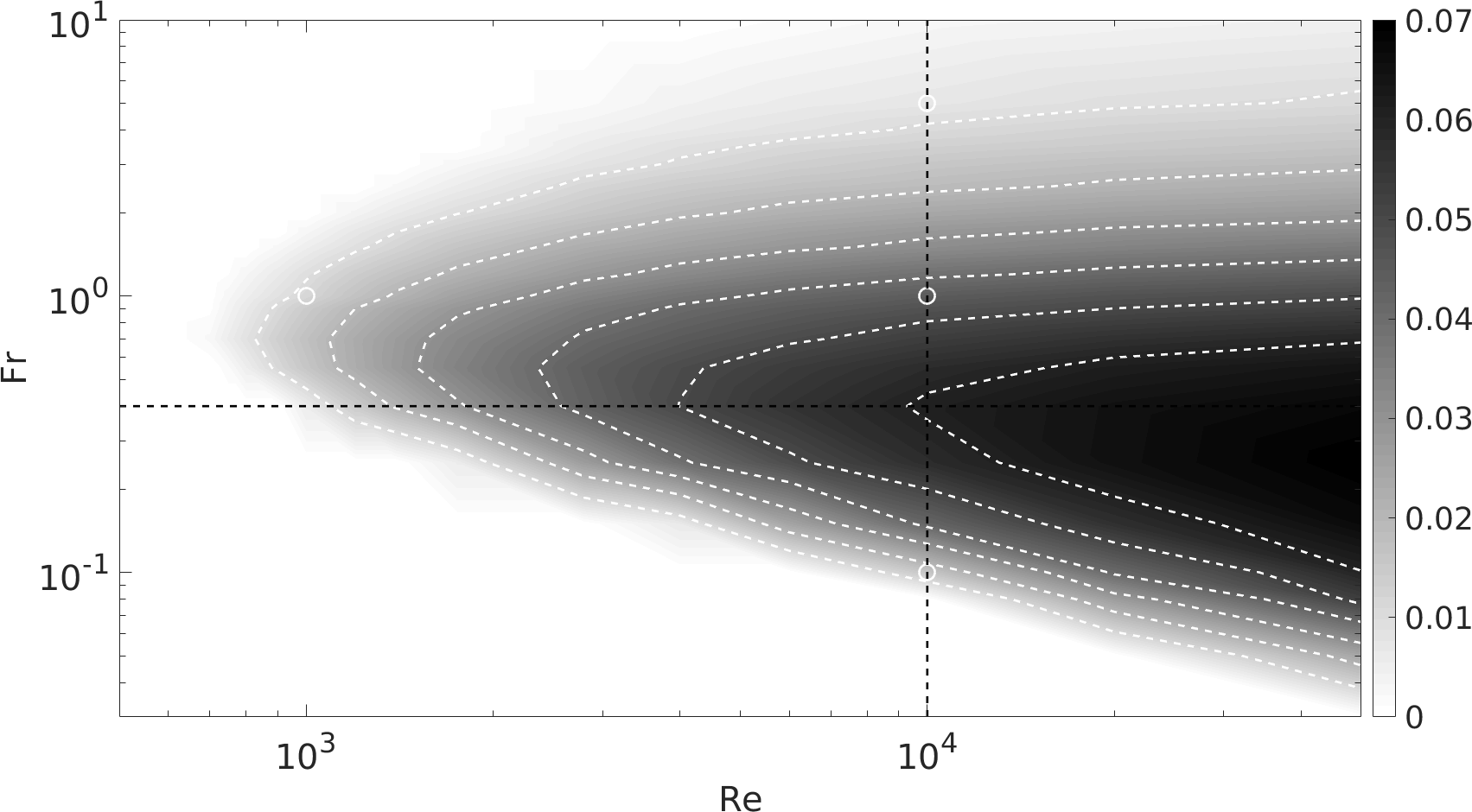}
\end{overpic}
\caption{
\label{fig:evalFrRe_all} Growth rate $Im(\omega)$ of the most unstable mode in the space
$(Re,Fr)$. Each point is obtained taking the maximum value of $Im(\omega)$ over a collection
of runs at fixed $(Re,Fr)$ and variable wave numbers $(k_x,k_z)$.
White dashed contours correspond to $Im(\omega)=0.01$,$0.02$,$0.03$,$0.04$,$0.05$,$0.06$.
Black dashed lines correspond to $Re=10000$ and $Fr=0.4$.
White circles correspond to the points of the diagram
analysed in figure \ref{fig:evalFR1Re1000} and \ref{fig:evalFR02_1_5Re1000}.}
\end{figure}

We explore the $(Re,Fr)$ parameter space over two decades around $Fr=1$ and for $Re$ from $500$ to $50000$.
For each combination $(Re,Fr)$, we solve the system (\ref{eq.eigen_all})-(\ref{eqb})
in the discretised wavenumber space $k_x\in [0,2],k_z \in [0,30]$, and look for all the possible linear growing ($Im(\omega)>0$) modes.
The ($k_x$, $k_z$) domain is suitably moved toward lower (higher) wave numbers when the $Fr$ number is significantly 
higher (lower) than $1$.  
In figure \ref{fig:evalFrRe_all} we report the 
stability diagram. 
Each point in the diagram corresponds to the most unstable mode,
whose relative $k_x$ and $k_z$ generally vary.
One observes that at $Re=1000$ the unstable region is relatively constrained around $Fr=1$ (i.e. $0.5\lesssim Fr\lesssim 2$) 
but already covers two decades in $Fr$ at $Re=10000$.
This indicates that instability first (i.e. at low $Re$ number) appears where density stratification and horizontal shear are comparable, i.e. $N\sim\sigma$, but is likely to be observed in a sensibly wider range of the ratio $\sigma/N$ 
provided that the $Re$ number is large enough.

The critical Reynolds number ($Re_c\sim700$) appears quite  moderate compared to other unstratified parallel flows
like the plane Poiseuille flow ($Re_c=5772$ according to 
\cite{Orszag1971}).
The value we find is comparable with that found
by \cite{ChenThesis2016} for a plane Poiseuille flow in the presence of vertical stratification, but still sensibly
lower than that indicated by the same authors
\citep{Chen2016} for the boundary layer (vertically stratified) profile $Re_c\sim 1995$.
The growth rate is moderate even at high $Re$ number, 
indicating that the observed instability is not only
constrained in the $(k_x,k_z)$ but also relatively slow
to establish.

\begin{figure}
\subfigure
{
\begin{overpic}[height=.3\linewidth]{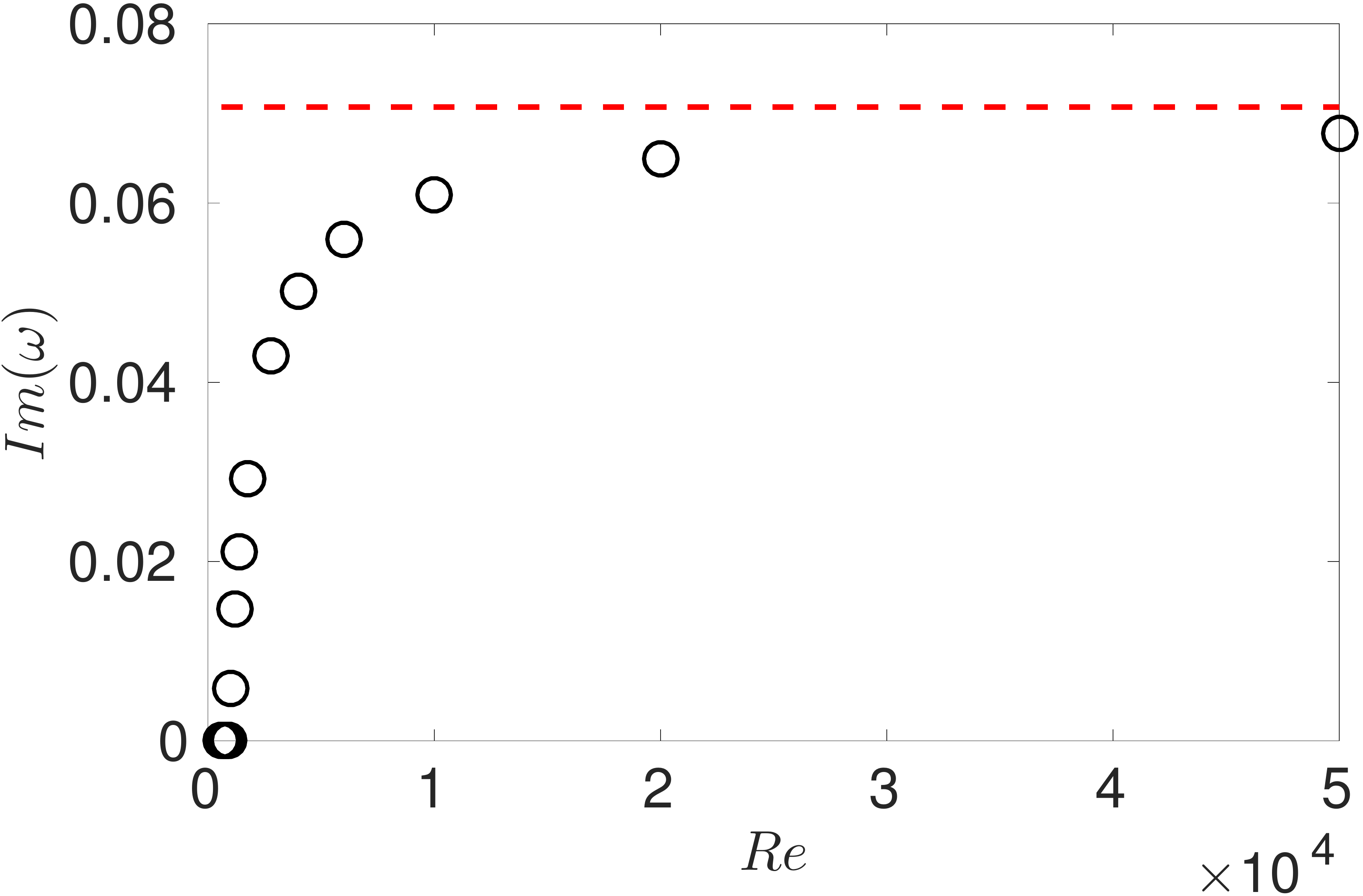}
\end{overpic}
}
\hfill
\subfigure
{
\begin{overpic}[height=.3\linewidth]{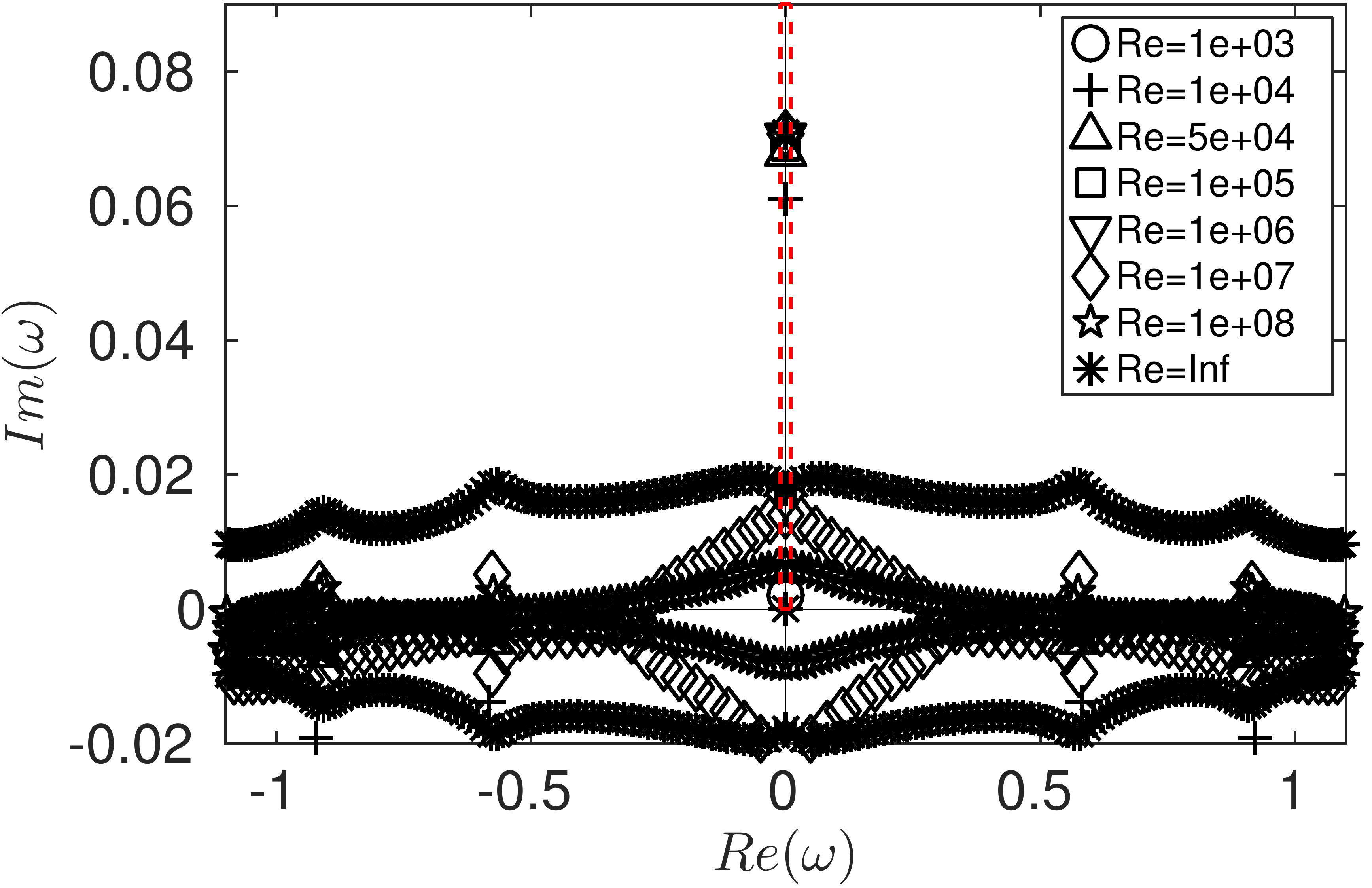}\rlap{\hspace{3.5em} \raisebox{6.6em}{\includegraphics[height=1.7cm]{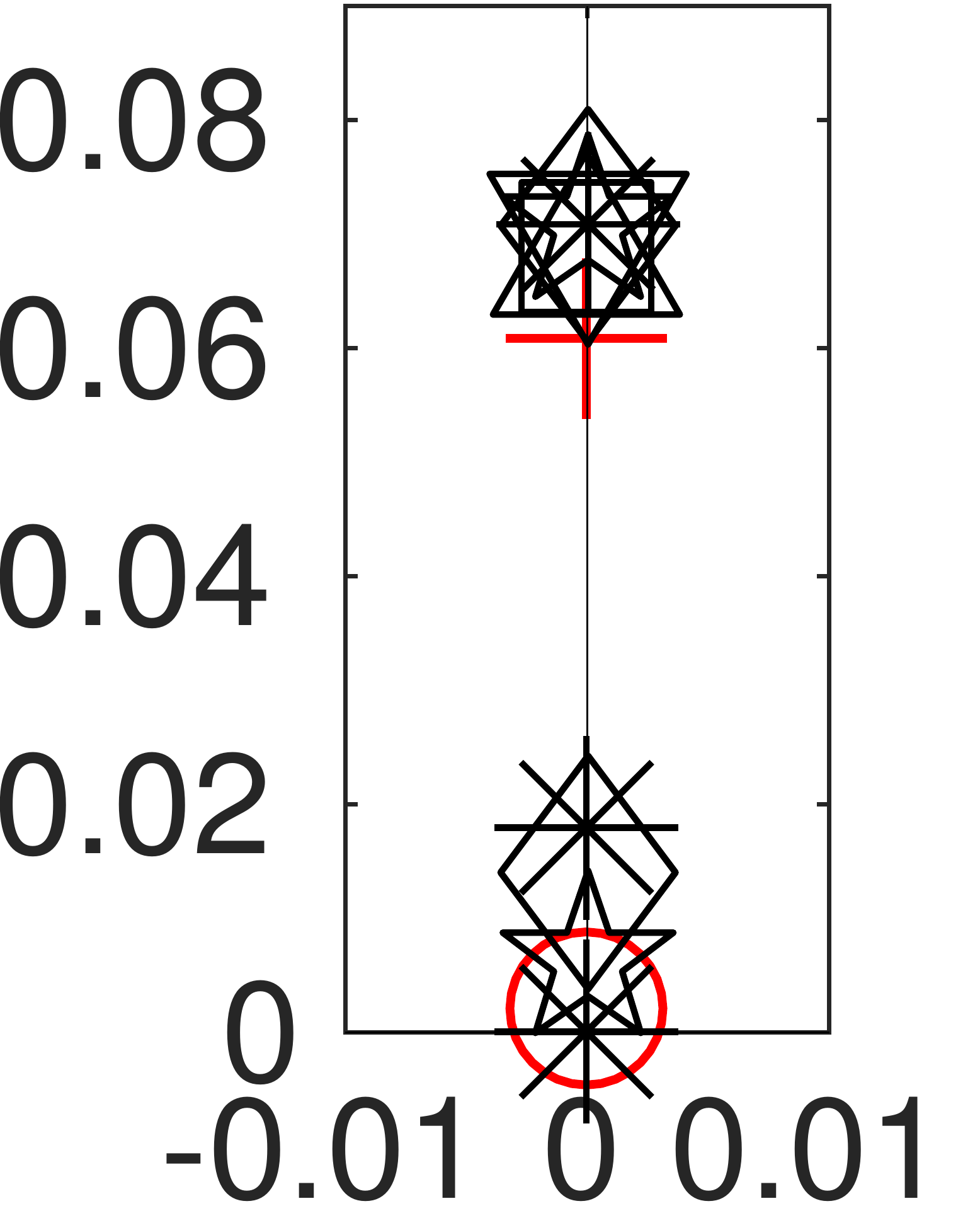}}}
\end{overpic}
}
\caption{\label{fig:Re_conv}
Left: $Im(\omega)$ (growth rate) of the most unstable mode
as a function of the Reynolds number at $Fr=0.4$.
The dashed line correspond to the inviscid solution
as obtained solving the eigenvalue problem.
Right: solutions of the eigenvalue problem at $Fr=0.4$ with fixed $k_x=1.29$ and $k_z=8.53$. Different symbols correspond
to different $Re$ numbers.
The inset corresponds to the thin rectangular region
indicated by the red dashed line in the main graph.
We highlight in red the two lowest $Re$ numbers. 
One sees that starting from 
the third one ($Re=50000$) the value of $Im(\omega)$ saturates
to an asymptotic value.}
\end{figure}

\begin{figure}
\subfigure
{
\begin{overpic}[width=.22\linewidth]{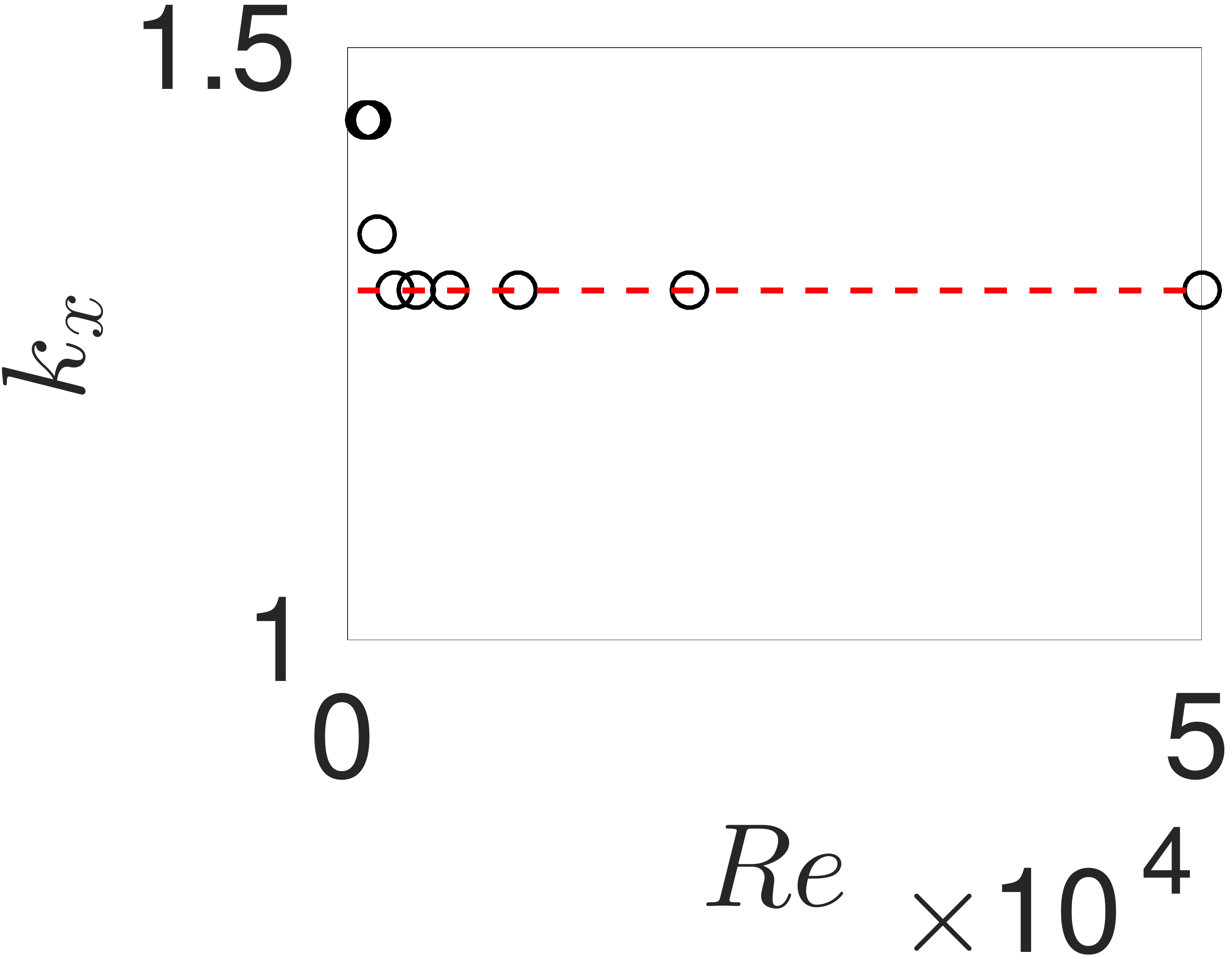}
\end{overpic}
}
\subfigure
{
\begin{overpic}[width=.22\linewidth]{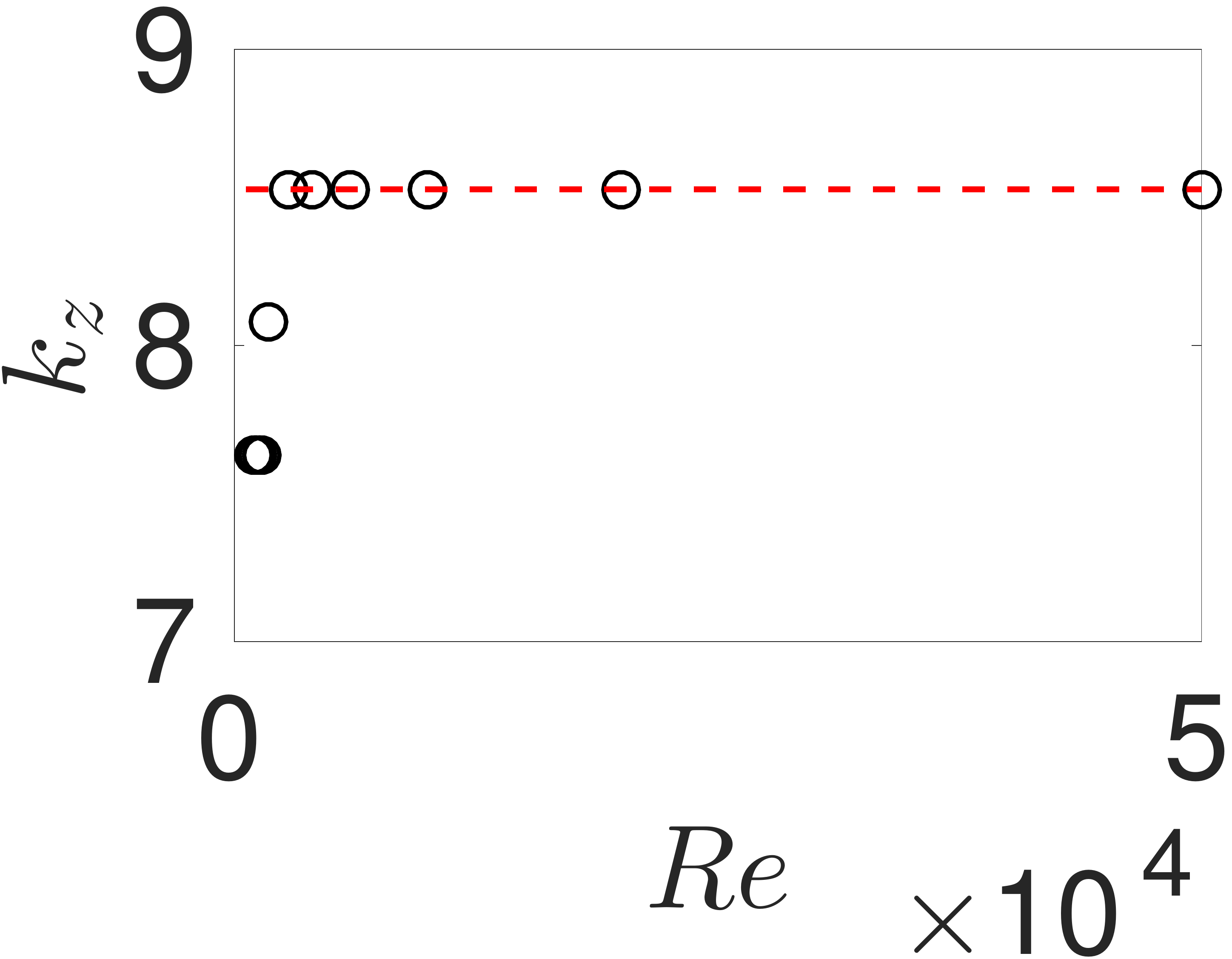}
\end{overpic}
}
\subfigure
{
\begin{overpic}[width=.22\linewidth]{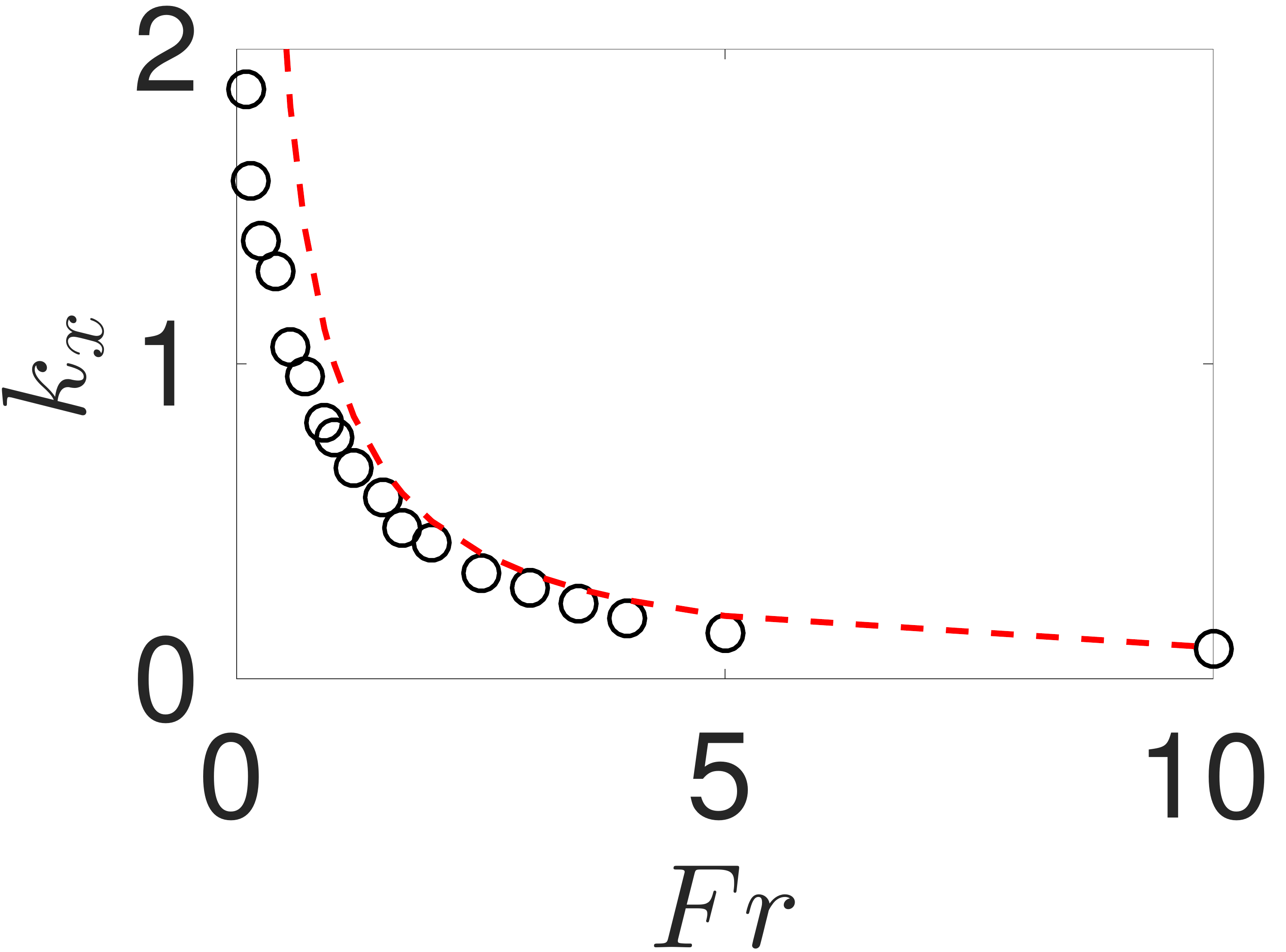}
\end{overpic}
}
\subfigure
{
\begin{overpic}[width=.22\linewidth]{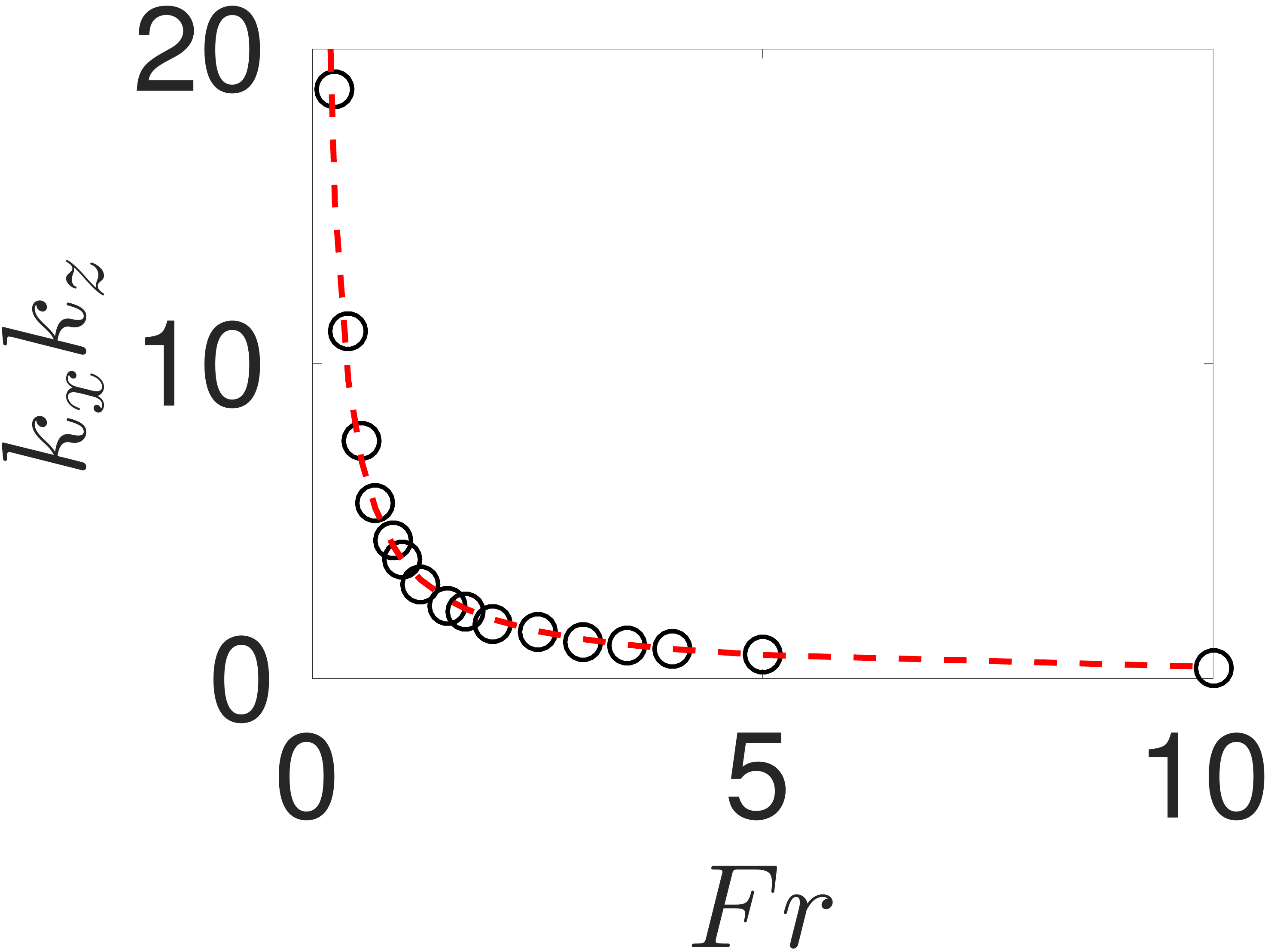}
\end{overpic}
}
\caption{\label{fig:ReFr_cut}
From the left: $k_x$ as a function of $Re$ at $Fr=0.4$, $k_z$ as a function of $Re$ at $Fr=0.4$, $k_x$ as a function of $Fr$ at $Re=10000$ and $k_xk_z$ as a function of $Fr$ at $Re=10000$.
In all the panels circles correspond to the most unstable mode. The dashed line corresponds to a constant
in the first two graphs and to $1/Fr$ in the last two.}
\end{figure}

Finally we want to discuss how the most unstable
mode changes as a function of $Re$ and $Fr$ numbers separately.
In figure \ref{fig:Re_conv} we analyse how the growth rate
$Im(\omega)$ of the most unstable mode (left) changes with the $Re$ number at fixed $Fr=0.4$.
One sees that $Im(\omega)$ rapidly saturates to a constant value. 
This result was confirmed by solving the eigenvalues problem (right) at very high $Re$ number (up to $10^8$) with $(k_x,k_z)$ fixed.
In figure \ref{fig:ReFr_cut} (left) we report the value of $k_x$
and $k_z$ for the most unstable mode as a function of $Re$ at 
fixed $Fr=0.4$.
One sees that both $k_x$ and $k_z$ tend to a constant value.
Let us recall that our approach demands to discretize  
the space $(k_x,k_z)$ which explains why the rate of
this convergence may appear disturbingly abrupt.
Thus we conclude that the observed instability must rely on
an inviscid mechanism and that the inviscid approximation 
is sufficient to capture the spatial $(k_x,k_z)$
and temporal $\omega$ feature of the most unstable mode.

In figure \ref{fig:ReFr_cut} (right) we report the value
of $k_x$ and $k_xk_z$ for the most unstable mode
as a function of the $Fr$ number at $Re=10000$.
The first panel shows that $k_x$ is always
slightly lower than $1/Fr$ (dashed line) which means 
that, for the most unstable mode,
baroclinic critical layers (i.e. $y=\pm 1/kFr$) 
fall close to the boundaries but slightly 
outside the domain boundaries $y=\pm1$, and
are likely not involved in the instability
mechanism.
In the second panel we see that all solutions seem to 
collapse on
the curve $A/Fr$ where $A=\left.k_xk_z\right|_{Fr=1}$
which provides a rule for the spatial pattern
of the most unstable mode and an interesting
limit for further analysis of the pressure equation (\ref{eq:pressure_spectral}).
Finally one should remark that, according to this relationship,
in exploring the stability diagram $(Re,Fr)$, 
the discretization of the wave number (i.e. the step size of the grid $kx,kz$) becomes critical at low $Fr$, while 
the size of the domain $(k_x,k_z)$ becomes critical at high $Fr$.

\subsection{The instability mechanism}\label{sec:inst_mechanism}
\begin{figure}
\centering
\subfigure
{
\begin{overpic}[height=.29\linewidth]{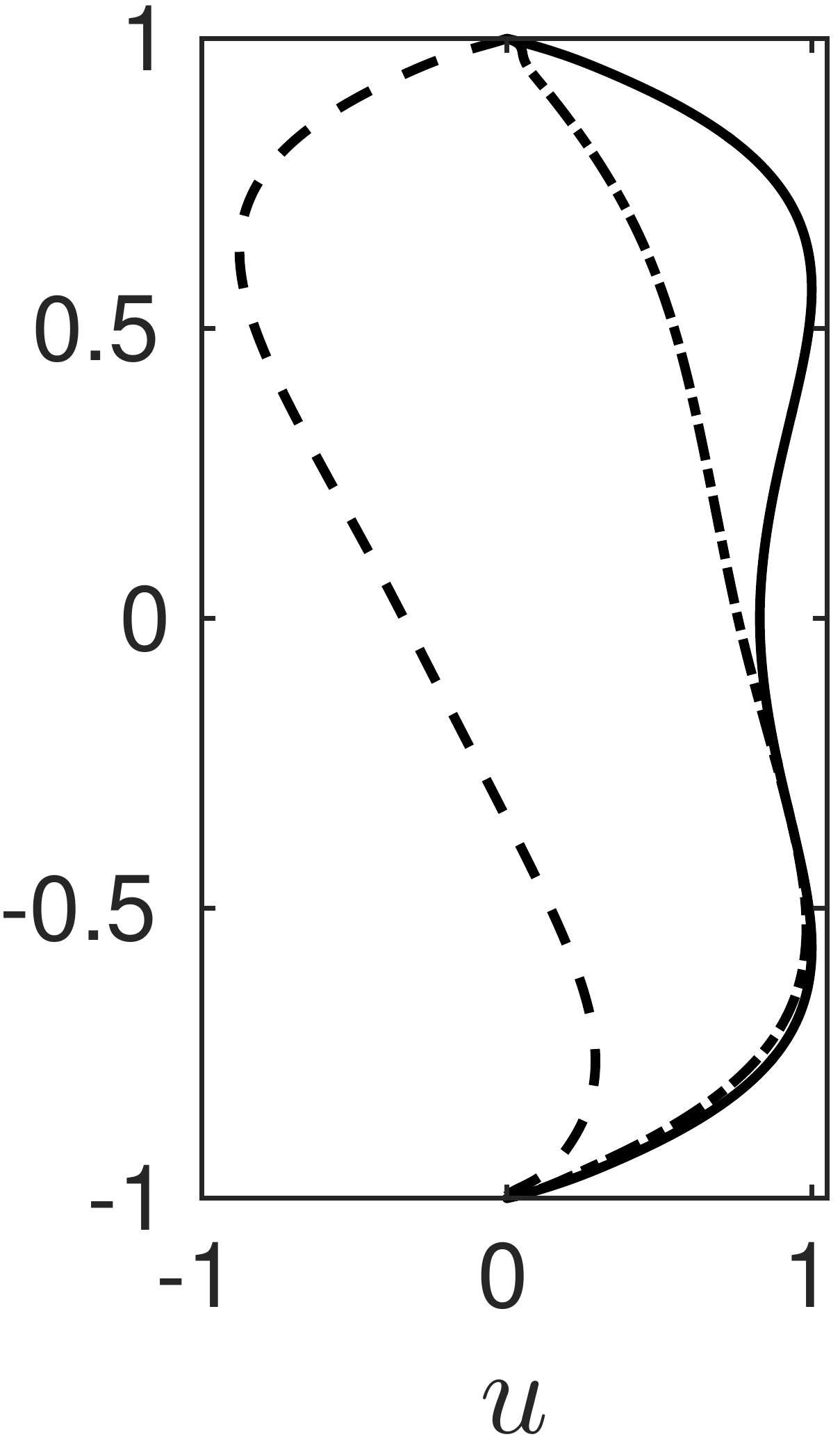}
\put (-5,56) {\rotatebox{90}{$y$}}
\end{overpic}
}
\hfill
\subfigure
{
\begin{overpic}[height=.29\linewidth]{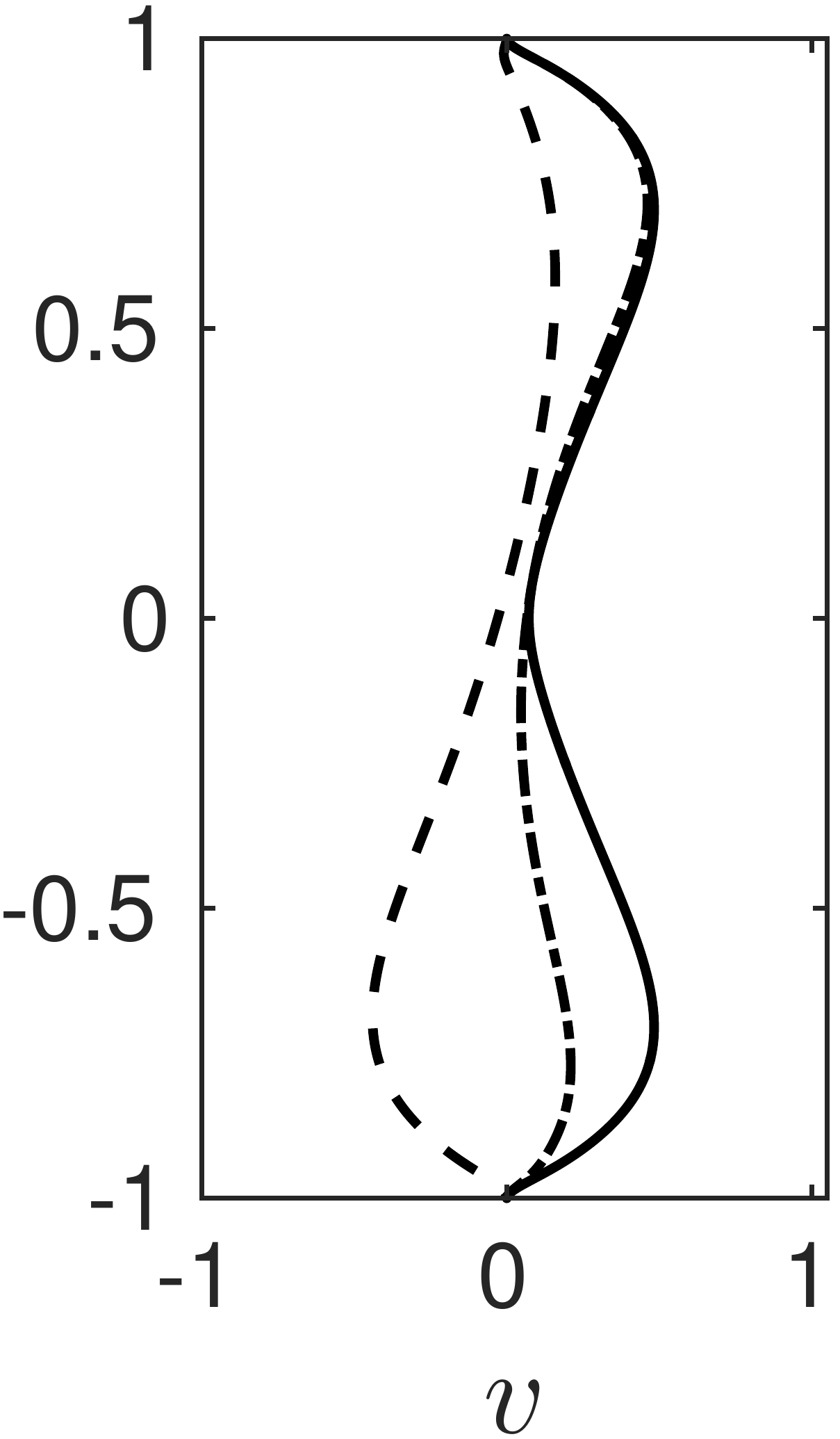}
\end{overpic}
}
\hfill
\subfigure
{
\begin{overpic}[height=.29\linewidth]{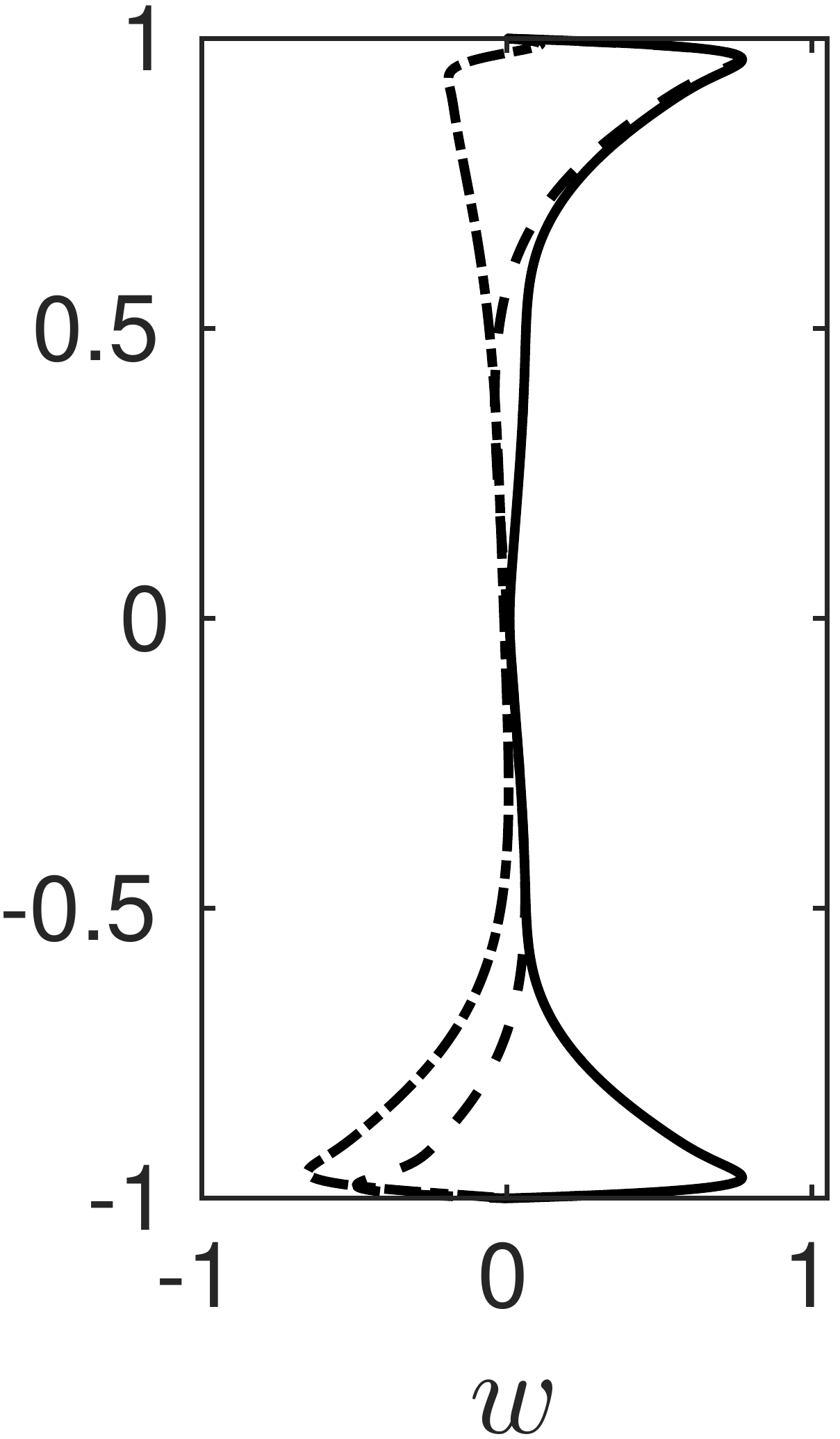}
\end{overpic}
}
\hfill
\subfigure
{
\begin{overpic}[height=.29\linewidth]{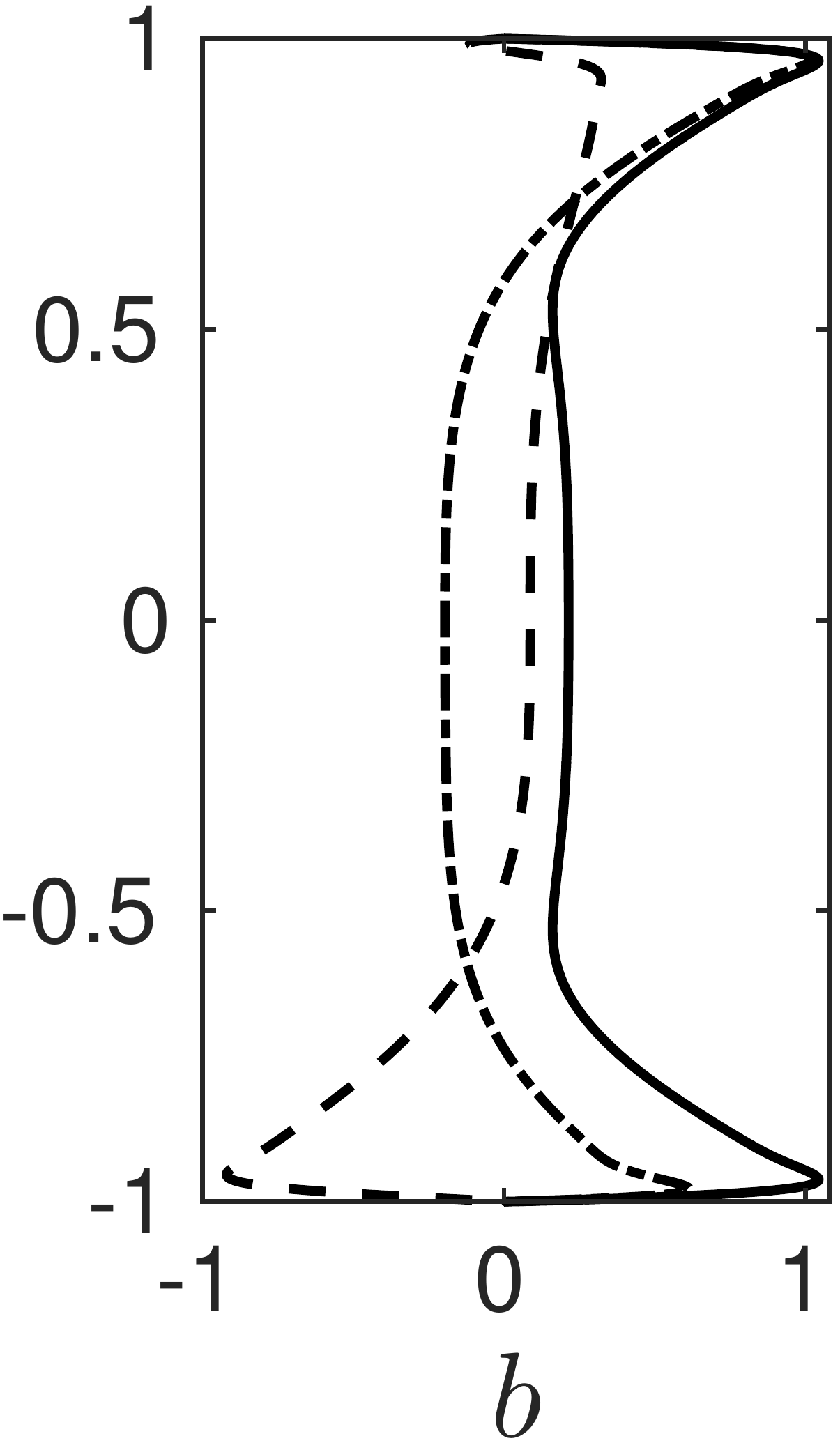}
\end{overpic}
}
\hfill
\subfigure
{
\begin{overpic}[height=.29\linewidth]{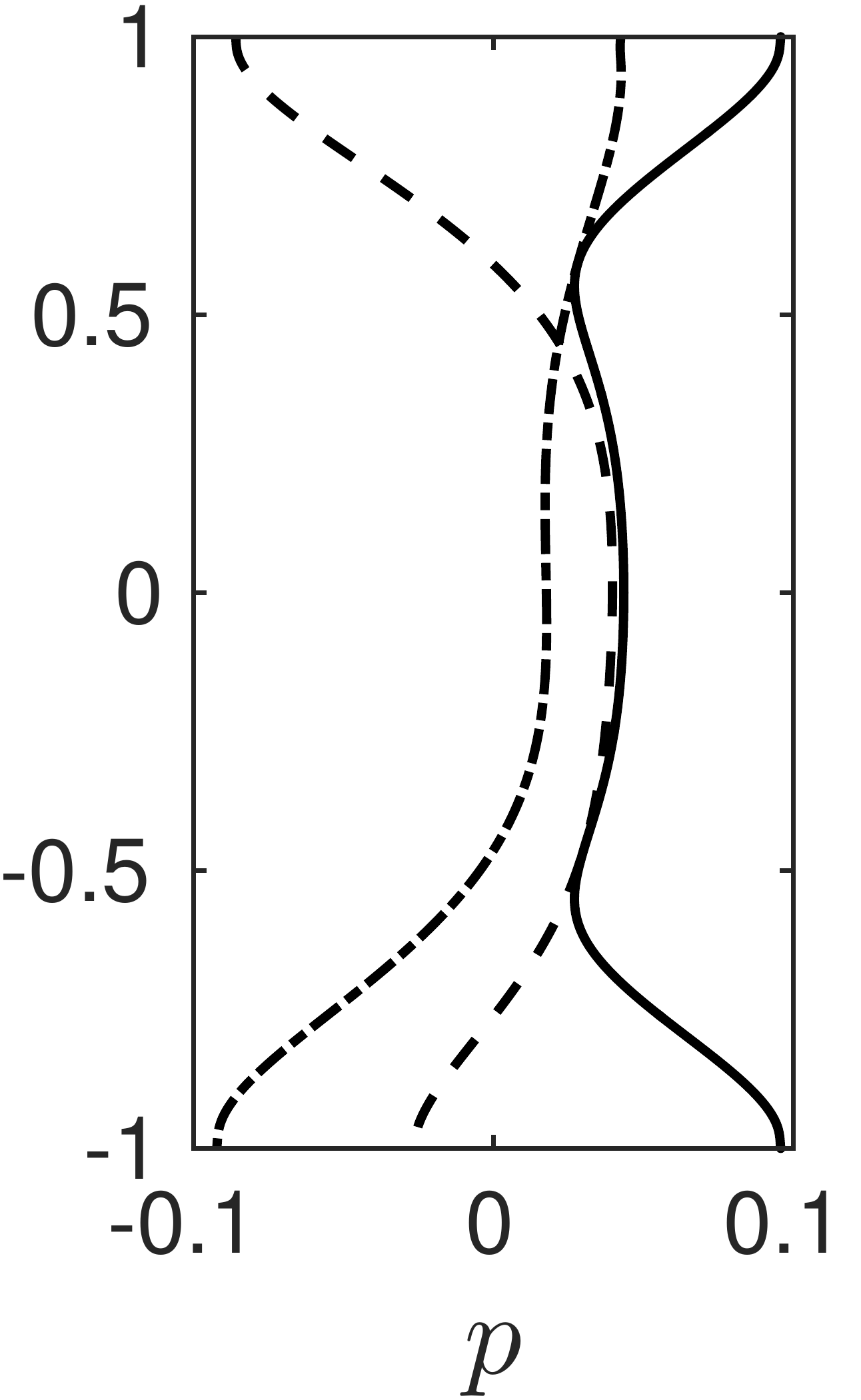}
\end{overpic}
}
\caption{ \label{fig:fields_pbuvw}
Eigenfunctions of the most unstable mode at $Fr=1$, 
$Re=10000$ and wave numbers $k_x=0.767$,
$k_z=4.937$.
Velocity fields and buoyancy are rescaled dividing by the 
maximum value of the perturbation $u$.
Solid lines refer to the absolute value while
dashed and dashed dotted lines refer to the real 
and imaginary part respectively.
}
\end{figure}

\begin{figure}
\centering
\subfigure
{
}
\hfill
\subfigure
{
\begin{overpic}[height=.22\linewidth]{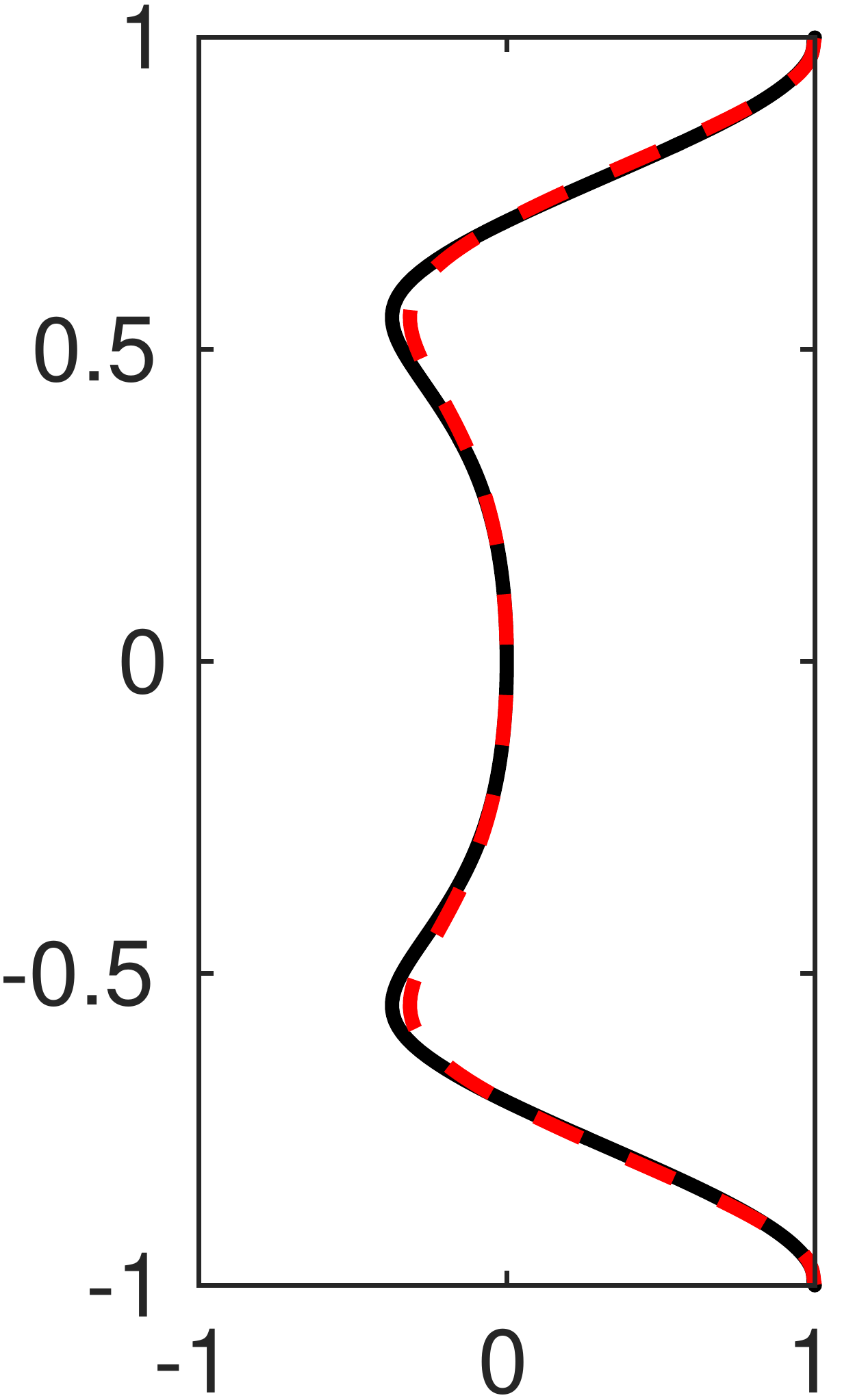}
\put (34,100) {\small a}
\put (34,-7) {\small p}
\put (-5,50) {\rotatebox{90}{$y$}}
\end{overpic}
}
\hfill
\subfigure
{
\begin{overpic}[height=.22\linewidth]{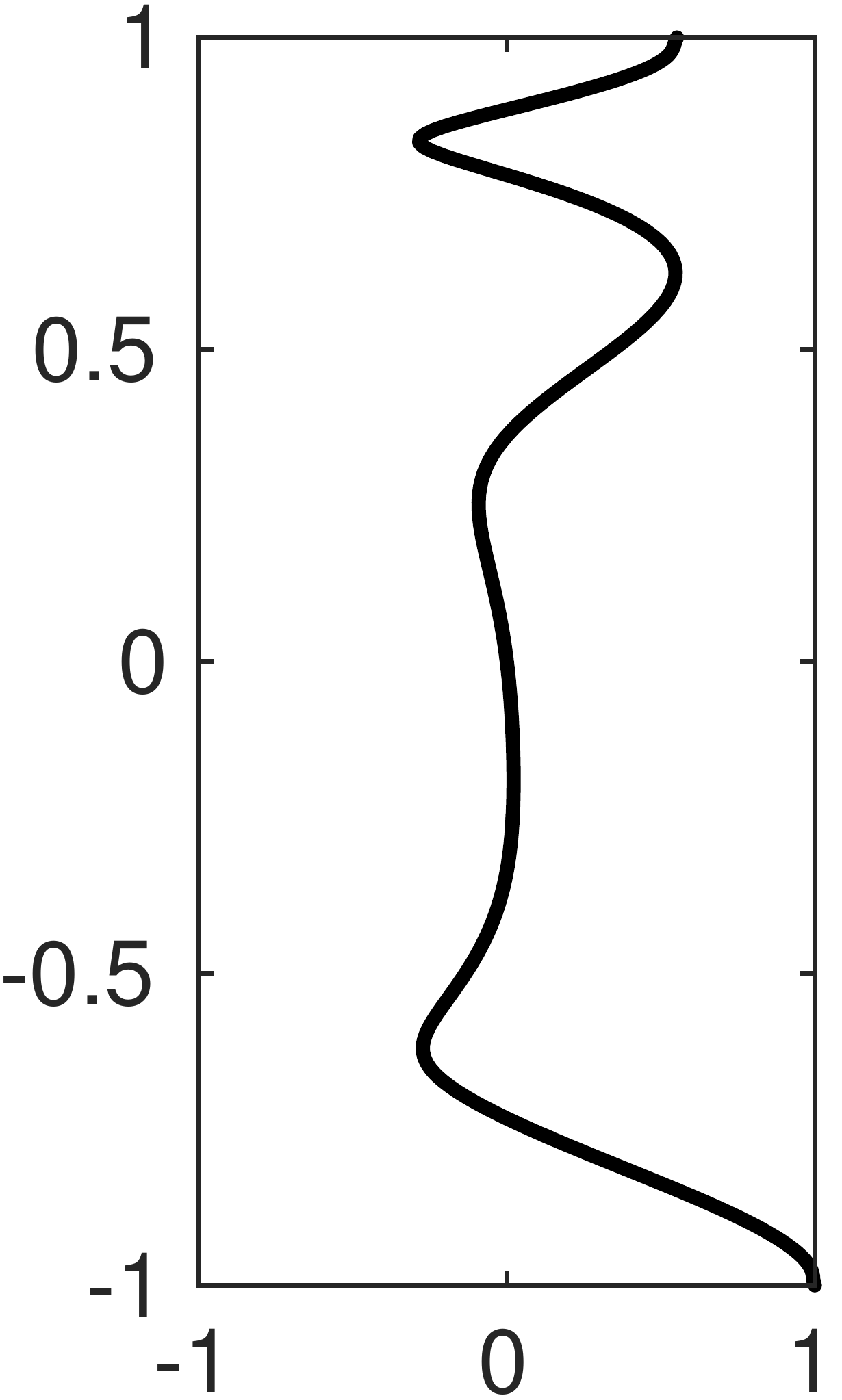}
\put (34,100) {\small b}
\put (34,-7) {\small p}
\end{overpic}
}
\hfill
\subfigure
{
\begin{overpic}[height=.22\linewidth]{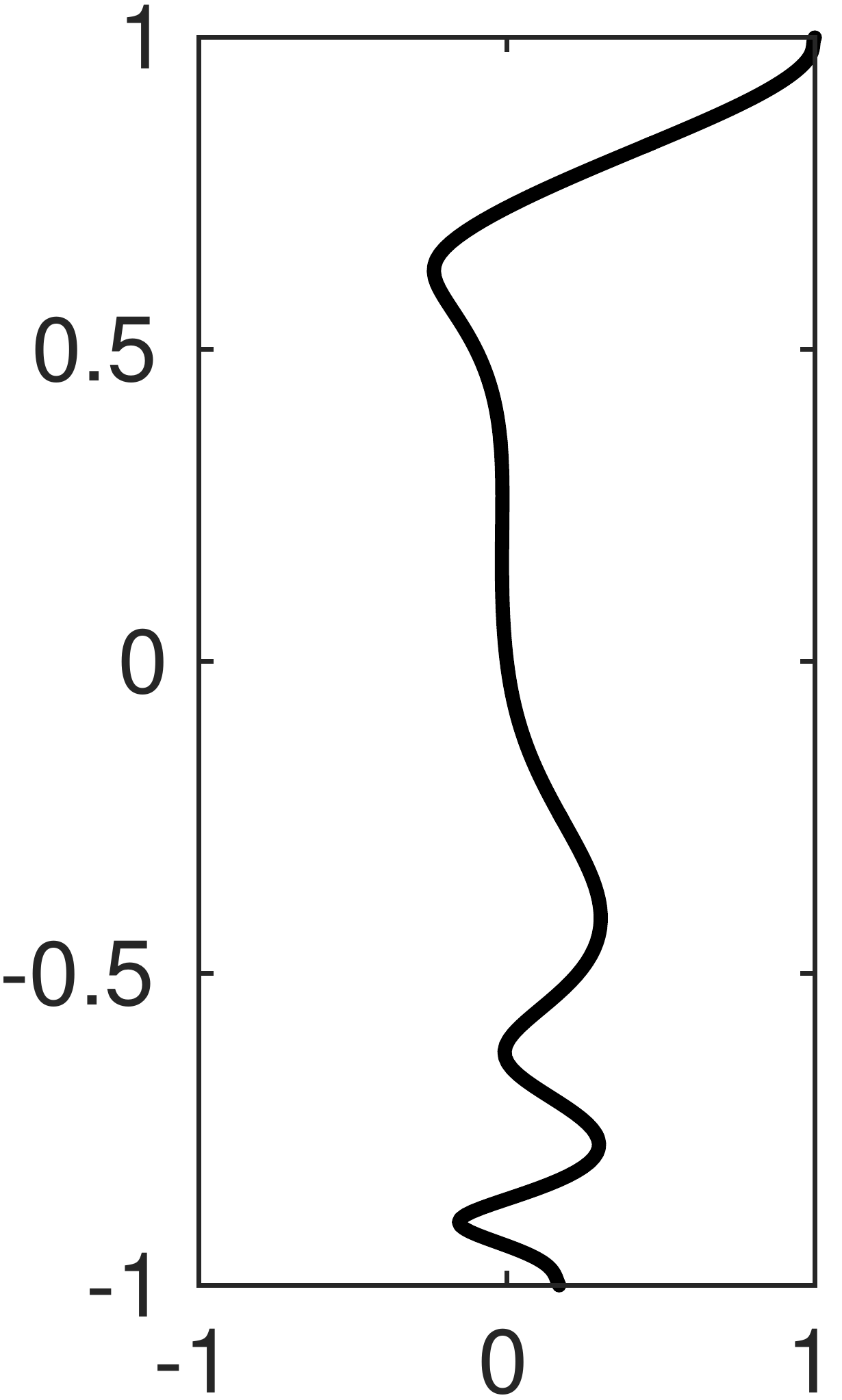}
\put (34,100) {\small c}
\put (34,-7) {\small p}
\end{overpic}
}
\hfill
\subfigure
{
\begin{overpic}[height=.22\linewidth]{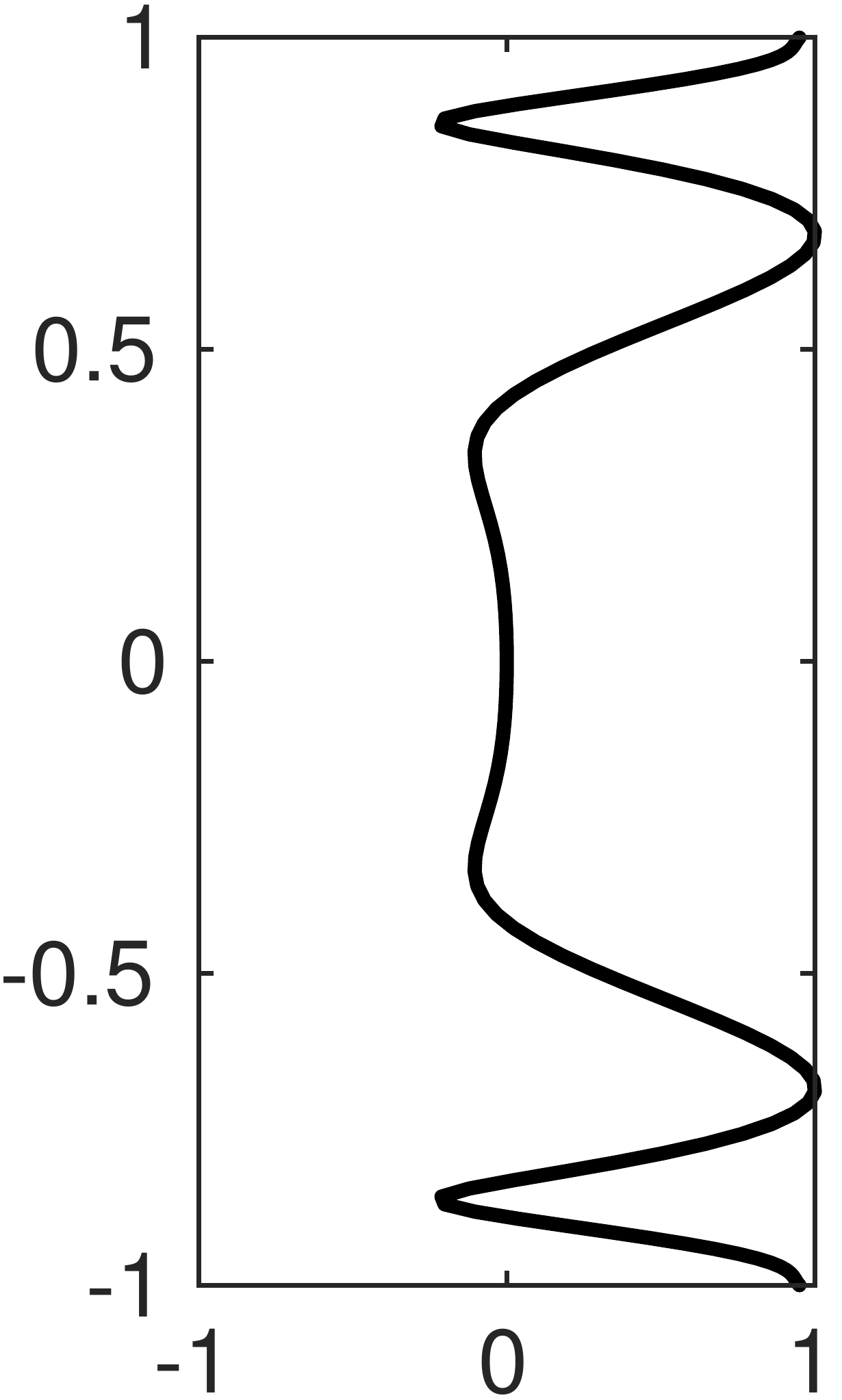}
\put (34,100) {\small d}
\put (34,-7) {\small p}
\end{overpic}
}
\subfigure
{
\begin{overpic}[height=.22\linewidth]{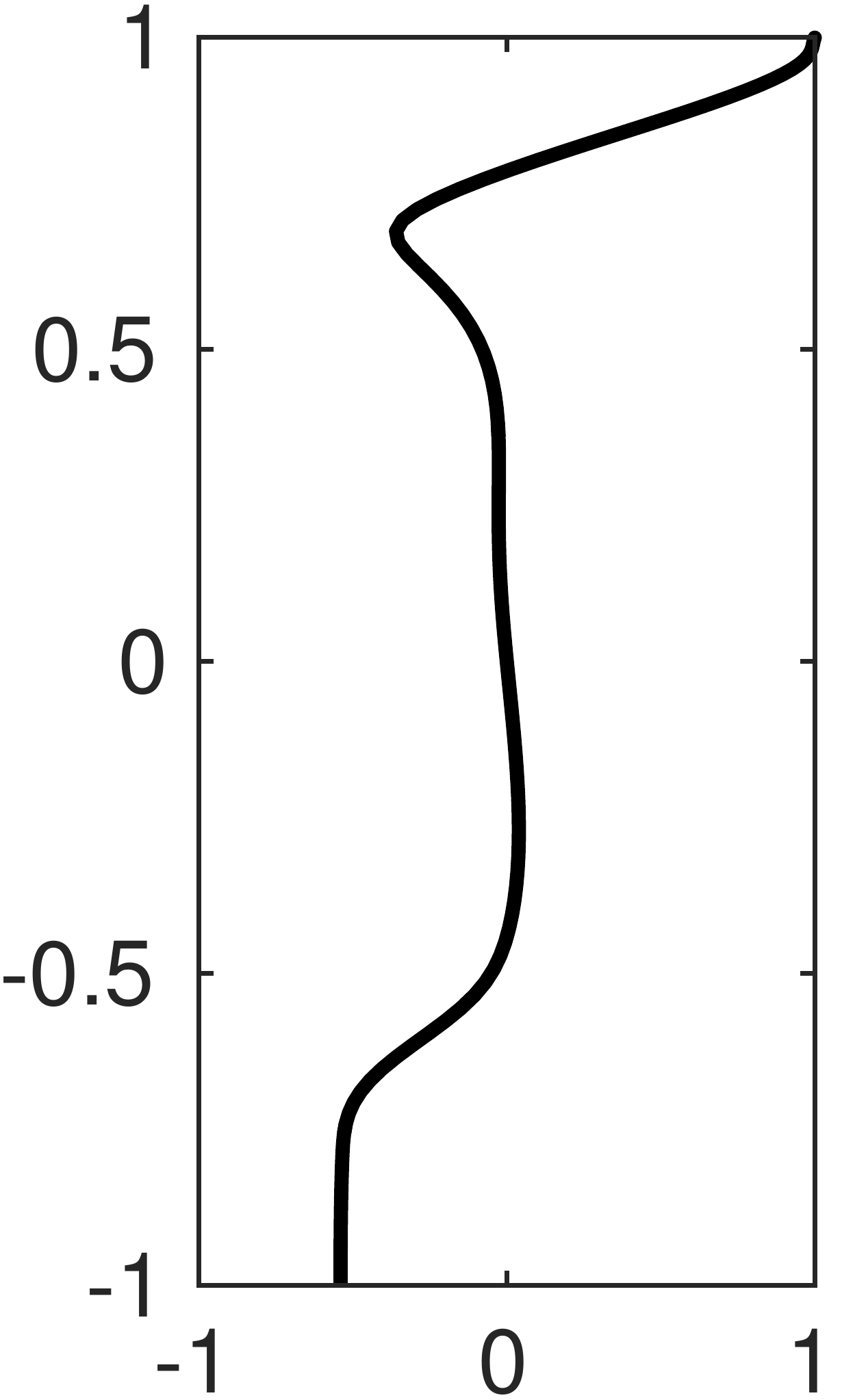}
\put (34,100) {\small e}
\put (34,-7) {\small p}
\end{overpic}
}
\subfigure
{
\begin{overpic}[height=.22\linewidth]{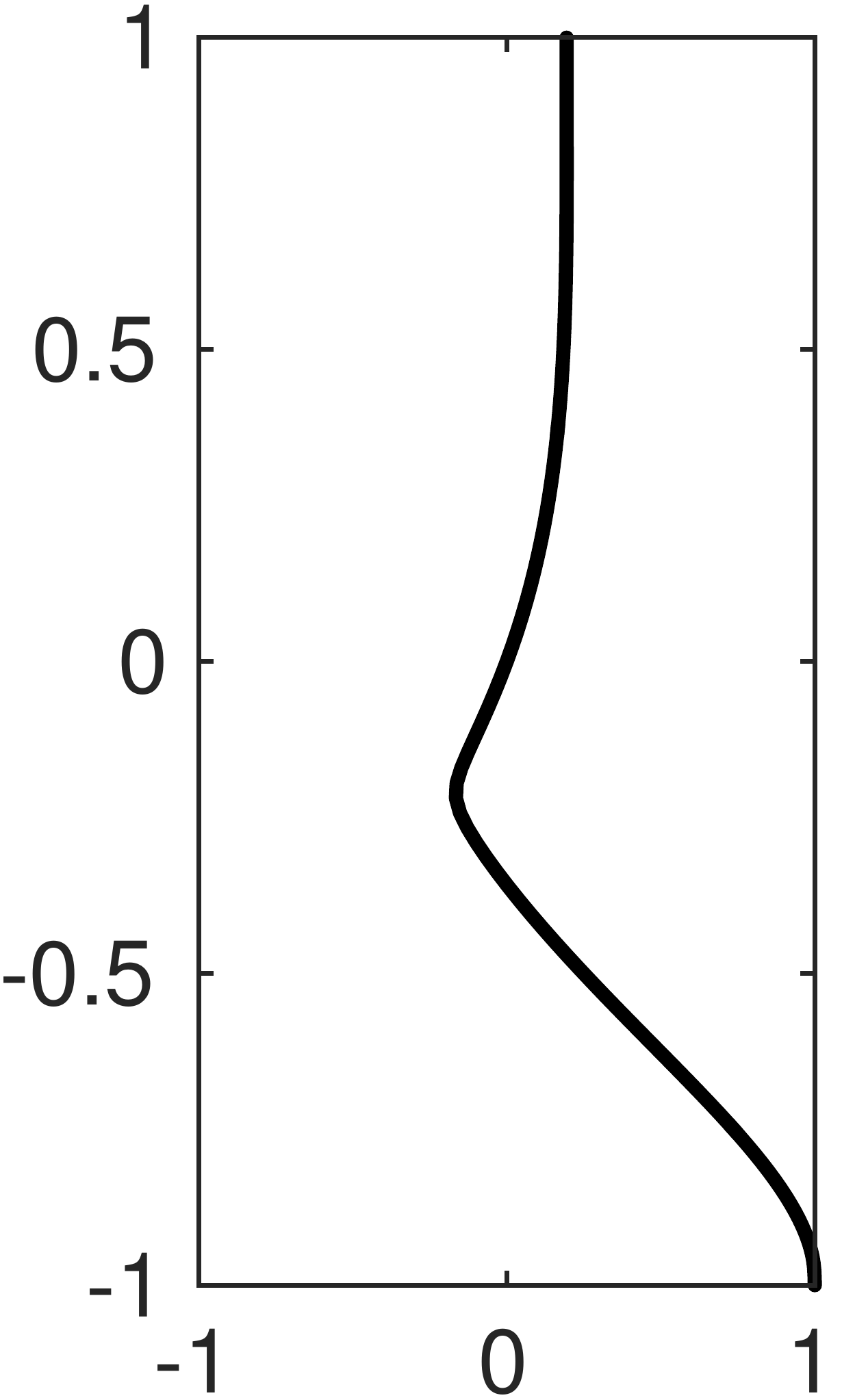}
\put (34,100) {\small f}
\put (34,-7) {\small p}
\end{overpic}
}
\vspace{2.0em}
\caption{\label{fig:pressure_modes_1234}
From the left: 
module of the pressure eigenmodes for $Re=10000$, $Fr=1$
 at $k_z=10.77$ and
$k_x=0.432$ (a), $k_x=0.624$ (b), $k_x=0.719$ (c),
$k_z=0.815$ (d).
The mode (e) correponds to  $k_x=1.055$ and $k_z=8.078$
while the mode (f) is taken at $k_x=0.432$ and $k_z=2.693$.
On (a) the red dashed line corresponds to
the most unstable mode, 
i.e. the same as in the first panel of figure
\ref{fig:fields_pbuvw}. One observes that two modes belonging
to the same branch slightly differ. 
}
\end{figure}
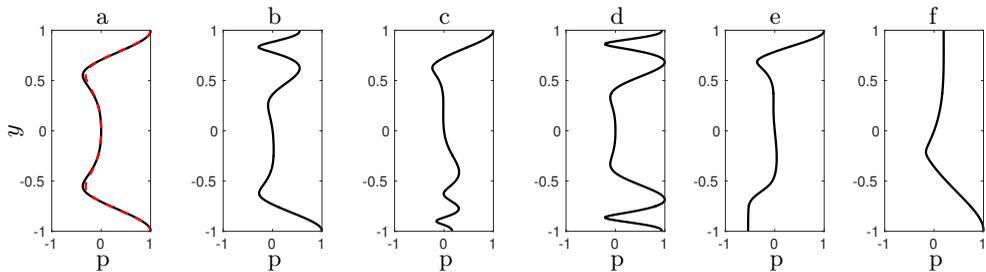

So far we have only focused on the features of the
most unstable mode for a given combination of the
dimensionless numbers $Re$, $Fr$ and a typical domain
in the wave number space ($k_x$,$k_z$).
This characterizes the instability from an operational
point of view but does not say anything about the underlying 
mechanism.
To this end we now analyse the shape of unstable modes.
We have seen that the asymptotic behaviour of the instability
at large $Re$ number
indicates that it relies on an inviscid mechanism.
In the inviscid limit the pseudo-spectral approach 
is far less intelligible because the solution of the
eigenvalue problem contains a large number
of spurious modes with $Im(\omega)>0$,
which makes the detection of genuine
unstable modes extremely difficult.
The idea is then to consider a finite $Re$ number
to keep the eigenvalue problem manageable
but also large enough to capture all
the possible features of the instability diagram.
It turns out that the choice $Re=10000$ fairly responds
to these criteria, thus we focus on the case $Fr=1$ and
$Re=10000$ as a reference one.
In figure \ref{fig:fields_pbuvw} we report the eigenfunctions
of the most unstable mode at $Fr=1$ and $Re=10000$,
which corresponds to the wave numbers $k_x=0.767$ and 
$k_z=4.937$.
One observes that the perturbations of the vertical
velocity $w$ and buoyancy $b$ are more important close
to the boundaries $y=\pm 1$ while 
at the center of the domain $y=0$, 
the velocity perturbation is mainly horizontal.
We consider now a sample mode for each different unstable
branch, for example corresponding to the red spots
we labelled with \textit{a},\textit{b},\textit{c},\textit{d},
\textit{e} and \textit{f} in figure \ref{fig:evalFR02_1_5Re1000}.
In figure \ref{fig:pressure_modes_1234} we compare the pressure 
eigenmode for all different branches.

One observes that the shape of the eigenmodes is significantly 
different in each panel.
Not surprisingly modes from the two stationary branches
\textit{(a)} and \textit{(d)} are symmetric in the
cross-stream direction $y$.
Conversely, travelling modes \textit{(b,c,e,f)} 
are asymmetric but always appear in pairs,
at $\omega_{\pm}=\pm Re(\omega)+iIm(\omega)$,
each mode in a pair being the $y$-mirrored of the
other one with respect to $y=0$.
Also in panel \textit{a} we superpose the pressure
eigenfunction of the most unstable
mode at $k_x=0.767$ and $k_z=4.937$, i.e. the
same as figure \ref{fig:fields_pbuvw} (last panel).
One remarks that two pressure eigenmodes
belonging to the same branch (i.e. the branch \textit{(a)}) 
have basically the same shape.
The scenario we described above is strikingly similar to that
presented by \cite{Satomura1981a} (see e.g. his figure 6)
who analysed the stability of a non-stratified PC 
flow in the shallow water approximation.
In this case the pressure $p$ is replaced by the elevation 
of the free surface $h$ in the analogous of
equation (\ref{eq:pressure_spectral}). 
The author suggested that the instability is produced by the resonance of two Doppler-shifted shallow water waves.
In this picture the wave (stream-wise) phase speed 
$C=\omega/k_x$ of a shallow water wave which 
travels close to one boundary
can be approximated to that of a shallow water wave 
in a fluid at rest plus a Doppler shift, say $U_d$,
which has the sign of the velocity of the considered
boundary.
Two distinct counter propagating waves situated at opposite
boundaries can then have the same phase speed and become
resonant. 
Moreover the resonant wave numbers constitute
a discrete spectrum because rigid walls
make the dispersion relation of (non sheared) waves discrete.
More recently, the same mechanism was also detailed to be responsible for linear instabilities in stratified, rotating plane Couette \citep{Vanneste2007} and stratified Taylor-Couette \citep{Park2013} flows.
We suggest and show below that this interpretation remains valid in our case if we replace
shallow water gravity waves with internal gravity waves.
The dispersion relation of the latter is also discrete
and one has:
\begin{equation}\label{eq:disp_relation_channel_IW}
C^{(n)}_{\pm}=\pm\frac{1}{k_xFr}\sqrt{1-\frac{k_z^2}{k_x^2+k_z^2+n^2\pi^2}},
\end{equation}
where we use the same notation as \cite{Satomura1981a},
i.e. $C=\omega/k_x$. 
Subscript $+$ ($-$) refers to waves propagating in the
positive (negative) direction of the $x$ axis, while
superscript $n$ labels different channel modes.
Note that here the velocity $C$ does not correspond
to the phase velocity of the wave nor to its horizontal component, but it is still the relevant quantity to describe the resonance mechanism.

\begin{figure}
\centering
\subfigure
{
\begin{overpic}[height=.35\linewidth]{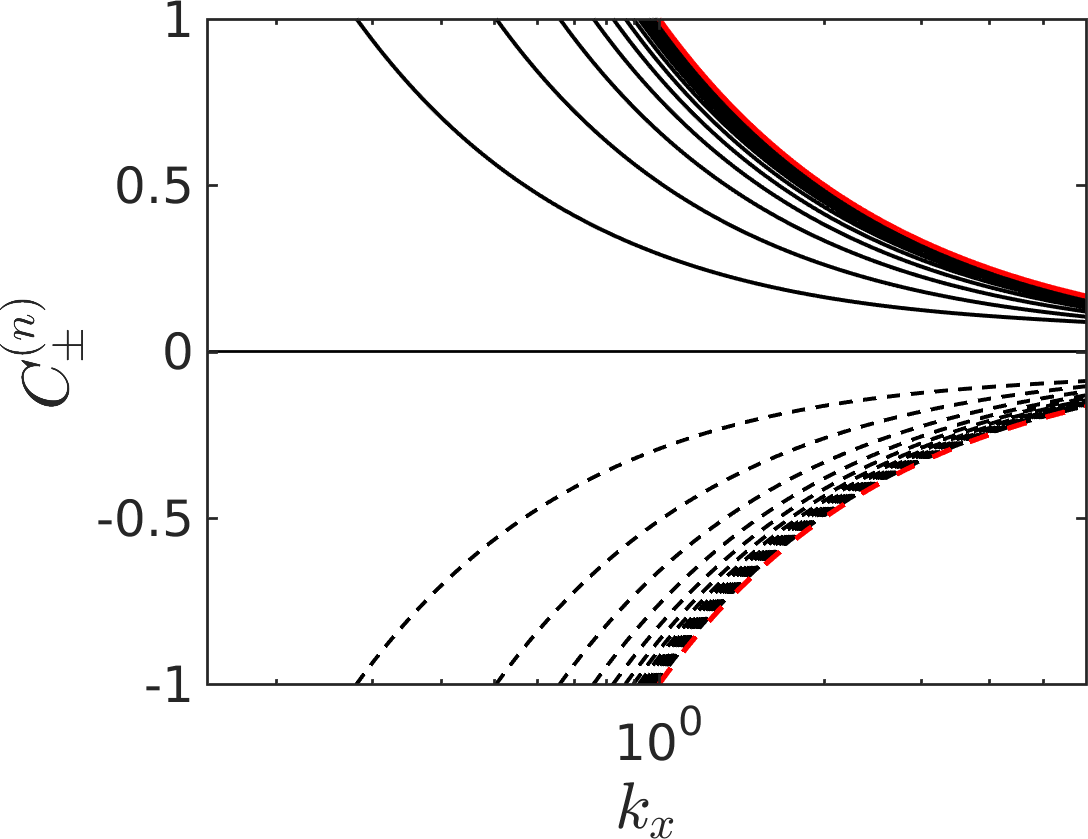}
\put (70,18) {\small $Fr=1$}
\end{overpic}
}
\hfill
\subfigure
{
\begin{overpic}[height=.35\linewidth]{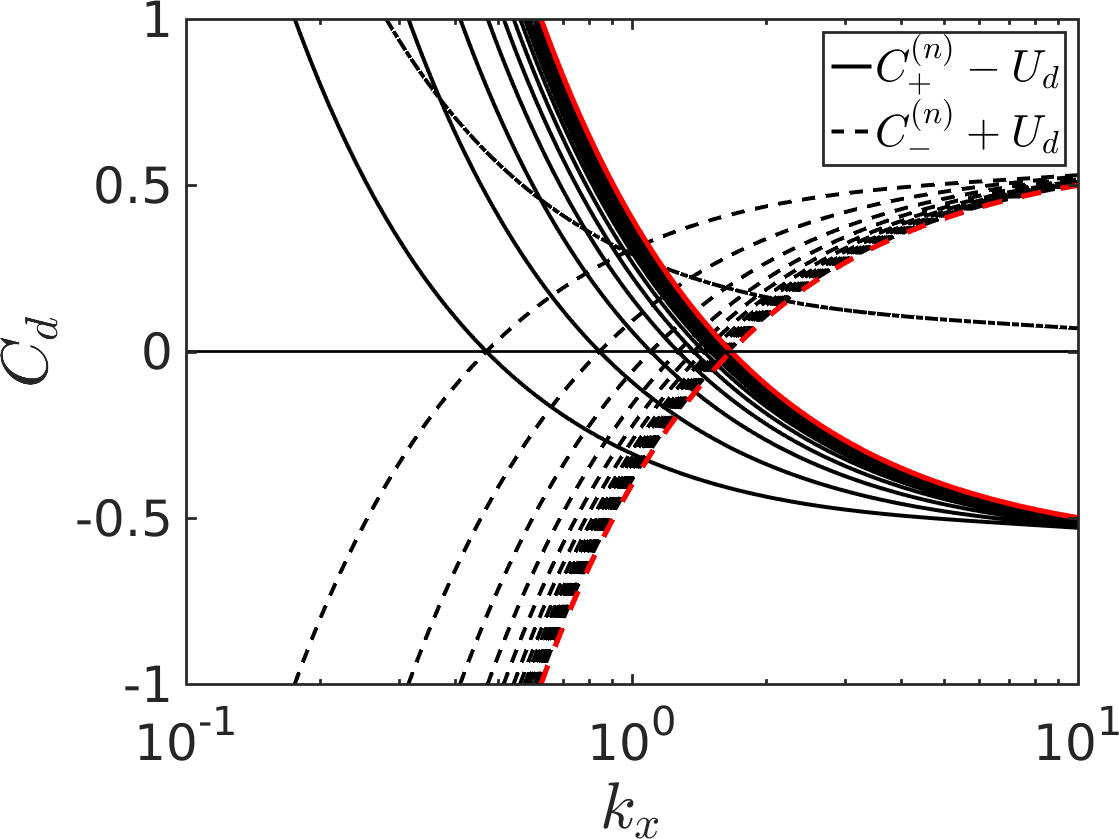}
\put (70,18) {\small $Fr=1$}
\put (41,46) {\scriptsize \textcolor{red}{\textbf{a}}}
\put (48,40) {\scriptsize \textcolor{red}{\textbf{b}}}
\put (48,50) {\scriptsize \textcolor{red}{\textbf{b}}}
\put (51,37) {\scriptsize \textcolor{red}{\textbf{c}}}
\put (51,51.5) {\scriptsize \textcolor{red}{\textbf{c}}}
\put (52,45) {\scriptsize \textcolor{red}{\textbf{d}}}
\put (40,65) {\scriptsize \textcolor{red}{\textbf{f}}}
\put (53,31) {\begin{tikzpicture}
\coordinate (O) at (13,8);
\draw[line width =1.4,-,red] (O) circle (.2);
\end{tikzpicture}}
\end{overpic}
}
\hfill
\subfigure
{
\begin{overpic}[height=.35\linewidth]{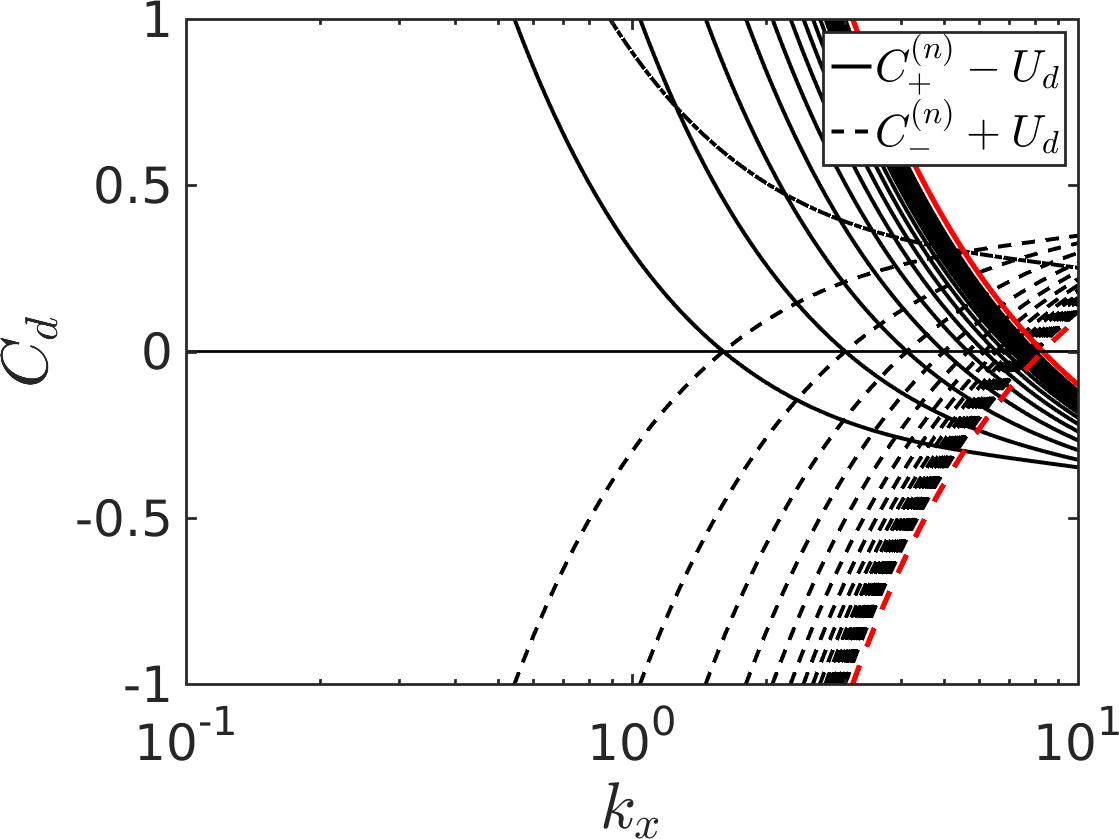}
\put (20,18) {\small $Fr=0.2$}
\end{overpic}
}
\hfill
\subfigure
{
\begin{overpic}[height=.35\linewidth]{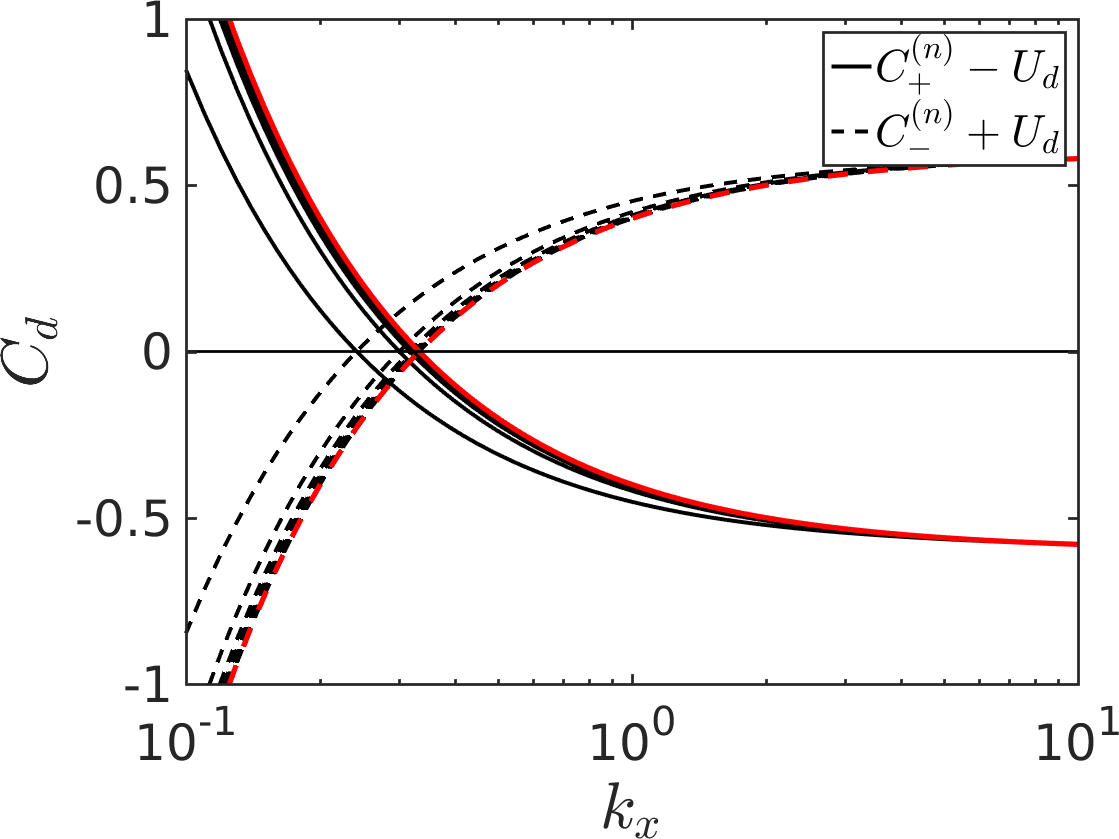}
\put (70,18) {\small $Fr=5$}
\end{overpic}
}
\caption{\label{fig:disp_shifted}
Top: velocity $C^{(n)}_{\pm}$ as a function of $k_x$ 
as given by equation (\ref{eq:disp_relation_channel_IW}) at $k_z=10.77$ and $Fr=1$ 
and  for the first $n=1$ to $20$ prograde (solid lines) and retrograde (dashed lines) 
confined internal gravity waves in the absence of 
mean flow (left), and the Doppler shifted velocities $C^{(n)}_{\pm}\mp U_d$, where we set arbitrarily $U_d=0.6$ (right).
The red lines indicate the limit of the dispersion relation 
for $n\rightarrow\infty$, the
crossing is situated at $k_x=1/Fr U_d$.
We indicate with letters the resonances (crossing points)
corresponding to different modes.
Note that resonances (b) and (c) come in symmetric pairs.
The red circle indicates the proximity of a degenerate
crossing point.
The single dashed-dotted line corresponds to $C^{(1)}_+$, i.e. a non transported prograde $n=1$ mode.
Bottom: Same as top right but at $Fr=0.2$, $k_z=18$ (left) and 
$Fr=5$, $k_z=3$ (right).
}
\end{figure}
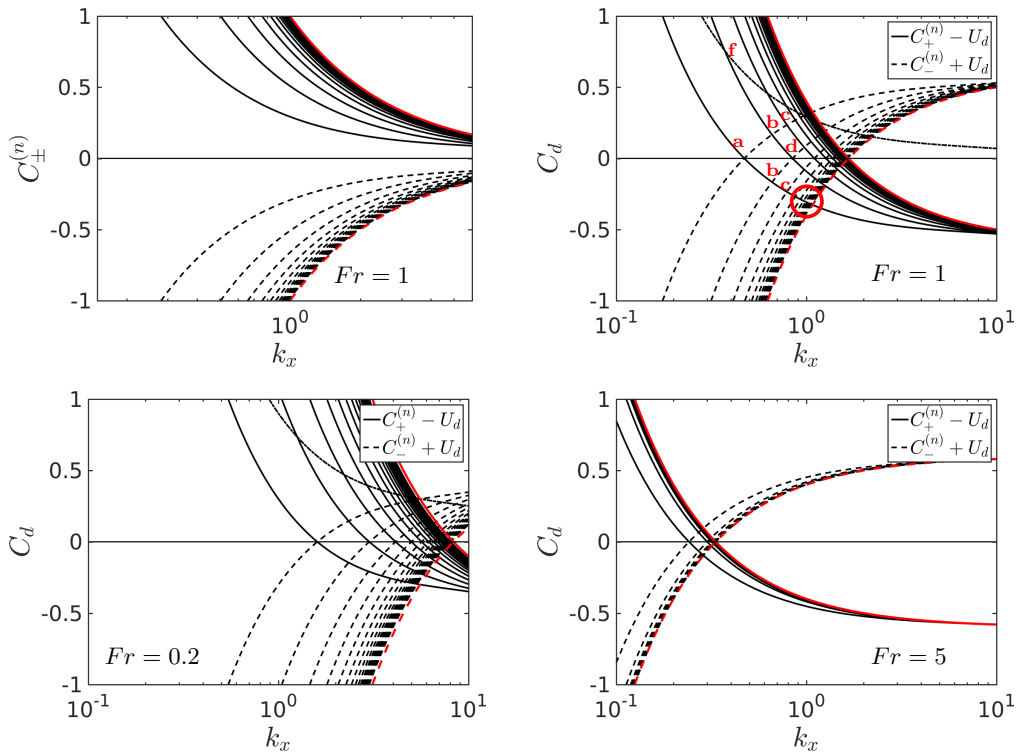

In the first panel of figure \ref{fig:disp_shifted} we
report the value of $C^{(n)}_{\pm}$
as a function of $k_x$ at $k_z=10.77$ and $Fr=1$.
In the second panel we show both Doppler shifted velocities 
$C_d=C^{(n)}_{\pm}\mp U_d$ where we consider prograde and retrograde waves moving upstream close to 
opposite boundaries and transported by the local mean flow. 
At this stage the value of $U_d$ is an adjustable 
parameter $-1\leq U_d \leq 1$
and was fixed to $U_d=0.6$.
One remarks that in the fixed frame (i.e $U_d=0$), prograde waves (solid lines) are well separated from retrograde waves (dashed lines).
On the contrary for Doppler shifted waves, there exists a discrete set of resonant $k_x$ where two curves of different
type cross each other.
At $Fr=1$ the first crossing (resonance) happens at $C_d=0$ and close to $k_x=0.4$, which is consistent with the appearance of
the first stationary mode (a) in the stability map of figure 
\ref{fig:evalFR02_1_5Re1000}. 
The next two resonances happen at a non zero 
value of $C_d$ which coherently recovers the appearance of the
first two oscillating modes (b) and (c) at larger $k_x$
in figure \ref{fig:evalFR02_1_5Re1000}.
The following crossing happens again at $C_d=0$, which confirms
the appearance of a fourth (stationary) unstable branch in
figure \ref{fig:evalFR02_1_5Re1000} when moving along $k_x$
and at constant $k_z=10.77$.
Looking back at figure \ref{fig:pressure_modes_1234} the mode (f) appears as a half of the mode (a) thus we speculate that
the corresponding resonance originates from the crossing
of a Doppler shifted wave and a non transported wave
(i.e. one for which $U_d=0$)
situated at the center of the domain (i.e. the dashed dotted
lines in figure \ref{fig:disp_shifted}).
Modes (e) do not originate from a resonance, consequently they
are not indicated in figure \ref{fig:disp_shifted}.
A closer inspection of the velocity field suggests that in this case
the baroclinic critical layers are excited,
and the instability relies on a different mechanism.
This hypothesis is consistent with the fact that the mode (e) 
(see figure \ref{fig:evalFR02_1_5Re1000}) belongs to a
region which mainly extends at $k_x>1/Fr$ where critical layers
can fit within the domain.
Now that we have possibly explained the origin of all the
 distinct modes as resulting from a degeneracy 
 of the Doppler shifted frequency, 
we want to show that this picture allows to fully capture 
the shape of
the unstable branches in figure \ref{fig:evalFR02_1_5Re1000}.
First one should recall that for a given channel mode 
(i.e. $n=const$), the dispersion relation of 
internal gravity waves
(\ref{eq:disp_relation_channel_IW}) is a function of 
two variables
$k_x$ and $k_z$, hence a surface.
It follows that degeneracy occurs indeed on the intersection 
of two surfaces (i.e. not two curves) which is a curve 
(i.e. not a single point).
The latter explains why the shape of the unstable branches
in figure \ref{fig:evalFR02_1_5Re1000} appears elongated in
one direction and constrained in the orthogonal one. 
In the particular case of a stationary mode, one can easily 
deduce the equation of such a curve from the dispersion
relation (eq. \ref{eq:disp_relation_channel_IW}) modified
by the Doppler shift $U_d$. We find: 
\begin{equation}\label{eq:disp_relation_sato}
k_z=\mathcal{F}(k_x,n,Fr,U_d)=\sqrt{\frac{n^2\pi^2}{U_d^2Fr^2k_x^2}-k_x^2-n^2\pi^2+\frac{1}{Fr^2U_d^2}}.
\end{equation}
In figure \ref{fig:evalFR02_1_5Re1000} we have superimposed the
value of $\mathcal{F}$ to the map of the growth rate 
$Im(\omega)$ at different $Fr$ numbers and for $n=1$ and $2$.
One observes the agreement is not only qualitative,
for example $\mathcal{F}$ reproduces the trend $k_xk_z\approx const$ observed before, but also quantitative, 
because fixing a unique value of $U_d=0.6$, we are able to predict
the position of almost all the unstable stationary branches. 

Finally we show that the mechanism we describe above allows
to predict the boundaries of the unstable region.
If we look back at figure \ref{fig:disp_shifted} one
observes that instability appears 
at a finite value of $k_x$, say $k^{inf}$, where 
\begin{equation}\label{eq:min_boundary}
C^{(1)}_{+}(k_x=k_x^{inf},n=1,k_z,Fr)=U_d
\end{equation}
and must disappear when the 
envelopes of prograde and retrograde modes (red lines)
cross each other, at $k_x^{sup}=1/FrU_d$.
Note that the latter upper boundary is independent of $k_z$. 
Conversely the lower boundary can be arbitrarily reduced,
for example $k_x^{inf}\rightarrow \infty$ provided that $k_z\rightarrow \infty$. Nonetheless any finite $Re$ number
will likely inhibit an instability happening
at large wave number $k_z$.
We conclude that according to the proposed resonance mechanism,
the instability is triggered by perturbations 
which are not stream-wise invariant
(i.e. $k_x\neq 0$), and at stream-wise wave number 
$k_x<1/FrU_d$.
Looking at the growth rate diagrams of figure 
\ref{fig:evalFR02_1_5Re1000} one actually sees that 
$k^{sup}_x$ tends to overestimate the upper bound
of the unstable region.
A closer estimation of the latter may then be given by
the crossing of the prograde (retrograde) $n=1$ Doppler shifted mode with the envelope of the retrograde (prograde) waves, 
which happens at $k_x<k^{sup}_x$, in the region we highlighted
with a red circle in figure \ref{fig:disp_shifted}.
The idea is that multiple degeneracy may inhibit 
the resonance mechanism.
Lower panels in 
figure \ref{fig:disp_shifted} illustrate the
same resonance mechanism for $Fr=0.2$ and $Fr=5$. 
The bottom left panel ($Fr=0.2$)
confirms that the instability range is extended and pushed 
at larger $k_x$ for small $Fr$ number (i.e high stratification).
Conversely the bottom right panel ($Fr=5$), shows that the
region where resonances take place,
both shrinks and is constrained to smaller $k_x$.
Ultimately $k^{inf}_x$ and $k^{sup}_x$ collide 
in the limit $Fr\rightarrow \infty$ and the
instability likely disappears or at least reduces to an
infinitely narrow range in $k_x$.
Note that the results above suggest that the upper boundary 
of the unstable region in figure \ref{fig:evalFrRe_all} is 
intrinsic to the instability mechanism, 
while lower boundary is controlled by the $Re$ number:
at small $Fr$ instability appears at larger $k_x$, thus
larger $k_z$ and is then more sensitive to viscous dissipation.
To conclude this section we recall that 
if the growth rate varies
with the $Re$ number and different branches appear at different 
$Re$ numbers, the value of $C=\omega/k_x$ on a same branch
is approximately constant and almost does not vary with the 
$Re$ number.
This supports the hypothesis of a resonance and confirms
that the appearance of the most unstable stationary and oscillating modes relies on an inviscid mechanism.
\subsection{Effect of the Schmidt number}\label{sec:Sc_effects} 
\begin{figure}
\subfigure
{
\begin{overpic}[height=.3\linewidth]{
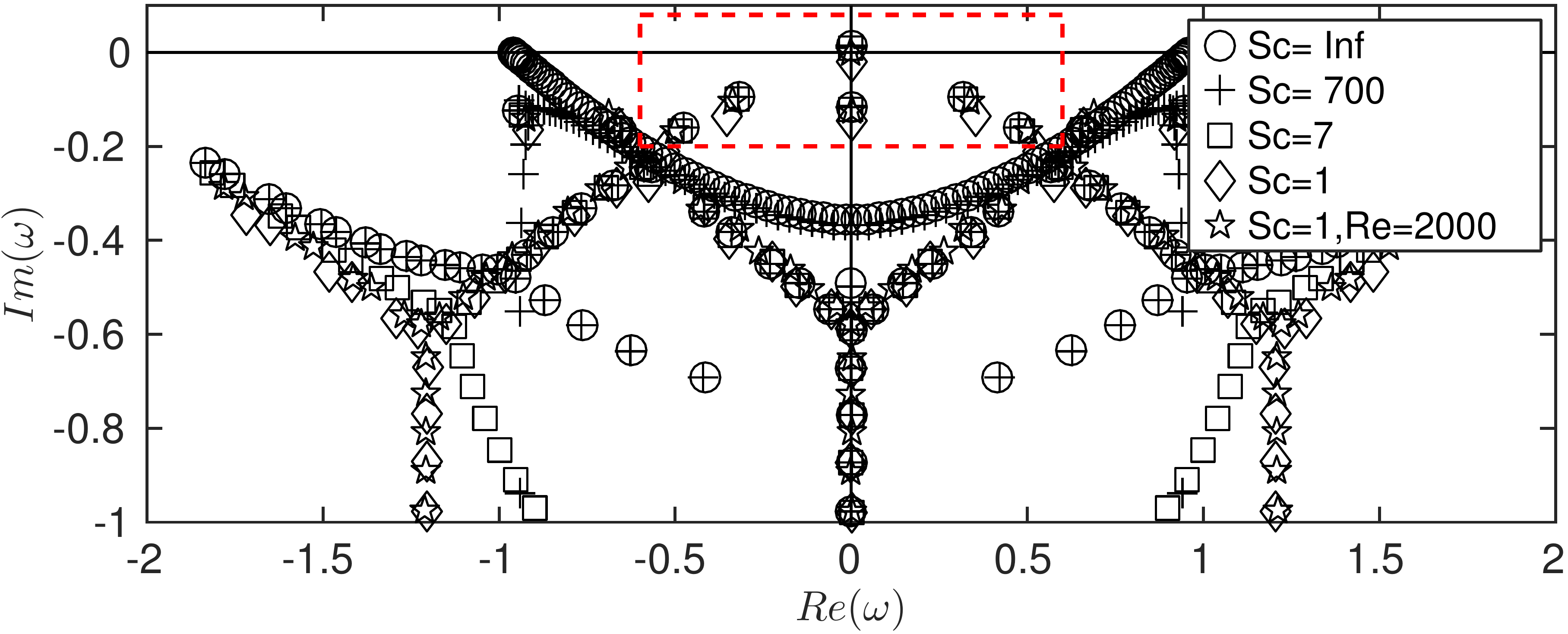}
\end{overpic}
}
\hfill
\subfigure
{
\begin{overpic}[height=.3\linewidth]{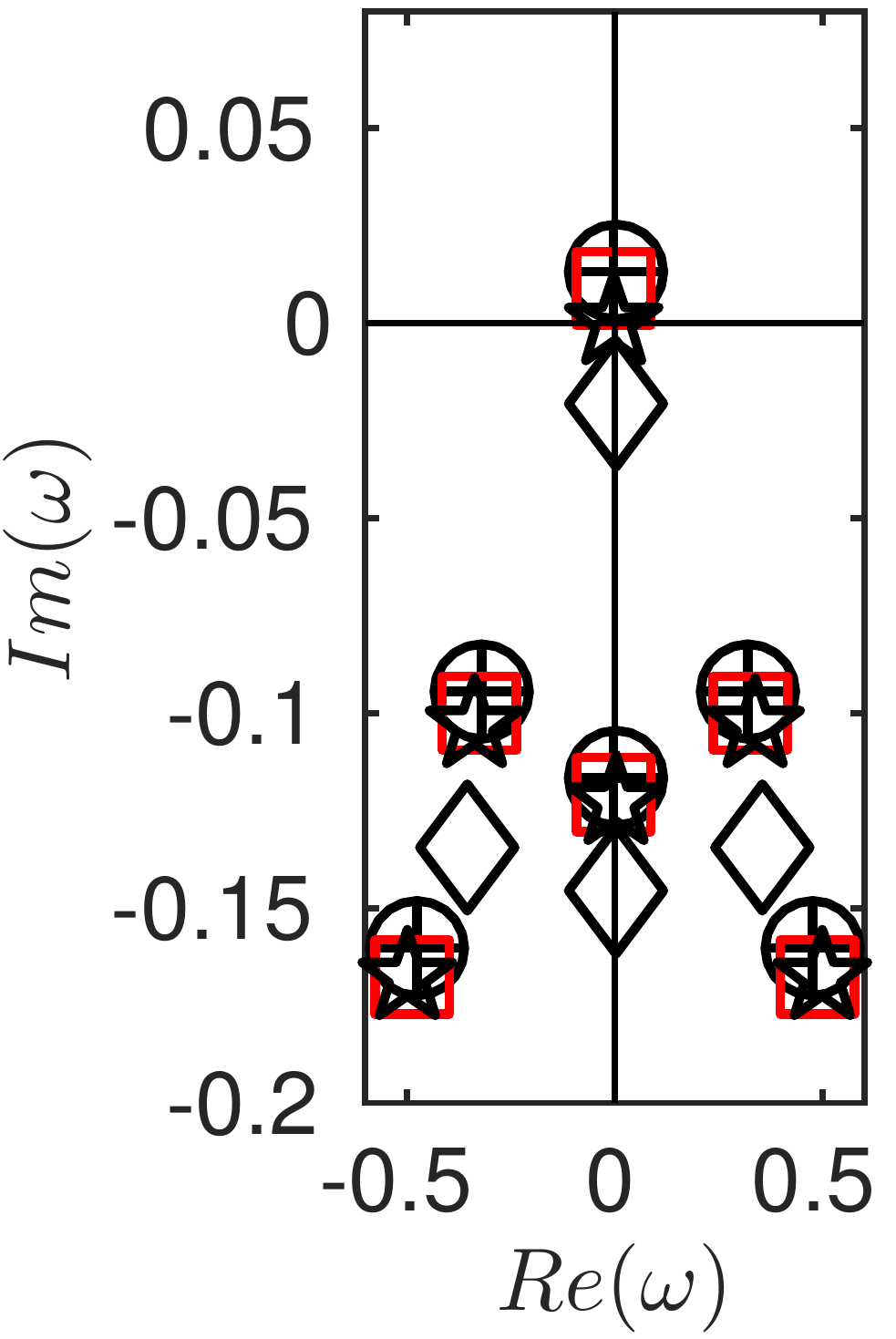}
\end{overpic}
}
\caption{\label{fig:Sc_effects}
Left: eigenvalues in the complex space for the reference case $Re=966$, $Fr=0.82$, $k_x=0.96$, $k_z=5.16$ and
$Sc=\infty$ (circles), $Sc=700$ (crosses), $Sc=7$ (squares),
$Sc=1$ (diamonds and stars). 
Right: zoom on the region contoured by the 
dashed red line in the left diagram, $Sc=7$ symbols
are reported in red.
}
\end{figure}

All the results we presented above correspond to solutions
of the eigenvalue problem where mass diffusivity was
completely neglected, that is $Sc=\infty$.
We have modified the eigenvalue problem and tested the relevance
of a finite $Sc$ number for the reference case 
$Re=966$ and $Fr=0.82$ which will serve as a comparison
between linear analysis, experiments and direct numerical
simulations.
All the simulations are performed at the wave numbers 
$k_x=0.96$, $k_z=5.16$, where the most unstable
mode appears in the $Sc=\infty$ case.
The results are reported
in figure \ref{fig:Sc_effects}.
First we report that at $Sc=\infty$ (circles) and
$Sc=700$ (crosses) the eigenvalues are well
superimposed. This suggests that our non diffusive
approximation is qualitatively and quantitatively
adequate to compare linear theory
with experiments performed with salty water, for which $Sc=700$.
Second we remark that at $Sc=7$ (squares) there is still an unstable mode and close to the origin the distribution
of eigenvalues has the same form. 
For example
looking at the close up on the right
of figure \ref{fig:Sc_effects}, one sees that all the
eigenvalues at $Sc=7$ (in red) are located close to
a non diffusive eigenvalue.
This result makes possible the comparison between
linear analysis, experiments, and direct numerical
simulations which will be performed at $Sc=7$.
Finally we observe that at $Sc=1$ (diamonds and stars) the
eigenvalues are distributed on three distinct Y-shaped
branches which is consistent with the previous
study of \cite{Bakas2009b} and \cite{ChenThesis2016} 
who found analogous branches at $Sc=1$
in the case of the PC flow and the 
plane Poiseuille flow, respectively.
We also remark that at $Sc=1$, $Re=966$ (diamonds) there
is no unstable mode, nonetheless instability is promptly
recovered at $Re=2000$ (stars).
We conclude that increasing mass diffusion, the threshold
of the instability is not severely affected as long as
$Sc\gtrsim7$
while it may change when $Sc$ is of the order of unity.

\section{Direct Numerical Simulations}\label{sec:DNS}
In addition to the linear stability analysis, we have 
performed Direct Numerical Simulations (DNS) 
of the full set of equations (\ref{eq.NS_full})-
(\ref{eq:NSgeneral1}).
The aim of a complementary DNS approach is to validate the linear theory and characterize the flow when retaining all the non-linearities.
Equations are solved in a rectangular box of dimensions ($L_x,L_y,L_z$).
The boundary conditions are periodic in both the stream-wise and vertical directions and rigid no-slip insulating 
boundaries in the cross-stream direction,
i.e. $\boldsymbol{u}=\boldsymbol{0}$ and $db/dy=0$ at $y=\pm1$.
In order to keep the computational time reasonable, while still focusing on the high $Sc$ number regime of the experiment
described in section \ref{sec:experiments}, we fix $Sc=7$.
We have seen in section \ref{sec:Sc_effects}
that this particular choice does not affect qualitatively the results, and in any case, ad-hoc solutions of the linear problem at $Sc=7$ can be considered for a quantitative comparison.
In order to ensure that the linear instability is well captured by the numerical simulation, we choose a box of size $\left(L_x=2\pi/k_x, L_y=2, L_z=2\pi/k_z\right)$, where $k_x$ and $k_z$ are the most unstable wave numbers as predicted by the linear stability analysis presented above.

We performed DNS using the spectral element solver Nek5000 \citep{FISCHER1997,FISCHER2007,nek5000-web-page}.
The use of spectral elements instead of more classical pseudo-spectral methods will be justified later (see section \ref{sec:experiments}) where we add the effect of the stream-wise confinement to mimic the experimental setup.
The global geometry is partitioned into hexahedral elements, with refinement close to the moving boundaries.
Velocity, buoyancy and pressure variables are represented as 
tensor product Lagrange polynomials of order $N$ and $N-2$ 
based on Gauss or Gauss-Lobatto quadrature points.
The total number of grid points is given by $\mathcal{E}N^3$ where $\mathcal{E}$ is the number of elements.
For all the results discussed in this paper, the number of elements is $\mathcal{E}=6336$ and we use a polynomial order from $N=7$ up to $N=11$ for the highest Reynolds number case.
Time integration is performed with a third-order explicit scheme for the advection and buoyancy terms while viscous and dissipative terms are integrated using an implicit third-order scheme.
The simulations are initialized with a small amplitude buoyancy perturbation and with an established linear PC flow.

\begin{figure}
\noindent
\begin{minipage}{0.48\linewidth}
\centering
\begin{overpic}[width=.9\linewidth]{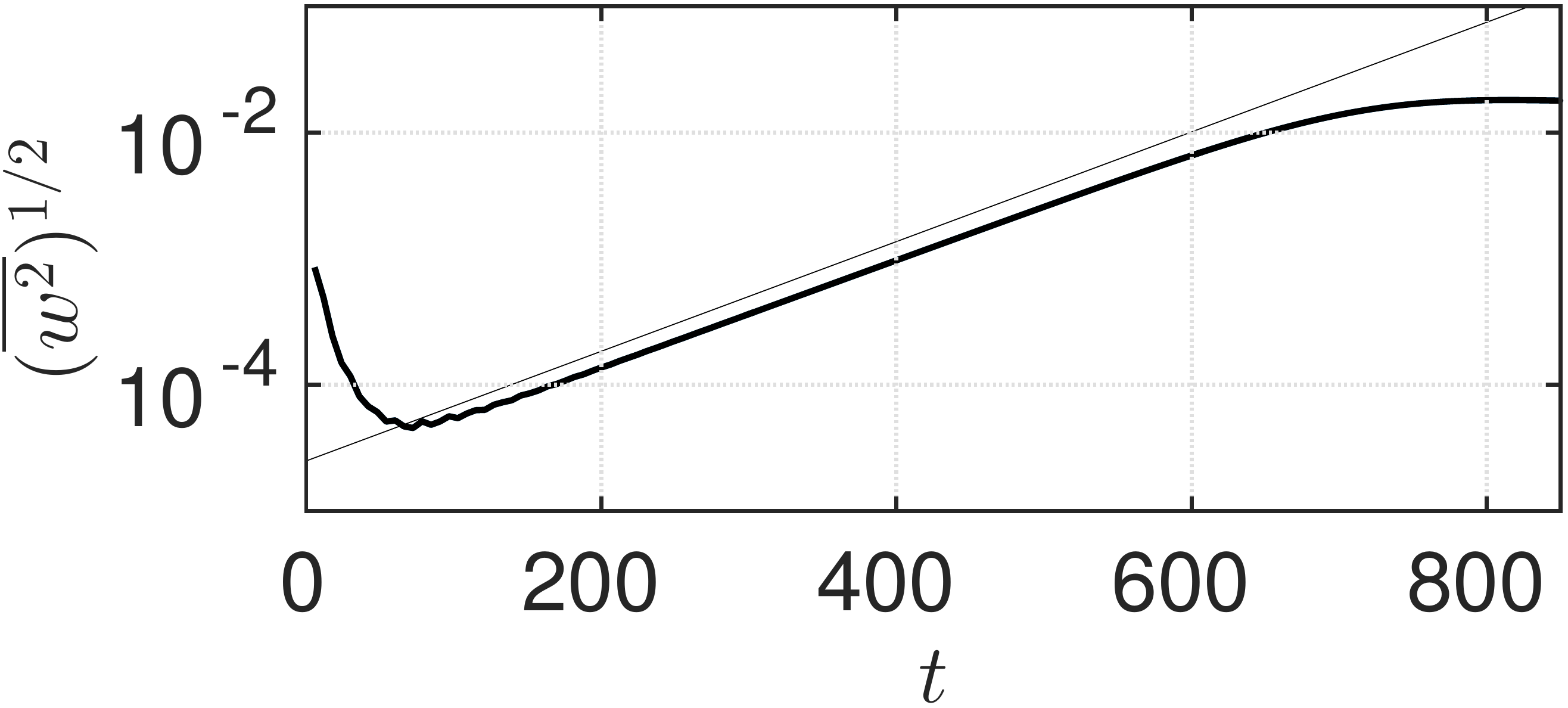}
\end{overpic}
\vspace{.5em}

\noindent
\begin{overpic}[width=.9\linewidth]{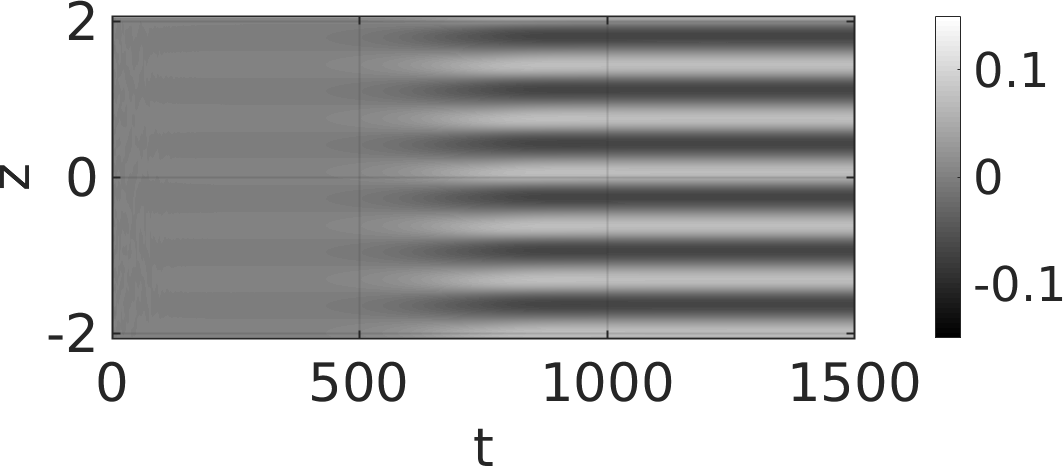}
\end{overpic}
\end{minipage}
\hfill
\begin{minipage}{0.48\linewidth}
\begin{overpic}[height=.9\linewidth]{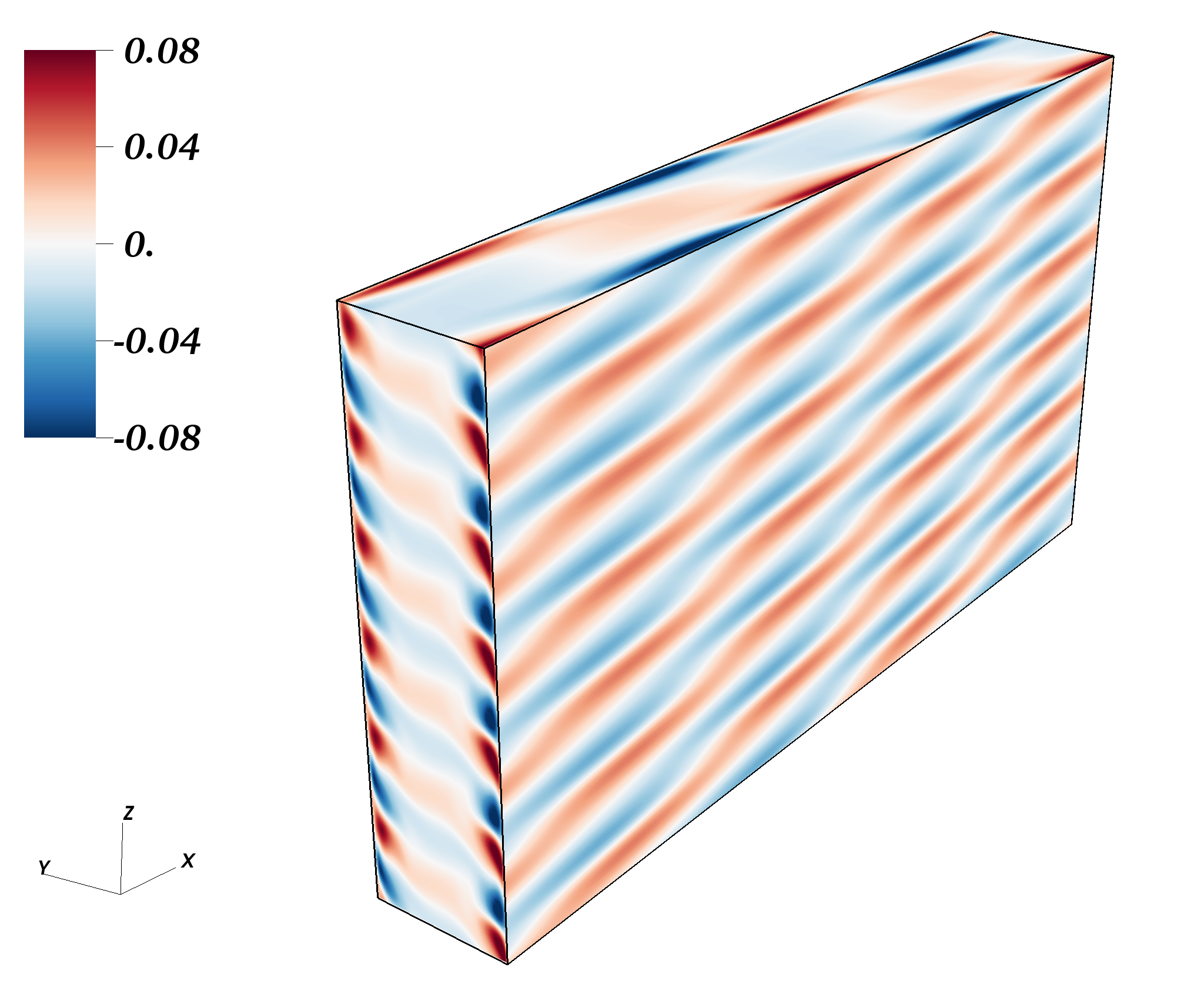}
\end{overpic}
\end{minipage}
\caption{\label{fig:DNS_GR_pattern}
Left: at the top, vertical kinetic energy 
$({\overline{{w}^2}})^{1/2}$ as a function of time
at $Re=966$, $Fr=0.82$ and $Sc=7$.
The thick line refers to the DNS simulation while the thin line
refers to the growth of the most unstable mode
as indicated by linear stability analysis.   
At the bottom, horizontal velocity perturbation $u$ 
at $x=y=0$ as a function of $t$ and $z$ for the
same DNS.
One observes that a stationary pattern appears close to $t=500$.
Right: instantaneous 3D map of the buoyancy perturbation once the flow has become unstable (DNS). 
As predicted by the linear analysis the selected mode
is mainly modulated in the vertical direction but still
not stream-wise invariant (i.e. $k_x\neq 0$).
Note also that perturbations are
concentrated near the boundaries $y=\pm 1$.
}
\end{figure}

In order to validate the eigenvalue problem we choose the reference case $Re=966$, $Fr=0.82$ which will serve
later as a comparison to experiments.
In figure \ref{fig:DNS_GR_pattern} (top left) 
we report the time
evolution of the vertical kinetic energy density 
(thick line) which is defined as:
\begin{equation}\label{eq:K_density_def}
({\overline{{w}^2}})^{1/2}=\left(\frac{1}{V}\int_V{w^2dV}\right)^{1/2}
\end{equation}
where $V$ refers to the volume of the simulation box.
The quantity $({\overline{{w}^2}})^{1/2}$ is appropriate 
since $({\overline{{w}^2}})^{1/2}_{t=0}=0$ for the base flow.
One observes that  $({\overline{{w}^2}})^{1/2}$ increases exponentially
and superposing the exponential growth predicted by the linear analysis one obtains an excellent agreement, 
with a relative discrepancy on the growth rate $\sigma_c$ of less than $1\%$.
In the same figure (bottom left) we also report the 
spatio-temporal diagram of the horizontal perturbation
$u$ at $x=y=0$.
One observes that a stationary pattern has established
around $t=500$ which has a well defined vertical wavelength. 
In figure \ref{fig:DNS_GR_pattern} (right) we 
report a visualisation
of the buoyancy perturbation $b$  
once the instability has saturated.
One observes an weakly inclined layering of the density field
which is a common feature in stratified turbulent shear flows
\citep[see][for a review]{Thorpe2016}.
Again we have a very good agreement with the linear theory:
a distinct spatial pattern appears and 
both vertical and horizontal wavelengths correspond to the 
predicted values.
One can notice that the spatial pattern perfectly fits in the
simulation domain. 
This condition is indeed necessary to observe the instability, and no relevant growth of the vertical 
kinetic energy is observed when none of the
unstable wave numbers fits inside the simulation domain.
%
%
\section{Experiments}\label{sec:experiments}
%
\subsection{Experimental apparatus}\label{subsec:Exp_App}
\begin{figure}
\begin{minipage}{0.45\linewidth}
\centering
\begin{overpic}[width=1.1\linewidth]{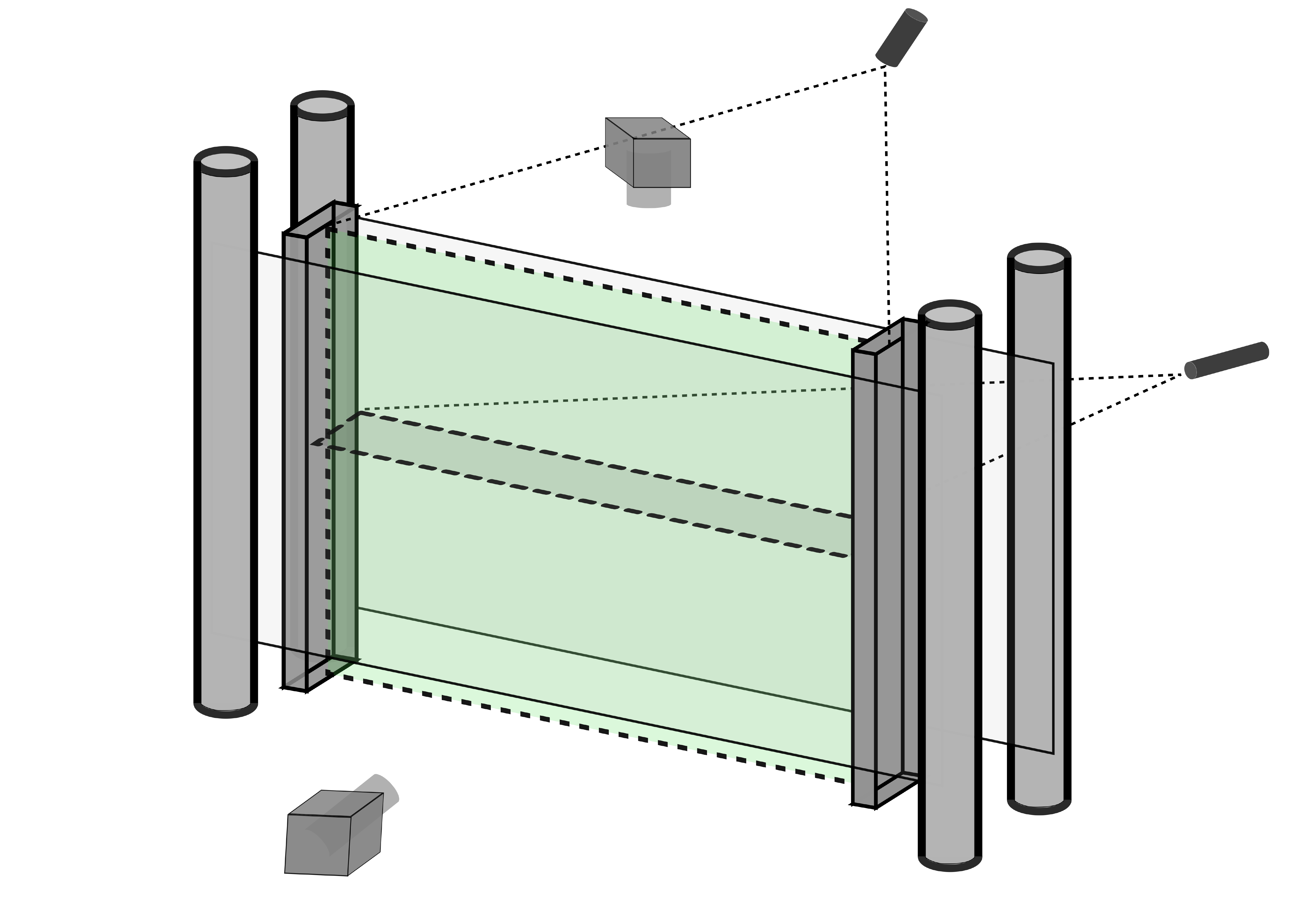}
\end{overpic}
\vspace{1 em}

\begin{overpic}[width=.9\linewidth]{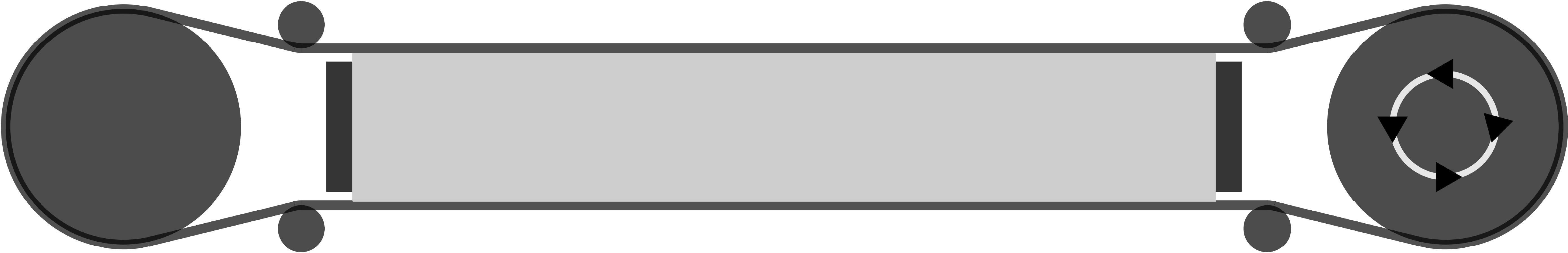}
\end{overpic}
\end{minipage}
\hfill
\begin{minipage}{0.45\linewidth}
\centering
\begin{overpic}[width=.75\linewidth]{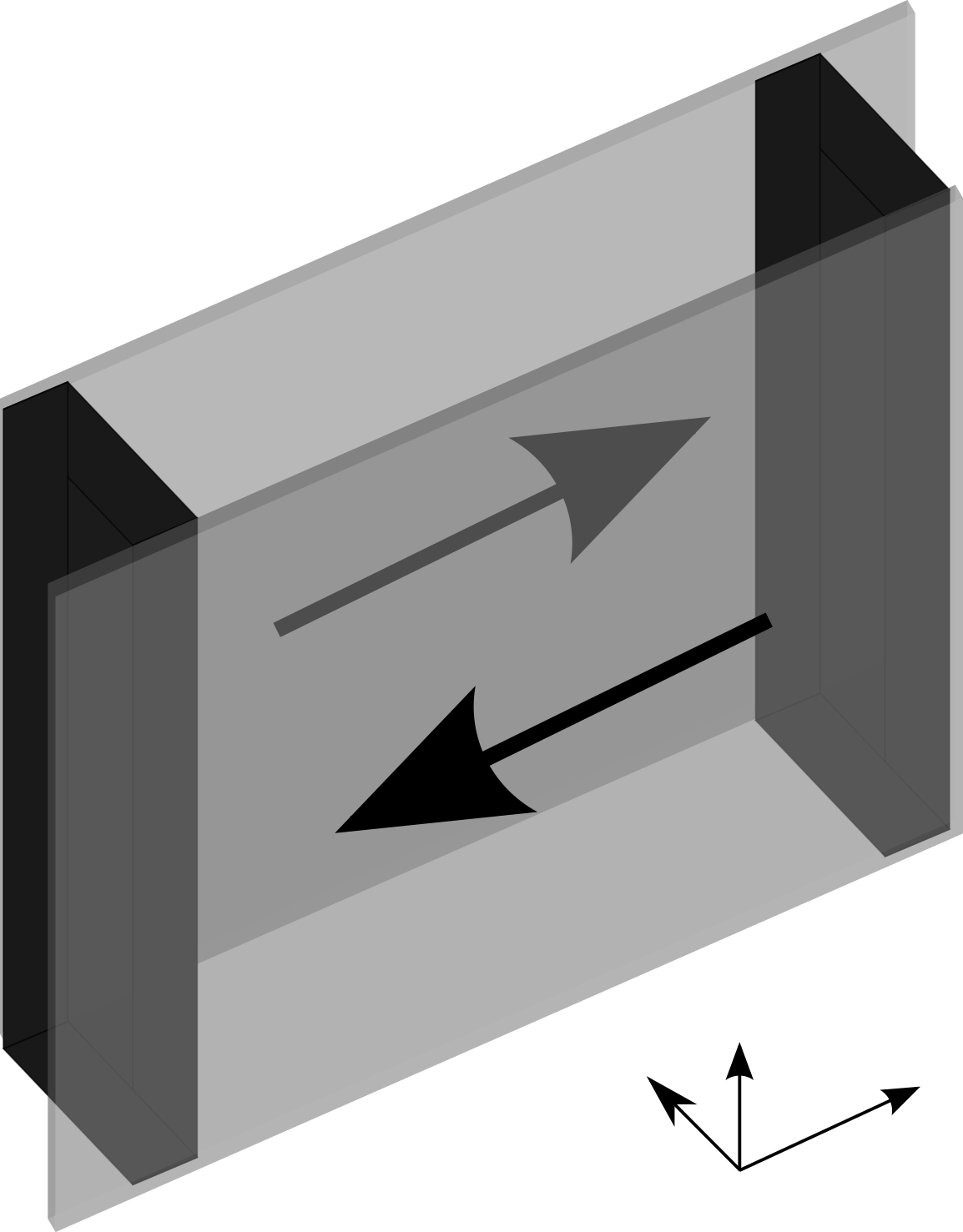}
\put (72,13) {$\boldsymbol{ \hat{x}}$}
\put (48,10) {$\boldsymbol{ \hat{y}}$}
\put (61.5,13) {$\boldsymbol{ \hat{z}}$}
\end{overpic}
\end{minipage}
\caption{\label{fig:sh_device}
Left: sketch of the experimental shearing device seen from the side (top) and from above (bottom). The two green shaded area correspond to two laser sheets which enlight the mid vertical plane (i.e. $y=0$) and the horizontal mid plane (i.e. $z=0$). Two cameras allow to image the flow in the enlighted areas.
Right: schematic of confined DNS experiments.
Two rigid lateral walls entrain the fluid at constant velocity,
and two rigid walls confine the flow in the stream-wise
direction. Vertical boundary conditions are periodic.} 
\end{figure}

Now, we want to study whether or not this linear instability of the stratified plane Couette flow does appear in a "real" configuration, and to do so, we look for some of its signatures in an experimental set-up, intrinsically limited in size.
The flow is produced with a shearing device which is placed inside a transparent tank ($\unit[50]{cm}$ x $\unit[50]{cm}$ x $\unit[70]{cm}$) made of acrylic. 
The tank is filled with salty water linearly stratified in density. 
The shearing device is sketched in figure \ref{fig:sh_device}
(left).
The device consists of a PVC transparent belt ($\unit[0.8]{mm}$ thick) which is closed on a loop around two vertical entraining cylinders made of dense sponge (we use standard spares entraining cylinders for commercial swimming-pool robots).
Two additional pairs of cylinders 
(inox, $\unit[2]{cm}$ diameter) 
constrain the two sides of the loop to be parallel and at a controlled distance $d$.
All cylinders are mounted on a system of acrylic plates which allows to vary the distance between the entraining cylinders (i.e. to tighten the belt) through two pairs of coupled screws (i.e. one pair for the bottom and one for the top).
The top acrylic plates also prevent the existence of a free surface which would affect any imaging from the top.
The motion of the belt is provided by a motor which is mounted on the top of the device and joined to the axis of one of the entraining cylinders.
Finally two PVC rigid plates are mounted vertically in front of the two entraining cylinders in order to reduce at most any perturbations coming from the entrainment system.
The distance between the edges of the plates and the belt
is a few $mm$.
Thus we look at the flow in the area shaded in light grey (figure \ref{fig:sh_device}, left).
In the present work we consider two values of the gap width $d=\unit[5.8]{}$ and $\unit[9.8]{cm}$, while the distance between the PVC plates $D$ was respectively $\unit[34]{cm}$ and $\unit[24]{cm}$, leading to a value of the aspect ratio $D/d$ of $5.7$ and $2.4$ respectively. 

The tank is filled with salty water of variable density.
As a general rule a water column of height  $H=10$ to $20$ $\unit[]{cm}$ linearly stratified in density always occupies the volume delimited by the belt and the confining barriers, while above and below the density stratification was generally weaker or negligible. 
The density profile is obtained by the double-bucket method \citep{oster1965bucket}. 
To measure the density profile we collect small samples of fluid ($\sim\unit[10]{ml}$) at different heights and analyze them with a density-meter Anton Paar DMA 35.
The Brunt-V\"ais\"al\"a frequency $N$ is constant for each
experiment with a value between $\unit[0.5]{rad/s}$ 
and $\unit[3.0]{rad/s}$. 
We measure the stratification before and after each experiment. 
The shearing motion clearly affects the stratification especially through the small scale features of the rotating part of the device which necessarily produces some mixing.
Also, in our highest $Re$ experiments we observe optical distortion which may indicate the presence of 
high density gradient zones and thus density layering, for example similarly to 
that observed in turbulent stratified experiments performed in Taylor-Couette devices \citep{Oglethorpe2013}.
Nevertheless we observe that the density profile at the end 
of an experiment is weakly perturbed and the relative
 discrepancy in the area of interest is around $5$\%.
Finally we assume the viscosity to be $\nu=\unit[10^{-6}]{m^2/s}$, and neglect any change associated to variable salt concentration.

The fluid is seeded with ($\unit[10]{\mu m}$ - diameter) hollow glass spheres and two laser sheets illuminate the particles in the vertical plane $y=0$ and the horizontal plane $z=0$ as shown in figure \ref{fig:sh_device}.
The flow is then recorded from the side by a $\unit[4]{Mpx}$ camera at a frame rate of $\unit[8]{fps}$ and from the top by a $\unit[2]{Mpx}$ camera at a frame rate of $\unit[30]{fps}$.
The velocity field is obtained with a Particle Image Velocimetry (hereafter PIV) cross-correlation algorithm \citep{meunier2003analysis}.
Note that the mid vertical plane $y=0$ is the appropriate place to detect the possible onset of an instability because in the ideal PC regime, the velocity should be zero there.
The current setup permits only one by one enlighting-recording of the flow, thus movies from the top and from the side are always taken at different times.

%
\subsection{Base flow}\label{sec:base_flow}
\begin{figure}
\begin{minipage}[c][][t]{.51\linewidth}
\begin{overpic}[width=1.\linewidth]{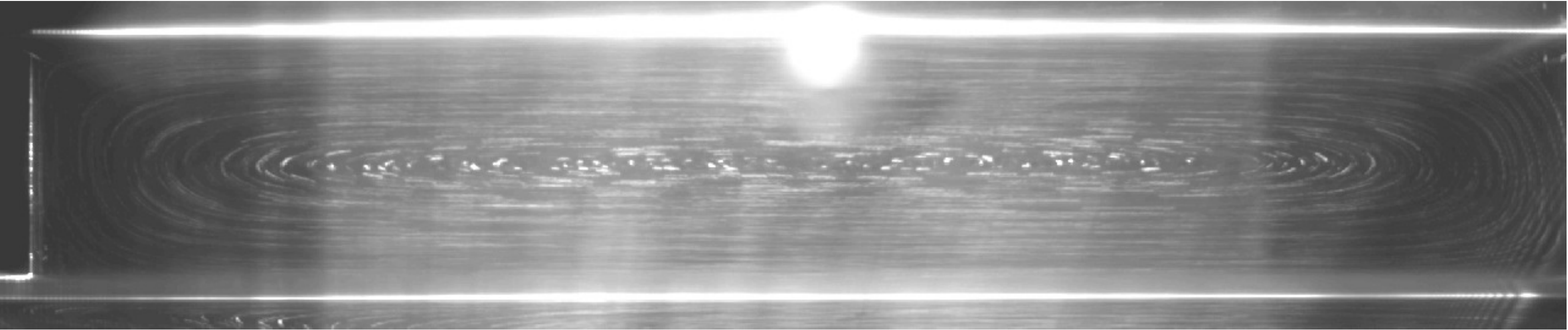}
\end{overpic}
\vfill
\vspace{1.em}
\hspace{-1.em}
\begin{overpic}[width=1.0\linewidth]{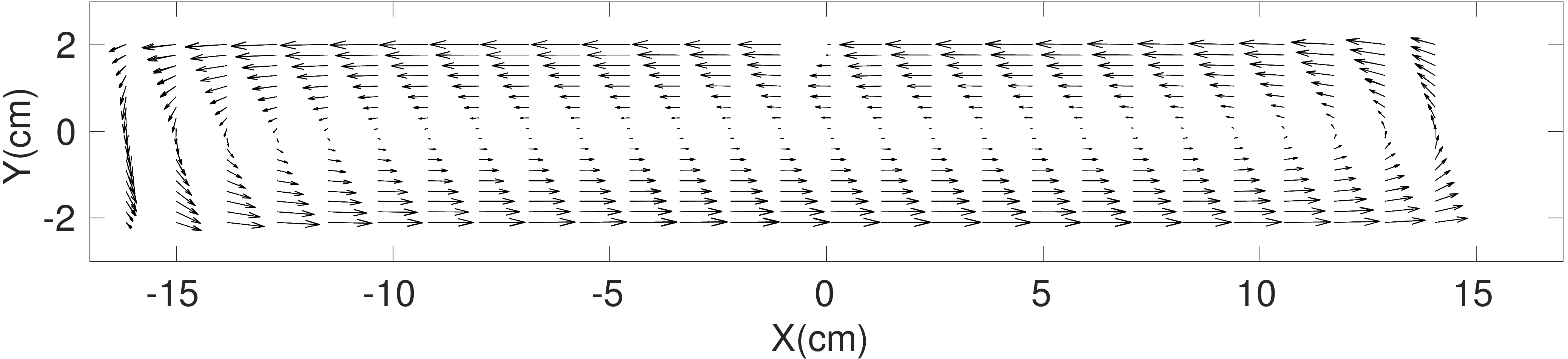}
\end{overpic}
\end{minipage}
\hfill
\begin{minipage}[c][][t]{.47\linewidth}
\begin{overpic}[width=1.\linewidth]{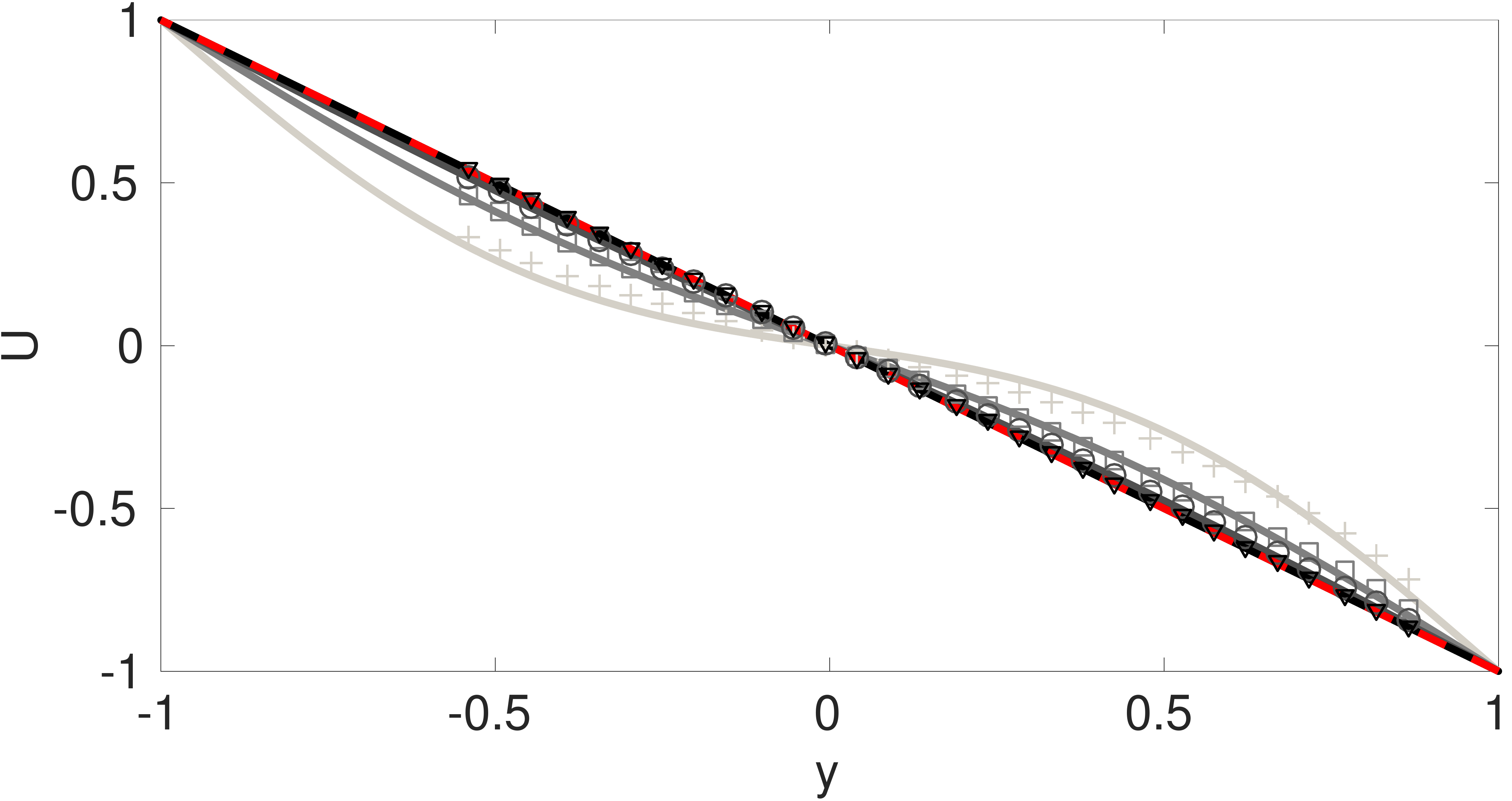}
\end{overpic}
\end{minipage}
\caption{\label{fig:cou_profile_transient}
Left: top view of the $z=0$ plane.
On the top we show a superposition of $40$ images (i.e. $\sim\unit[1.3]{s}$) as captured by the camera. Only light contrast was exalted to show PIV particle trajectories.  
On the bottom the velocity field $(u,v)$ as reconstructed via
the PIV algorithm. The velocity plot is obtained averaging over
$40$ PIV fields (i.e. $\sim\unit[1.3]{s}$). 
Right: horizontal velocity as a function of the cross-stream direction $y$ at four different times $T_{\nu}/10$ (crosses), $T_{\nu}/5$ (squares), $T_{\nu}/3$ (circles) and $T_{\nu}$ (triangles).
Symbols refer to experimental observations. 
Each profile corresponds to the time average of the horizontal velocity fields over $0.03T_{\nu}$.
Solid lines refer to the value of the expression (\ref{eq.cou_transient}) expected for the two infinite walls problem (increasing time from clear gray to black).
The dashed line refers to the asymptotic $t=\infty$ solution.
}
\end{figure}

First we report that a PC flow can be observed in the region confined between the belt and the PVC barriers.
On the top left of figure \ref{fig:cou_profile_transient} we superpose $40$ images of the $z=0$ plane exactly as captured by the camera. Only the contrast was altered to exalt streamlines.
Both the intersections of the belt with the laser sheet and the left barrier edge can be easily recognized as brighter lines.
One also sees that streamlines close up near the PVC barriers and recirculations are present.
This is confirmed by the velocity field given by the PIV algorithm and shown just below. 
The velocity plot is obtained by averaging over $40$ 
PIV fields ($\sim\unit[1.3]{s}$), also we plot only 
one arrow over four in the horizontal direction, to make the diagram readable.
One remarks that up to $\unit[10]{cm}$ far from the center the flow is nicely parallel and the velocity gradient is linear.
Both streamlines and PIV fields refer to an experiment where the base flow was already stationary.
Now, any experiment necessarily implies a transient phase where the flow evolves from a first stationary phase, 
e.g. the whole fluid is at rest, to a second stationary phase which is the forced parallel flow. 
We expect the base flow to establish via viscous entrainment starting from the fluid layers which are close to the walls, thus the viscous time $T_{\nu}=d^2/\nu$ seems to be an appropriate time scale for the transient.
In order to verify this we need some more quantitative prediction 
and consider the transient flow generated by two infinite walls treated by \cite{Acheson1990}.
As a first step, recirculations are neglected.  
If the flow is initiated at $t=0$ the horizontal velocity has the form  $U(y,t)=U_0(y)-U_T(y,t)$, where $U_0(y)$ is the asymptotic base flow and the transient part $U_T(y,t)$ reads:
\begin{equation}\label{eq.cou_transient}
U_T(y,t)=(U_0-U_i)\sum_{j=1}^{\infty}{
\frac{2}{\pi}
\frac{(-1)^{j}}{j}e^{-\pi^2 j^2 t/Re}\sin{j\pi y}}
\end{equation}
where $U_i$ is the velocity of the belt at $t=0$,
for example $U_i=0$ if the experiment is started with the fluid at rest.
In figure \ref{fig:cou_profile_transient} we compare the value
of $U(y,t)$ as expected from equation (\ref{eq.cou_transient}) with the average value of the horizontal velocity as observed in a typical experiment.
The value of $U$ is plotted as a function of $y$ at four different times. 
First, the velocity profile collapses on the expected PC flow (dashed line) around $t=T_{\nu}$, which confirms that the base flow establishes via viscous entrainment.
Also at $t=T_\nu/3$ (circles), the value of the average horizontal velocity is already very close to the PC flow.  
Secondly one remarks an excellent agreement of the experimental observations with the infinite walls approximation, 
which suggests that the recirculation does not affect significantly the shape of the transient flow.
With regard to this, one should notice that knowing the time the base flow needs to establish becomes crucial when determining the growth rate of the instability, which will be discussed later.

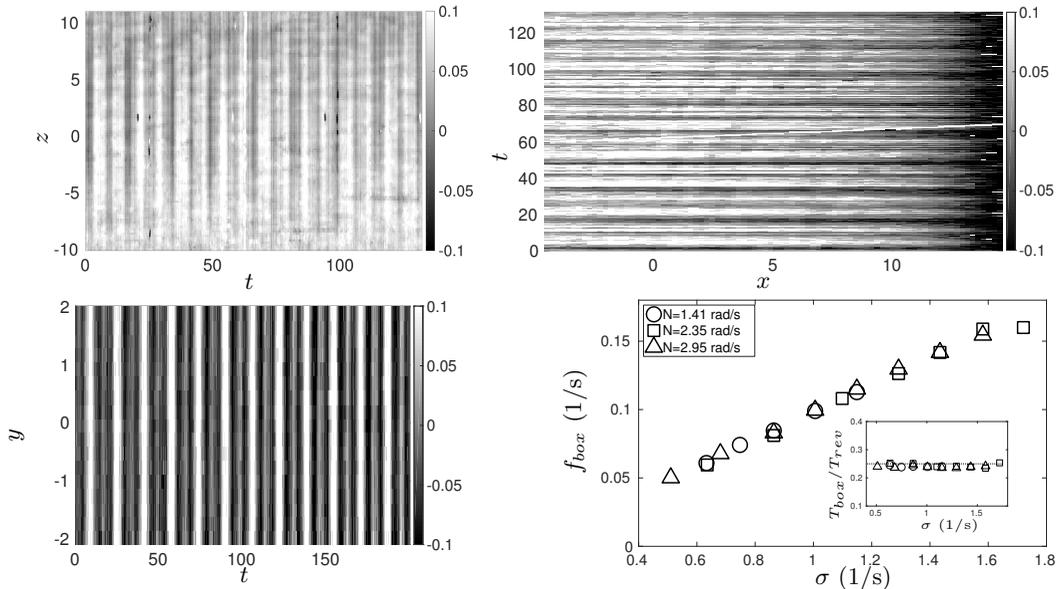
\begin{figure}
\subfigure
{
\begin{overpic}[height=.26\linewidth]{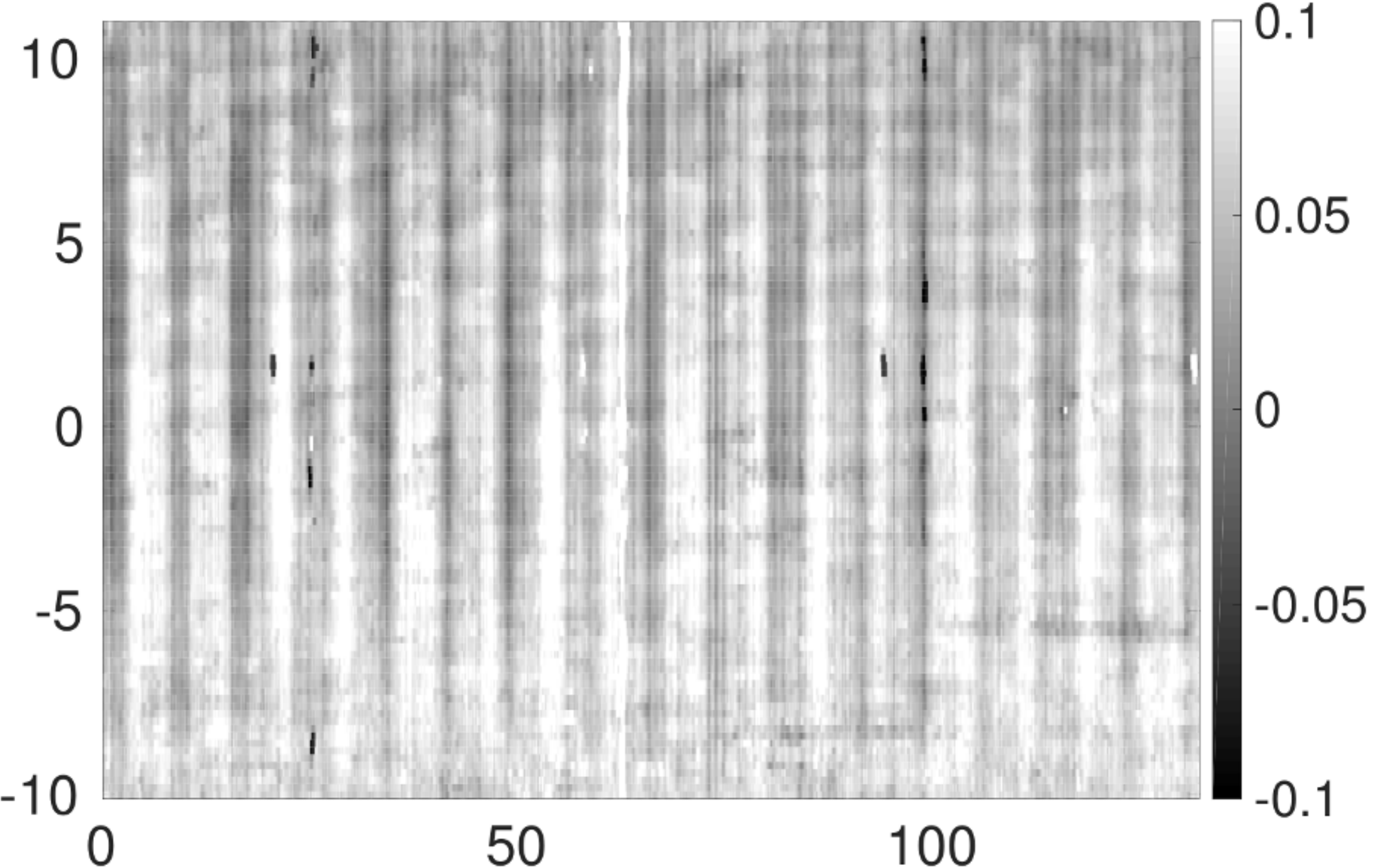}\\
\put (45,-4){\small$t$ }							
\put (-4,30){\rotatebox{90}{\small$z$}}		
\end{overpic}
}
\hfill
\subfigure
{
\begin{overpic}[height=.26\linewidth]{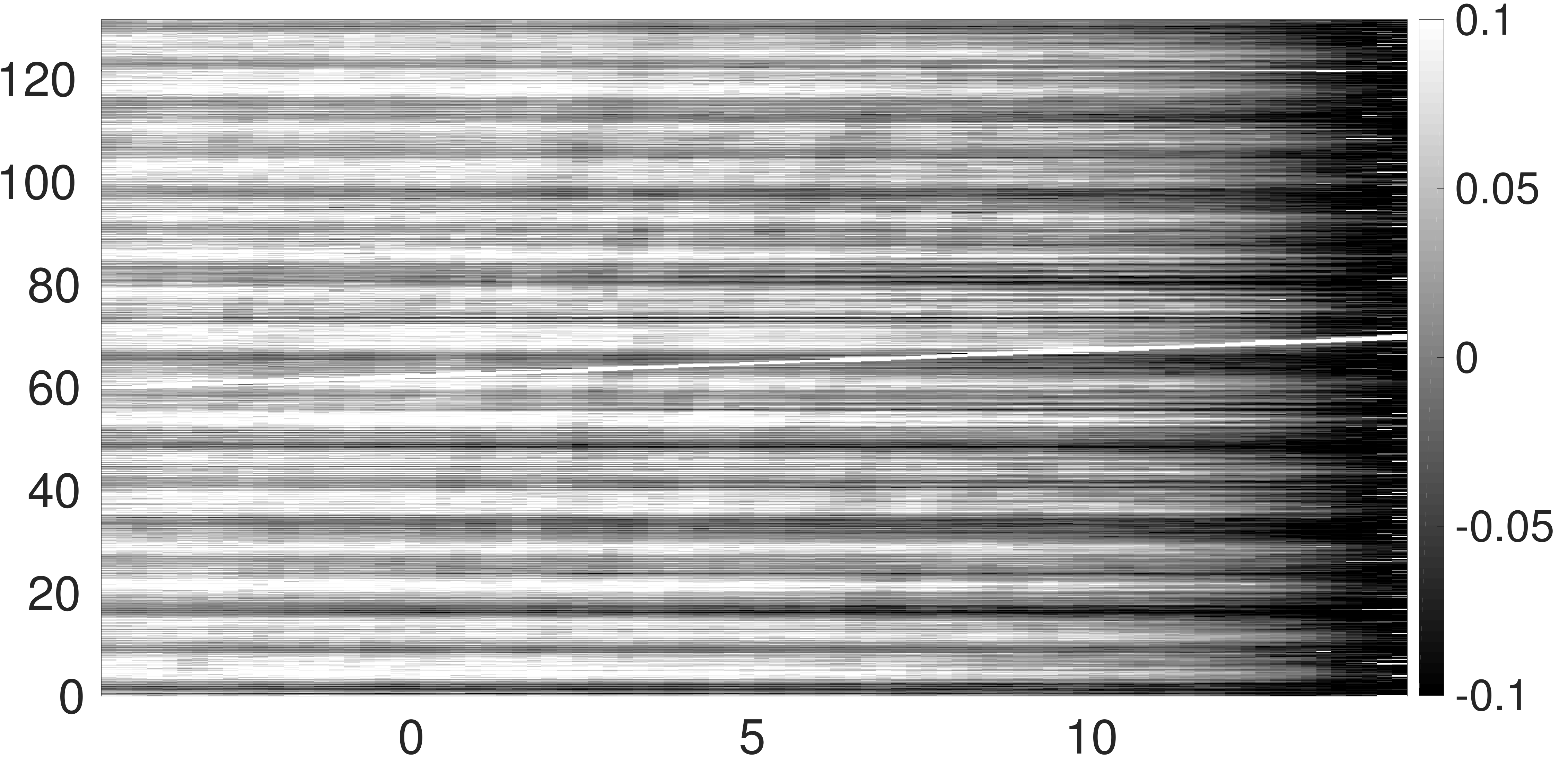}
\put (45,-3){\small$x$ }							
\put (-3,20){\rotatebox{90}{\small$t$}}		
\end{overpic}
}
\vfill
\subfigure
{
\begin{overpic}[height=.26\linewidth]{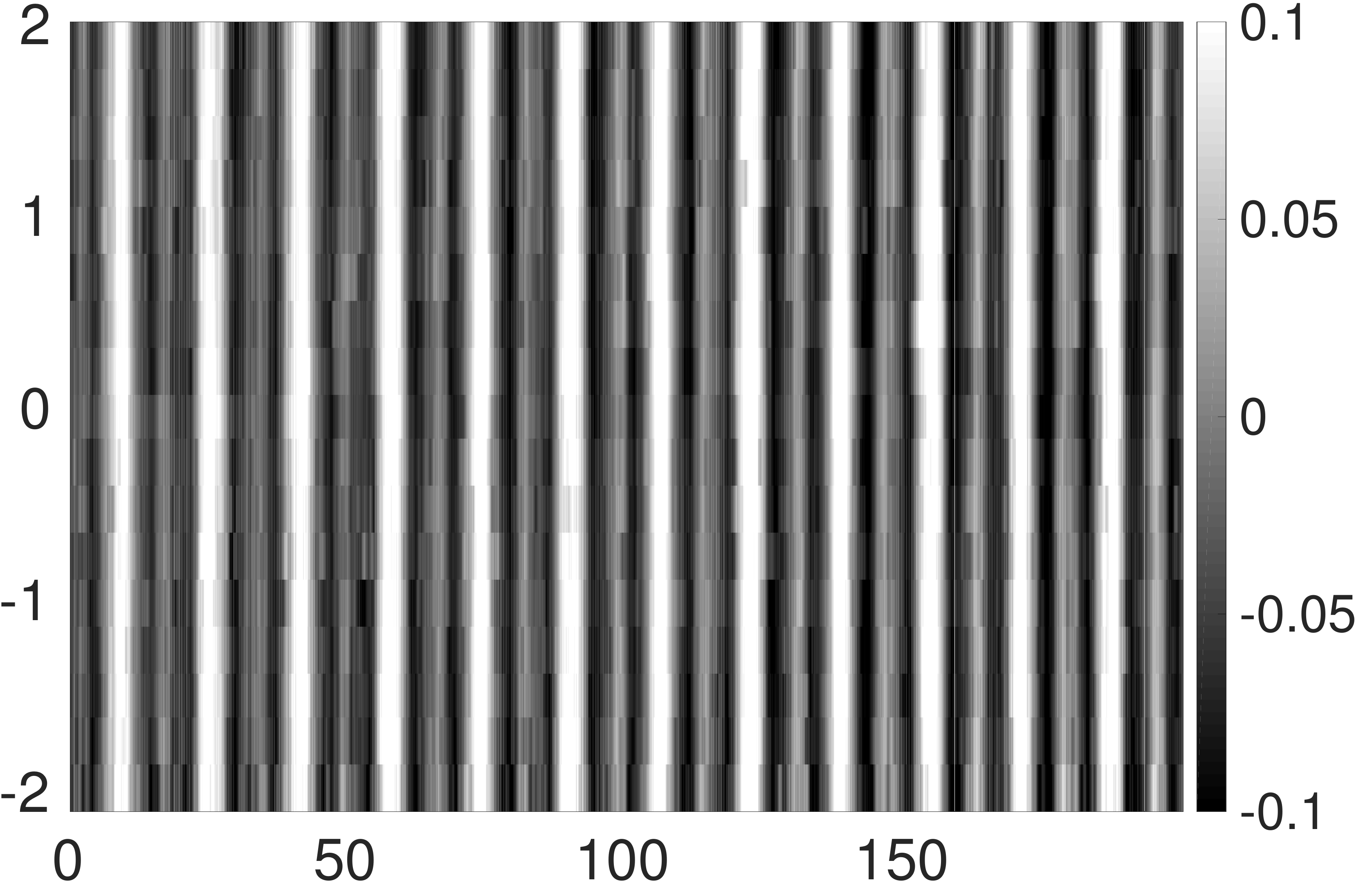}
\put (45,-4){\small$t$ }							
\put (-11,30){\rotatebox{90}{\small$y$}}		
\end{overpic}
}
\hfill
\subfigure
{
\begin{overpic}[height=.26\linewidth]{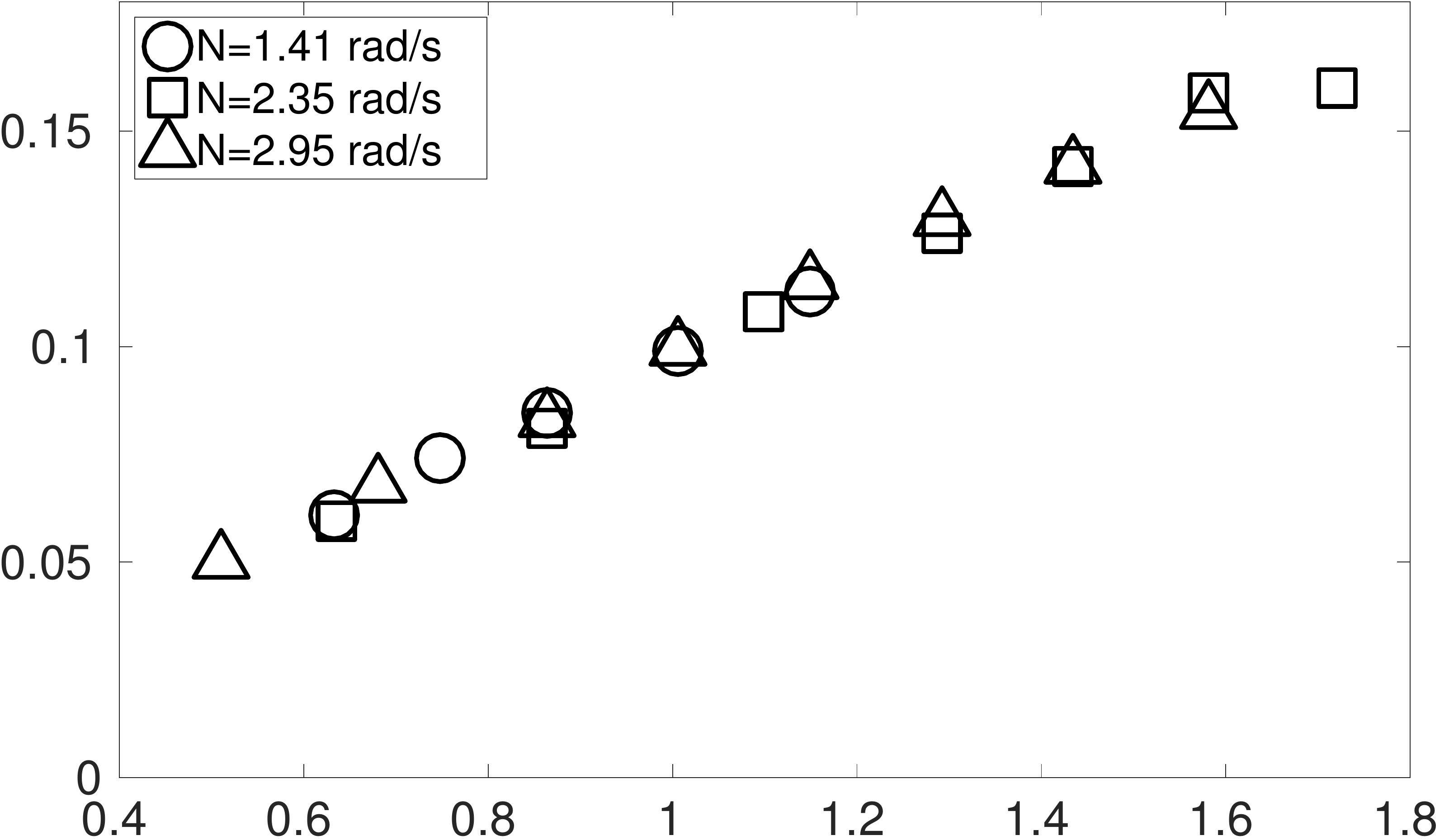}\rlap{\hspace{9.5em} \raisebox{2.em}{\includegraphics[height=1.3cm]{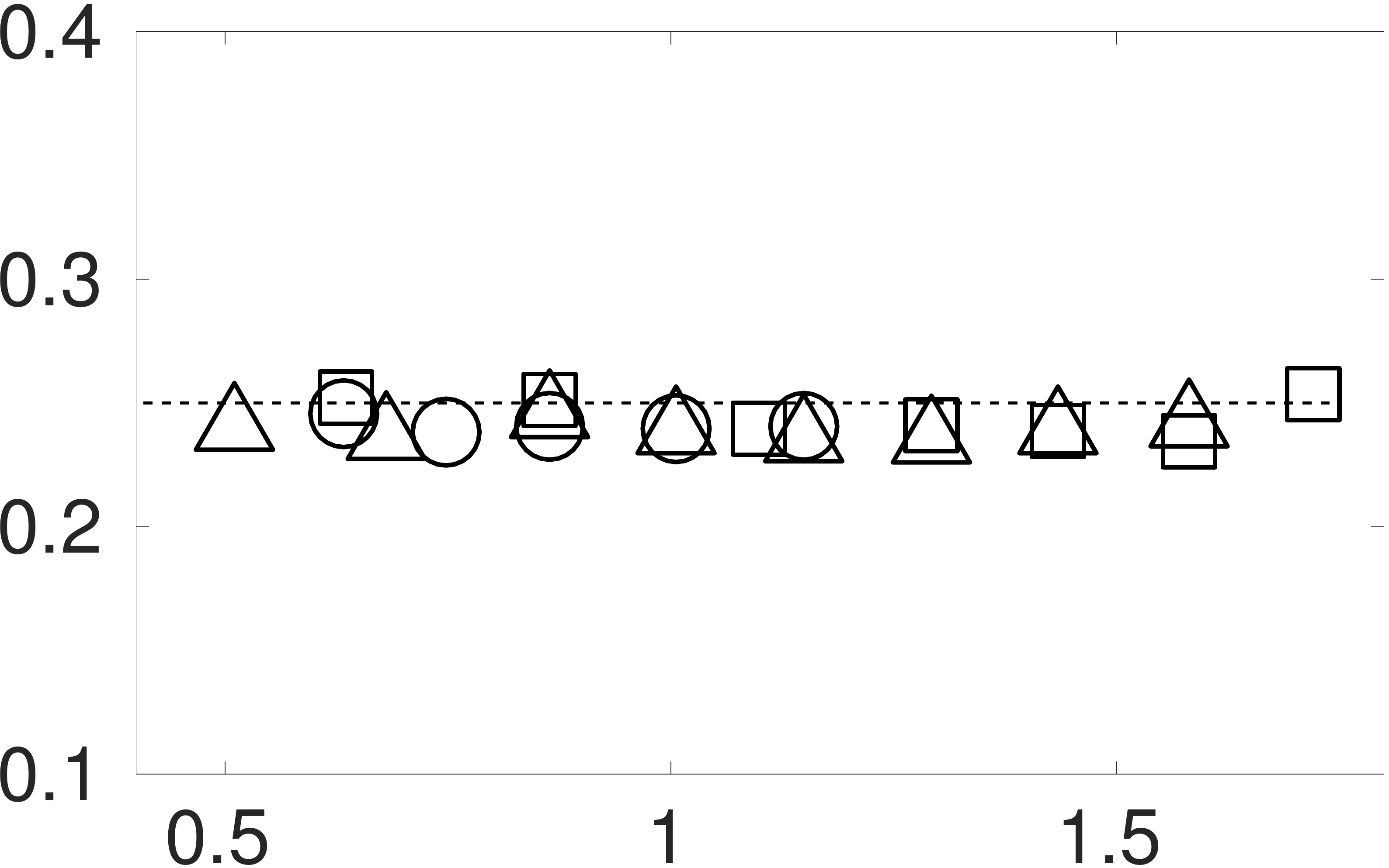}}}
\put (45,-4){\small$\sigma$ $(\unit[]{1/s})$}							
\put (-10,20){\rotatebox{90}{\small$f_{box}$ $(\unit[]{1/s})$}}		
\put (68,8){\tiny$\sigma$ $(\unit[]{1/s})$}							
\put (48,10){\rotatebox{90}{\tiny$T_{box}/T_{rev}$ }}		
\end{overpic}
}
\caption{\label{fig:bandes_initiales}
Top: horizontal velocity perturbation $u$ for a moderate $Re$ experiment ($Re=530$, $Fr=0.45$) in the mid vertical plan $y=0$. 
On the left, $u$ as a function of the time $t$ and the vertical coordinate $z$ ($x=0$). On the right, $u$ as a function of the horizontal coordinate $x$ and the time $t$ ($z=0$).
Bottom left: horizontal velocity perturbation $u$ 
for the same experiment in the mid horizontal plan $z=0$, 
as a function of time $t$ and $y$ ($x=-1$).
All the quantities are non dimensional.
Bottom right: global frequency $f_{box}$ as a function of the imposed shear $\sigma$ for a collection of experiments 
with different stratifications $N$ (different symbols).
$f_{box}$ was computed as the maximum in the temporal Fourier transform of $\bar{u}(t)$, which is the vertical average of $u$
at $x=y=0$.
In the inset we report the corresponding global period
in unit of revolution time of the belt as a function of
$\sigma$. 
}
\end{figure}

Once the PC profile is established we want to detect
possible deviations from the base flow.
For this aim we mainly focus on the mid plane $y=0$ where no motion is expected for the base flow.
As a standard protocol we initiate the flow at low shear rate
$\sigma$ and then increase $\sigma$ by a small fraction (typically $15\%$).
Top views of the plane $z=0$ are also taken to verify the shape of the parallel base flow.  
In each experiment the flow was observed for at least one 
viscous time $T_{\nu}=d^2/\nu$ which may be taken as an upper-bound for establishing the base flow.
First we report that starting from very moderate Reynolds number $Re\gtrsim 300$ the observed fluid oscillates coherently at a well defined frequency.
In figure \ref{fig:bandes_initiales} we report the spatio-temporal evolution of the horizontal velocity perturbation $u$
along the vertical line $y=0,x=0$ and the two horizontal lines
$y=0,z=0$ and $z=0,x=-d$.
As visible in each diagram, parallel periodic structures appear which are fairly homogeneous over the spatial domain.
This suggests that the entire fluid bulk oscillates in a coherent way. 
In addition all the diagrams show a well defined temporal frequency.
We then perform the temporal fourier transform of the vertical average of $u(x=y=0,z,t)$ and denotes with $f_{box}$ the peak in
the frequency spectrum.
In the last panel of figure \ref{fig:bandes_initiales} we 
report the value of $f_{box}$ as a function of 
the imposed shear $\sigma$
for a collection of three experiments at different $Fr$ numbers, 
where only the imposed shear $\sigma$ is changed.
One sees that the observed global frequency $f_{box}$ increases linearly with $\sigma$, and rescaling the observed period 
with the belt revolution time $T_{rev}$, we find (see inset) that the observed oscillation period $T_{box}$ is very close to $T_{rev}/4$.
A possible explanation for this unexpected observation
 is that the two pairs of confining cylinders (see figure \ref{fig:sh_device}) divide the 
path of the belt in four (almost) equivalent sections.
We stress that these cylinders constitute one of the biggest source of noise because they are rigid and each of them tends
 to perturb the belt once per revolution time, when they touch the roughness of the belt junction.
Correspondingly we expect that a $T_{rev}/4$ resonant periodic forcing may establish, and give the velocity field the observed 
temporal pattern. 
As a summary we report that deviation from a zero 
velocity field are observed in the horizontal velocity perturbation $u$, 
from a very moderate value of the $Re$ number.
Nonetheless the observed motion shows a trivial spatial pattern 
(i.e. $k_x=k_y=k_z=0$), looks like a bulk oscillation, and
seems to be connected to the shearing device. 
Thus, in the presence of only spurious bulk oscillations 
(even if robust), 
we assess that the corresponding $Re$, $Fr$ pair 
is a stable point in the stability diagram.
In the following we discuss how perturbations become more finely structured at higher $Re$ number, and we give a criterion
to distinguish these initial deviations from a truly unstable pattern.
\subsection{Instability}\label{sec:exp_instability}
\begin{figure}
\centering
\subfigure
{
\begin{overpic}[width=1.\linewidth]{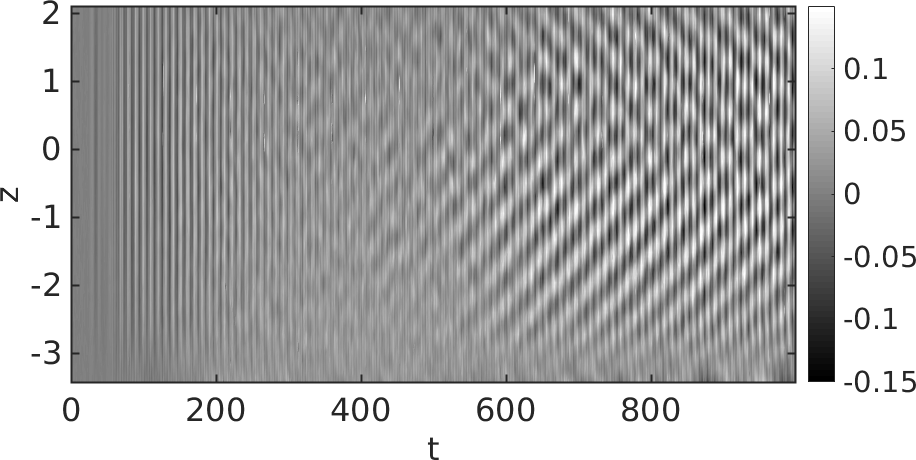}
\end{overpic}
}
\subfigure
{
\begin{overpic}[height=.31\linewidth]{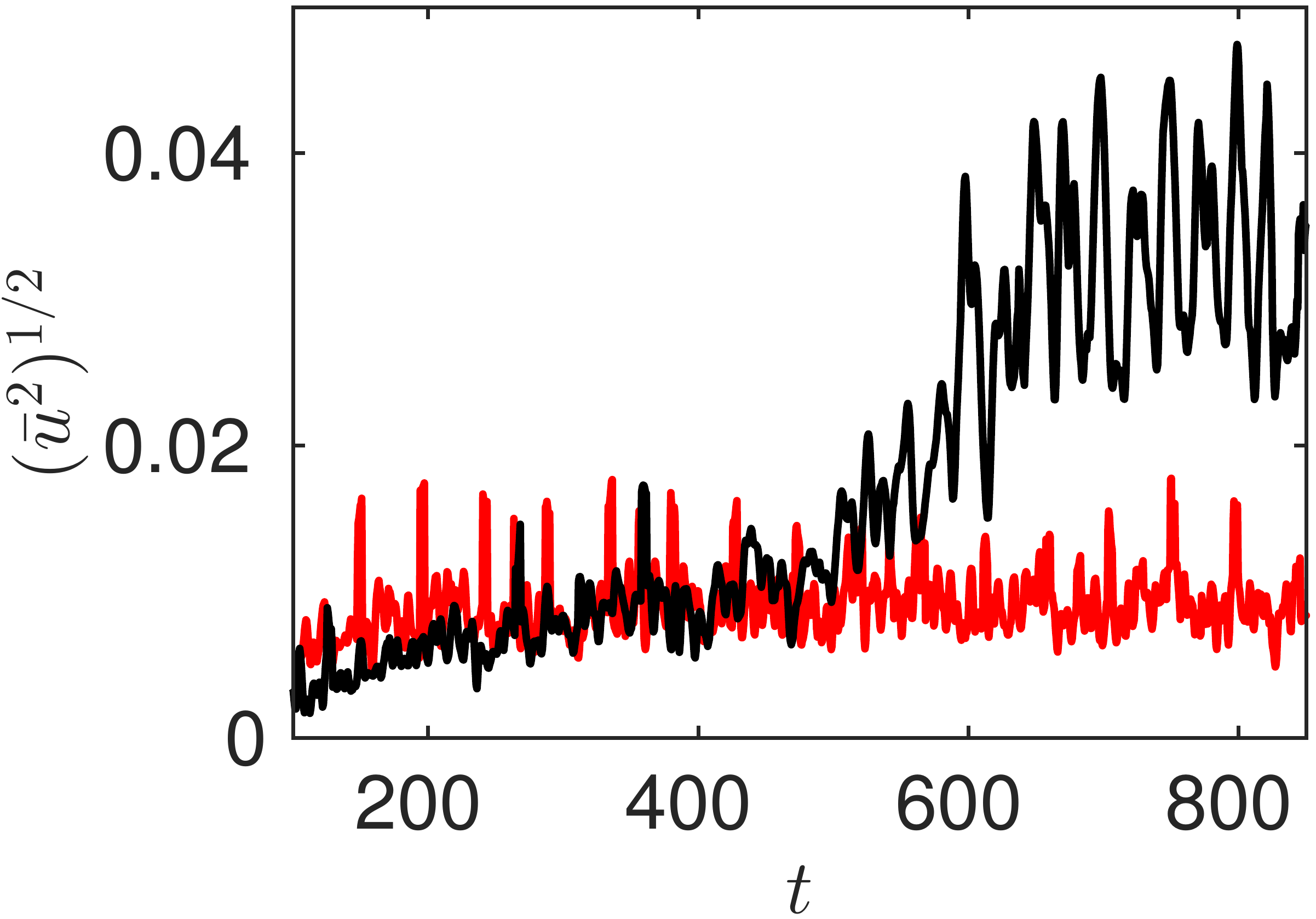}\rlap{\hspace{4.4em} \raisebox{6.8em}{\includegraphics[height=1.8cm]{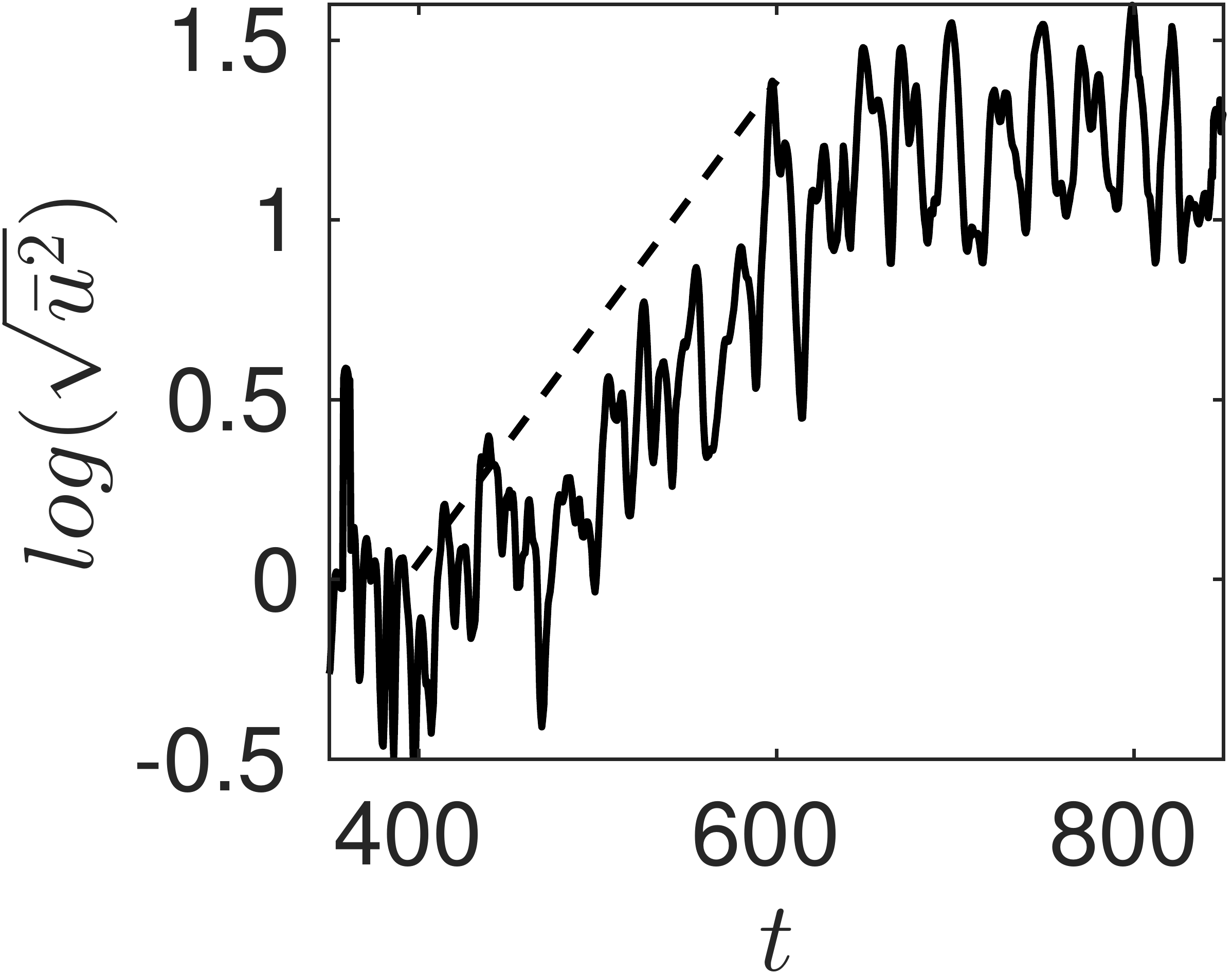}}}
\end{overpic}
}
\hfill
\subfigure
{
\begin{overpic}[height=.31\linewidth]{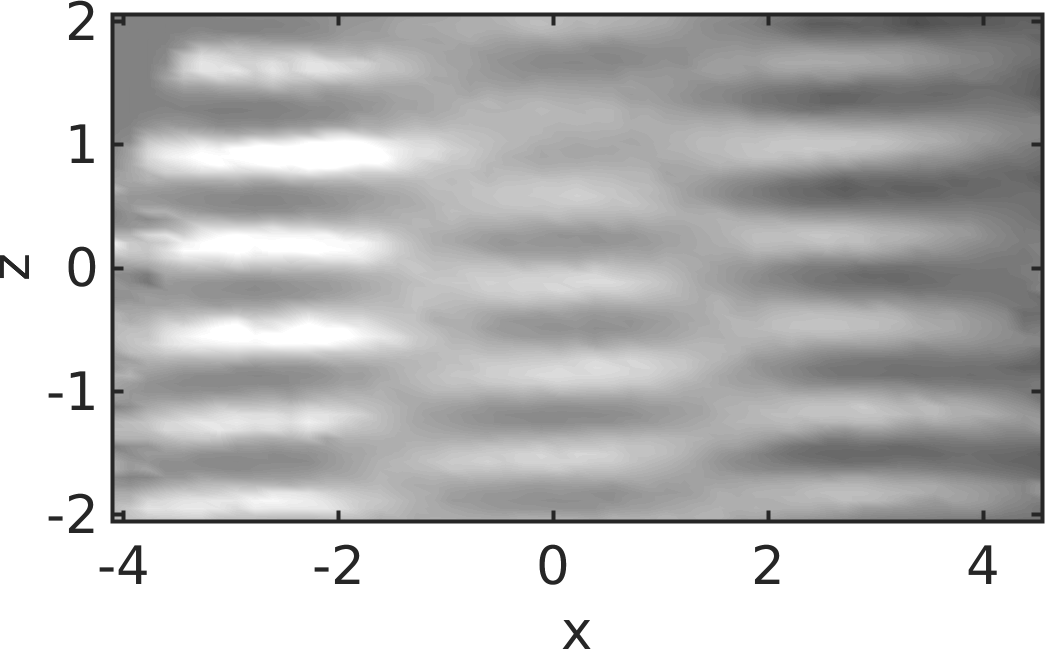}
\end{overpic}
}
\hfill
\caption{\label{fig:ExpRefCase} Top: Horizontal velocity perturbation $u$ at $x=y=0$ as a function of time. Colorbar is set to $\pm15\%$ of the wall speed. Here $\sigma=\unit[1.15]{s^{-1}}$, $Fr=0.82$, $Re=969$.
All quantities are dimensionless.
At 
$t=\unit[103.5]{}$  
the imposed shear switched from $\unit[0.34]{s^{-1}}$ to 
$\unit[1.15]{s^{-1}}$. 
Bottom left: Evolution of the mean horizontal
perturbation $({\overline{u^2}})^{1/2}$ as a function of time
for the same experiment (black line) and for a stable
experiment where both $Fr$ and $Re$ are diminished
by a fraction $1/8$ (red line). 
The inset shows $\log{({\overline{u^2}})^{1/2}}$ for the unstable case. 
At each time $\overline{u^2}$ is obtained  averaging over a short interval 
$\sim \unit[4]{}$
, taking the square and finally averaging over the vertical direction.
Bottom right: Snapshot of the horizontal velocity perturbation $u$ in the plane $y=0$, 
here 
$t\sim\unit[900]{}$.    
}
\end{figure}
\begin{figure}
\subfigure
{
\begin{overpic}[width=.3\linewidth]{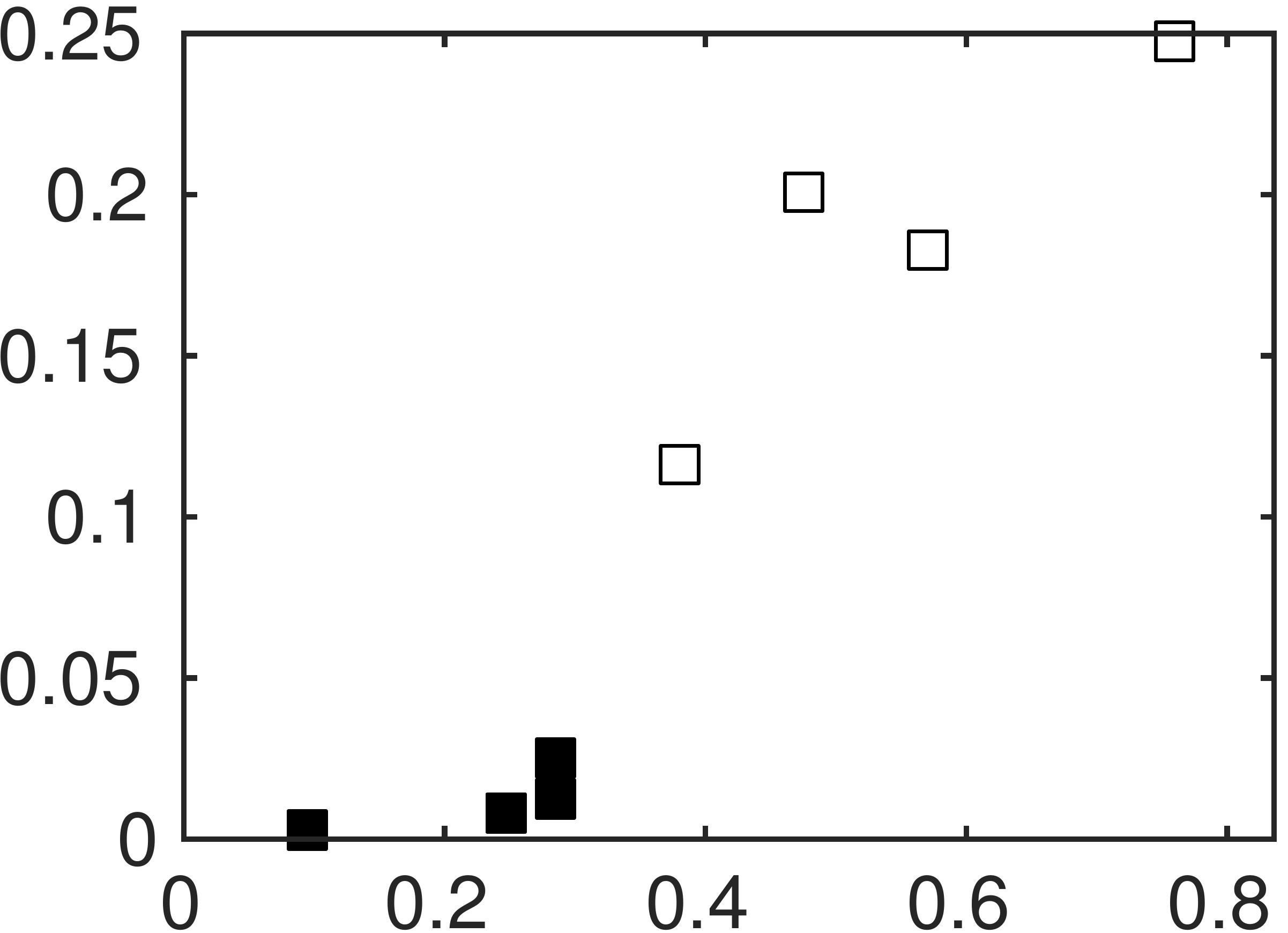}
\put (-15,-15) {\rotatebox{90}{\large 
$(\overline{u^2})^{1/2}(cm/s)$}}
\put (23,58) {\large\textbf{L1}}
\end{overpic}
}
\hfill
\subfigure
{
\begin{overpic}[width=.3\linewidth]{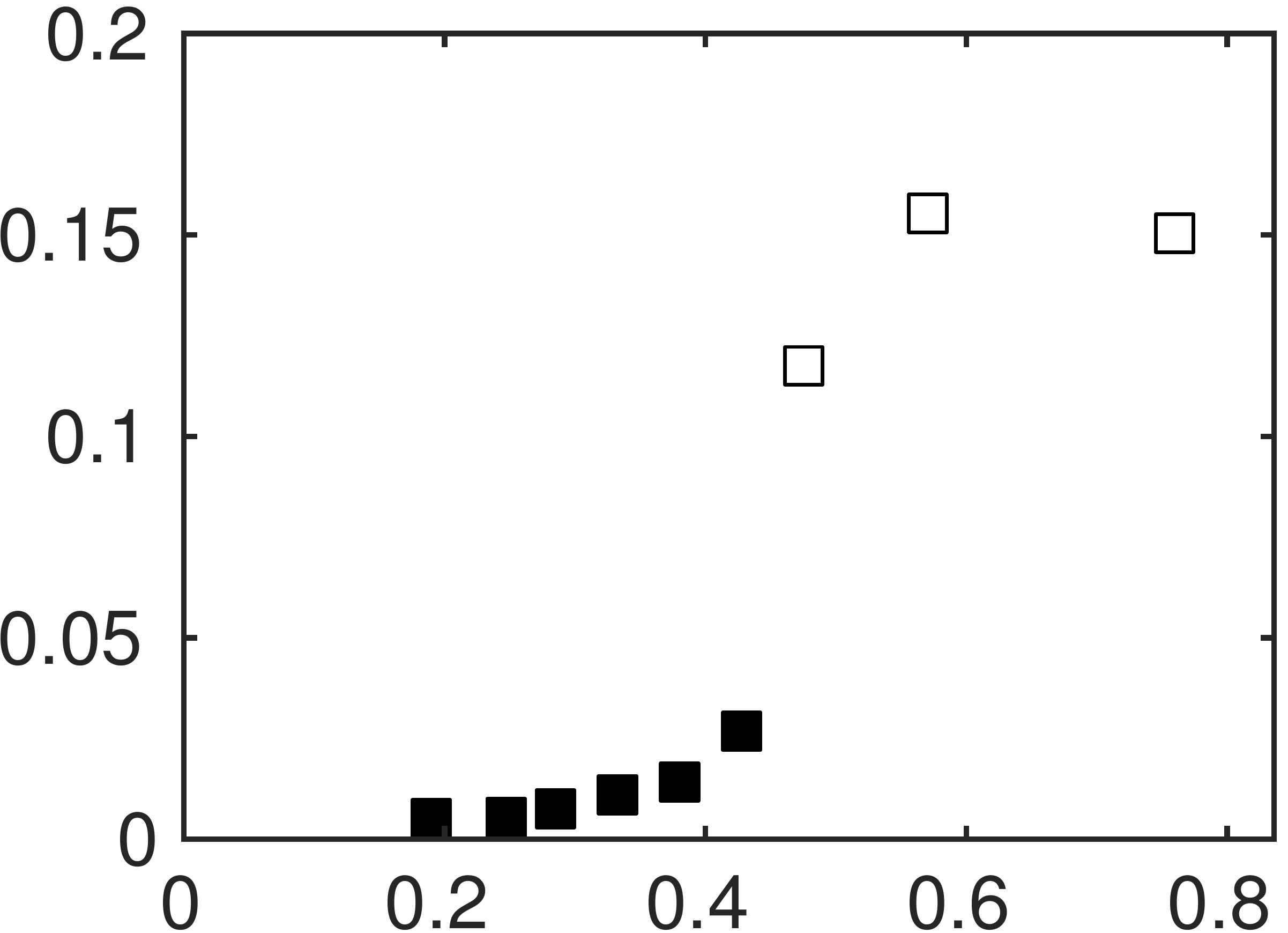}
\put (23,58) {\large\textbf{L2}}
\end{overpic}
}
\hfill
\subfigure
{
\begin{overpic}[width=.3\linewidth]{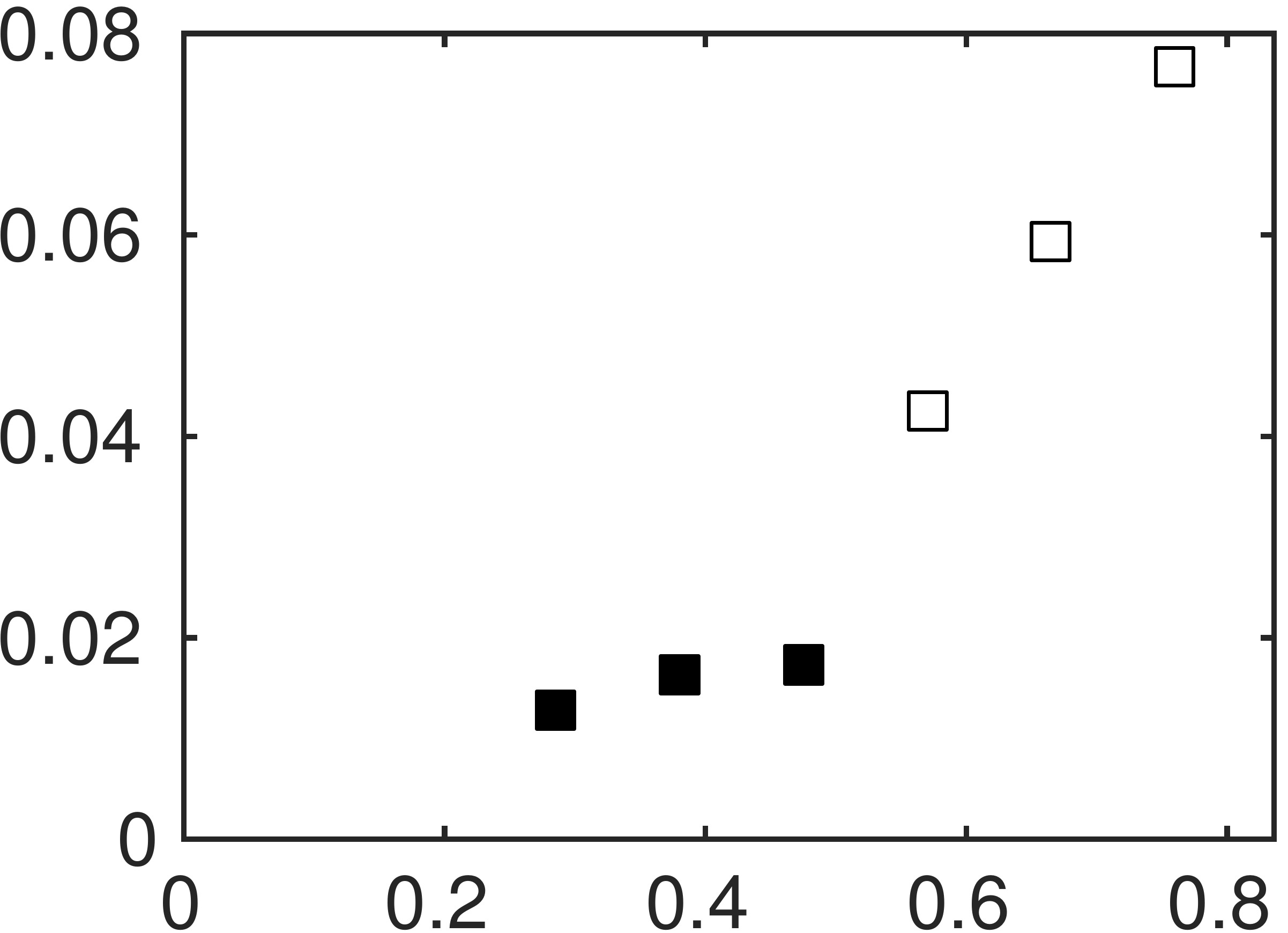}
\put (23,58) {\large\textbf{L3}}
\end{overpic}
}
\hfill
\subfigure
{
\begin{overpic}[width=.3\linewidth]{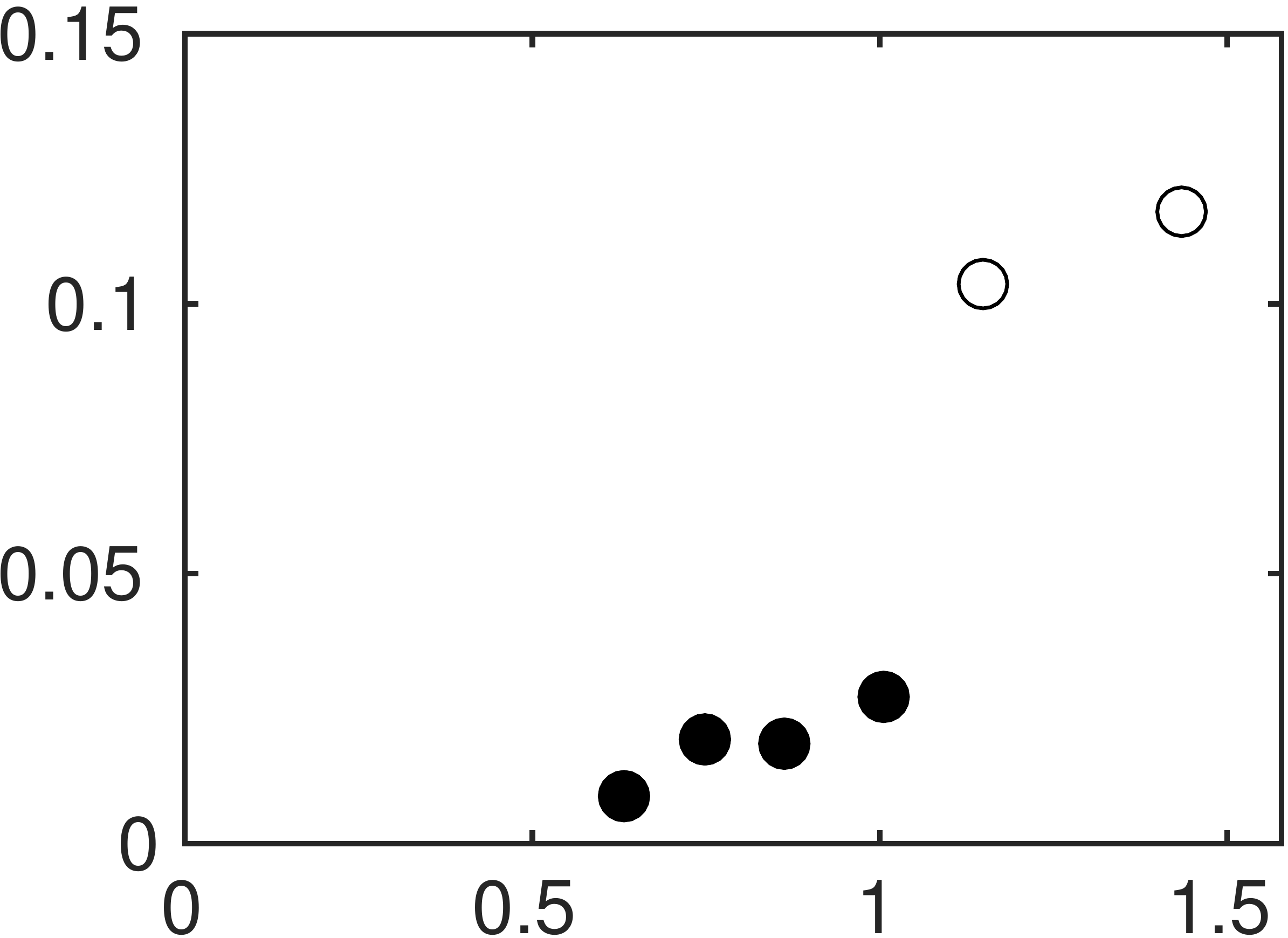}
\put (23,58) {\large\textbf{N1}}
\end{overpic}
}
\hfill
\subfigure
{
\begin{overpic}[width=.3\linewidth]{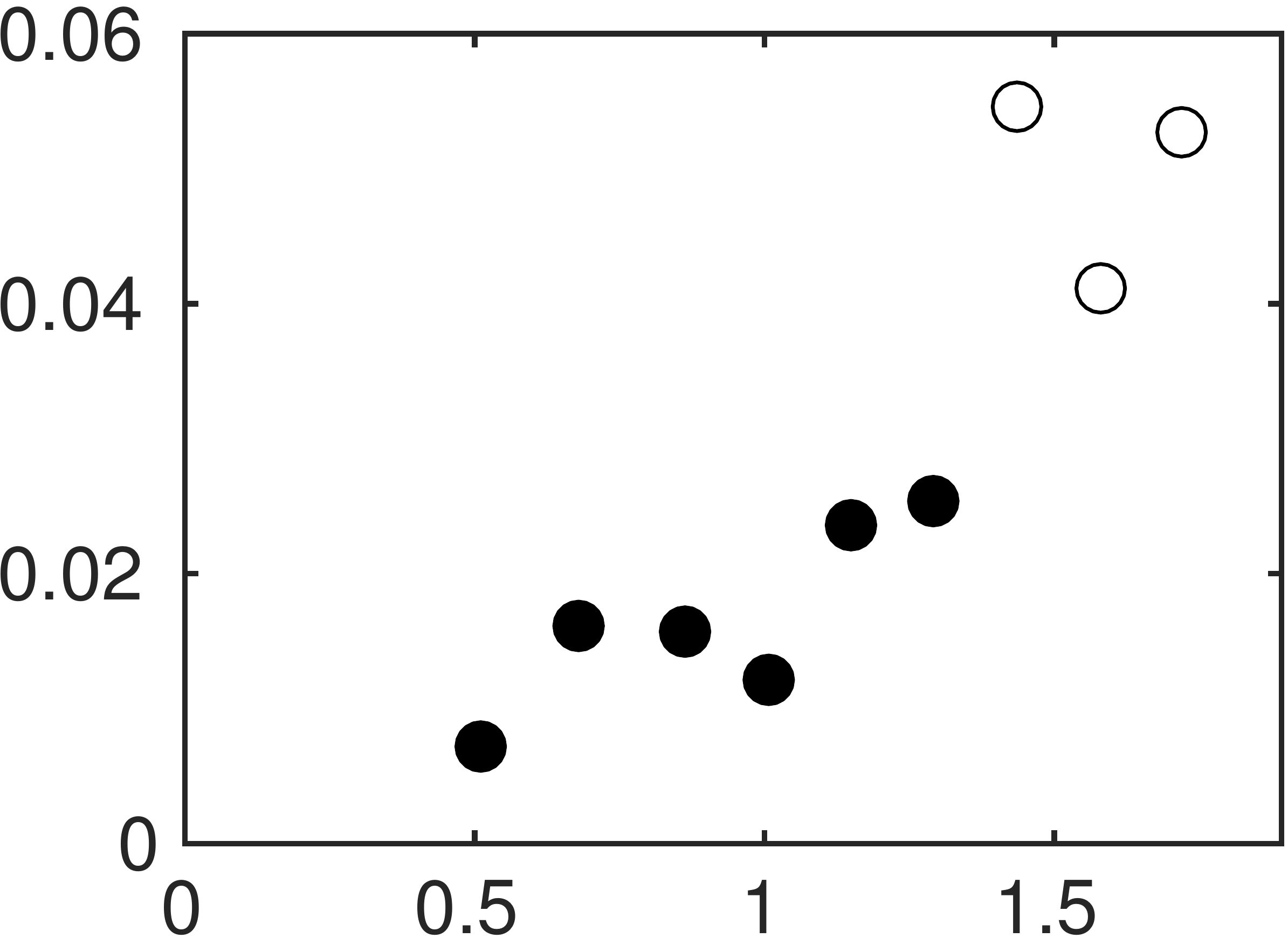}
\put (45,-9) {\large $\sigma(\unit[]{1/s})$}
\put (23,58) {\large\textbf{N2}}
\end{overpic}
}
\hfill
\subfigure
{
\begin{overpic}[width=.3\linewidth]{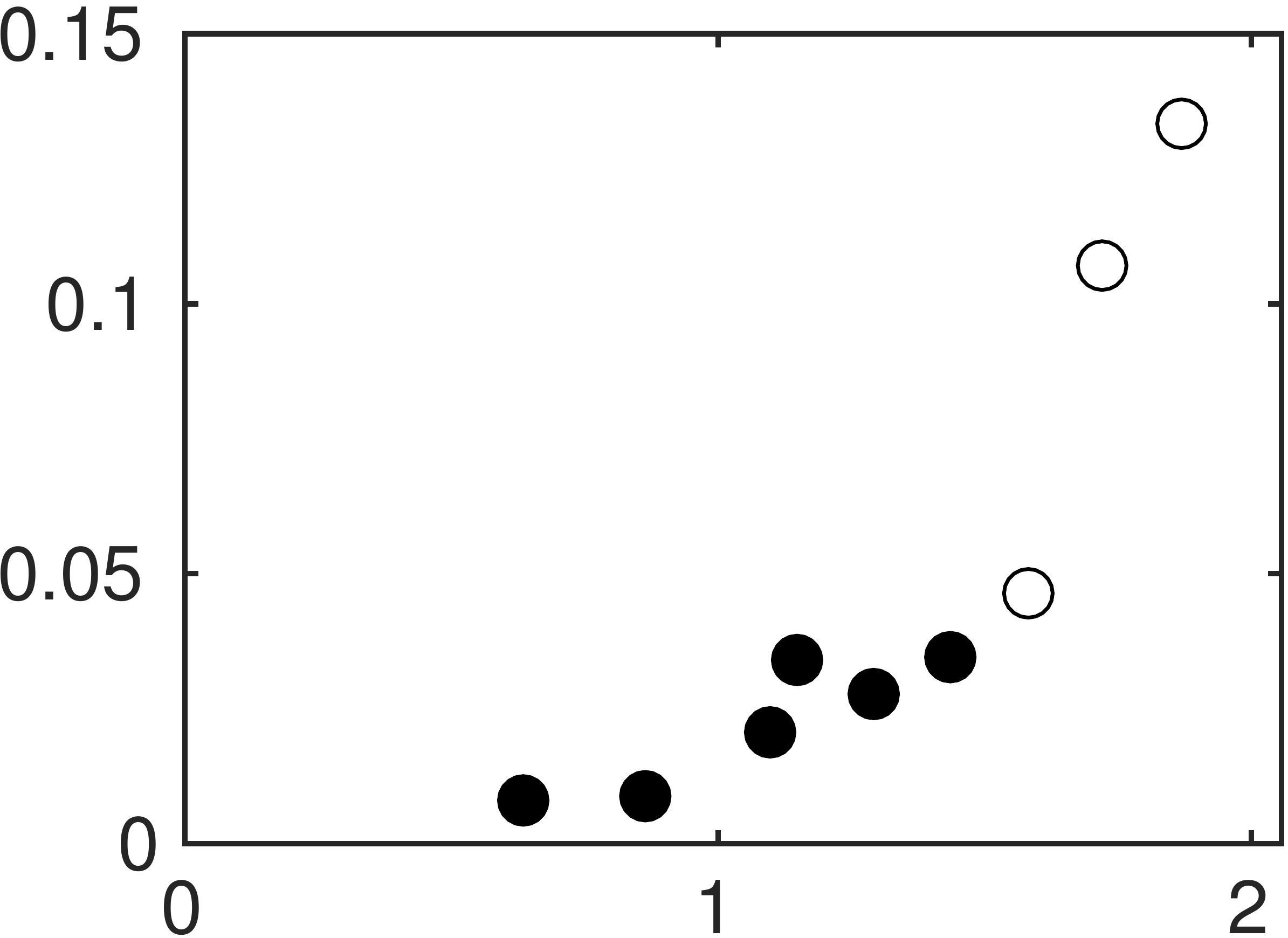}
\put (23,58) {\large\textbf{N3}}
\end{overpic}
}
\vspace{.8em}
\caption{\label{fig:Ampl_all}
Mean horizontal perturbation 
$({\overline{u^2}})^{1/2}$ as a function of the imposed shear $\sigma$.
The value of $\overline{u^2}$ is obtained averaging 
$u$ over the vertical direction $z$ and time.
Close and open symbols are assessed to be stable and unstable cases respectively.
Squares refer to large gap experiments (L) and
circles to narrow gap experiments (N).
Different experiments refer to different ratio $Fr/N$
as visible in figure \ref{fig:FrRe_LinExp}.}
\end{figure}

When the $Re$ number is sufficiently high $Re\gtrsim1000$ and for Froude number $Fr\sim1$, an exponentially growing motion is observed to form in the mid plane $y=0$.
The horizontal velocity perturbation $u$ shows a well defined spatial pattern where horizontal and vertical wavelengths $\lambda_x$, $\lambda_z$ can be fairly detected, with $\lambda_x/\lambda_z\sim 8$.
Results for this reference case are summarised
 in figure \ref{fig:ExpRefCase}.

On the top of figure \ref{fig:ExpRefCase} we plot the horizontal velocity perturbation $u$ at $x=y=0$ as a function of the time $t$ and vertical direction $z$ for a reference (unstable) experiment. 
At $t\sim\unit[60]{s}$ the imposed shear has changed from a lower value of $\sigma=\unit[0.34]{s^{-1}}$ to 
$\unit[1.15]{s^{-1}}$ and at $t\sim\unit[600]{s}$ one observes the appearance of a vertical wavelength.
One also observes that the $t-periodic$ and $z-invariant$ bulk motion described in the previous section is present since the very beginning and is still visible at large time,
superposed to the instability pattern.
In figure \ref{fig:ExpRefCase} (bottom left), we consider
the time evolution of  
the order parameter $({\overline{u^2}})^{1/2}$ for 
the same unstable case as above (black line)
and for another case where (red line) the imposed 
shear, $\sigma$ is
smaller by a fraction $1/8$. 
The average square of the horizontal perturbation 
$\overline{u^2}$ is computed at the center vertical line 
$x=0,y=0$ in three steps starting from the spatio-temporal diagram of $u$.
At each time we take the average of $u$ over a short interval 
$\sim \unit[4]{\sigma^{-1}}$, subtract 
the linear regression in $z$,
 compute the square and finally average over the vertical direction.
We stress that subtracting the linear regression allows us to get rid of the bulk oscillations and of any possible top-bottom anisotropy due to a non perfect verticality of the 
laser sheet.
First of all one remarks that the unstable case (black line) shows 
a clear growth event of the order parameter which does not
happen for the stable case (red line), thus
indicating the appearance of an instability.
Focusing on the unstable case one clearly sees that a first increase of $({\overline{u^2}})^{1/2}$ 
occurs during the interval of $\sim0.3T_{\nu}$ after the change in the imposed shear, where the value of the viscous time is $T_\nu=d^2/\nu
\sim\unit[1000]{\sigma^{-1}}$.
At larger time, $\overline{u^2}$ increases again, now in an exponential way (see the inset in semi-log scale), and finally saturates to a constant value.
The exponential growth rate is approximately
$\omega\sim\unit[0.06]{\sigma^{-1}}$ (although the noise 
makes difficult a precise measurement of the growth rate).
We claim that the first growing phase coincides with the progressive onset of the base PC flow at the imposed 
shear while the second growing phase corresponds to the onset of a linear instability.
Note that for the stable case 
(red line) the first growing phase is less visible, 
because the shear $\sigma$ is imposed 
starting from a slightly 
lower value.
Finally on the bottom right of figure \ref{fig:ExpRefCase} 
we present a snapshot of the $u$ field in the plane $y=0$.
One observes a regular periodic pattern characterized by 
a vertical wavelength $\lambda_z\sim\unit[0.7]{L_0}$ and
an horizontal wavelength $\lambda_z\sim\unit[5.5]{L_0}$,
where we recall that $L_0$ is half the width of the channel.

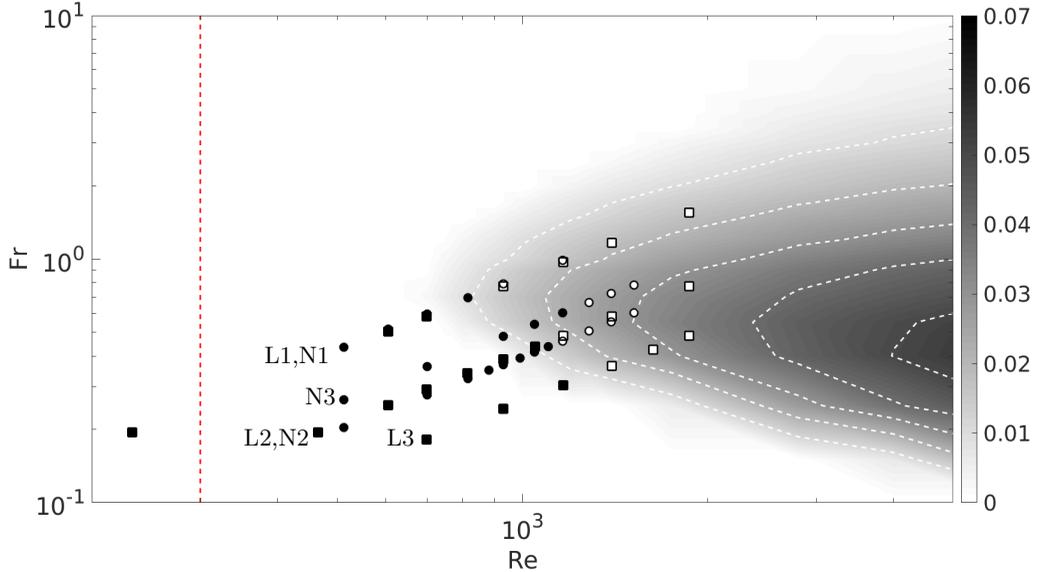
\begin{figure}
\begin{overpic}[width=1.\linewidth]{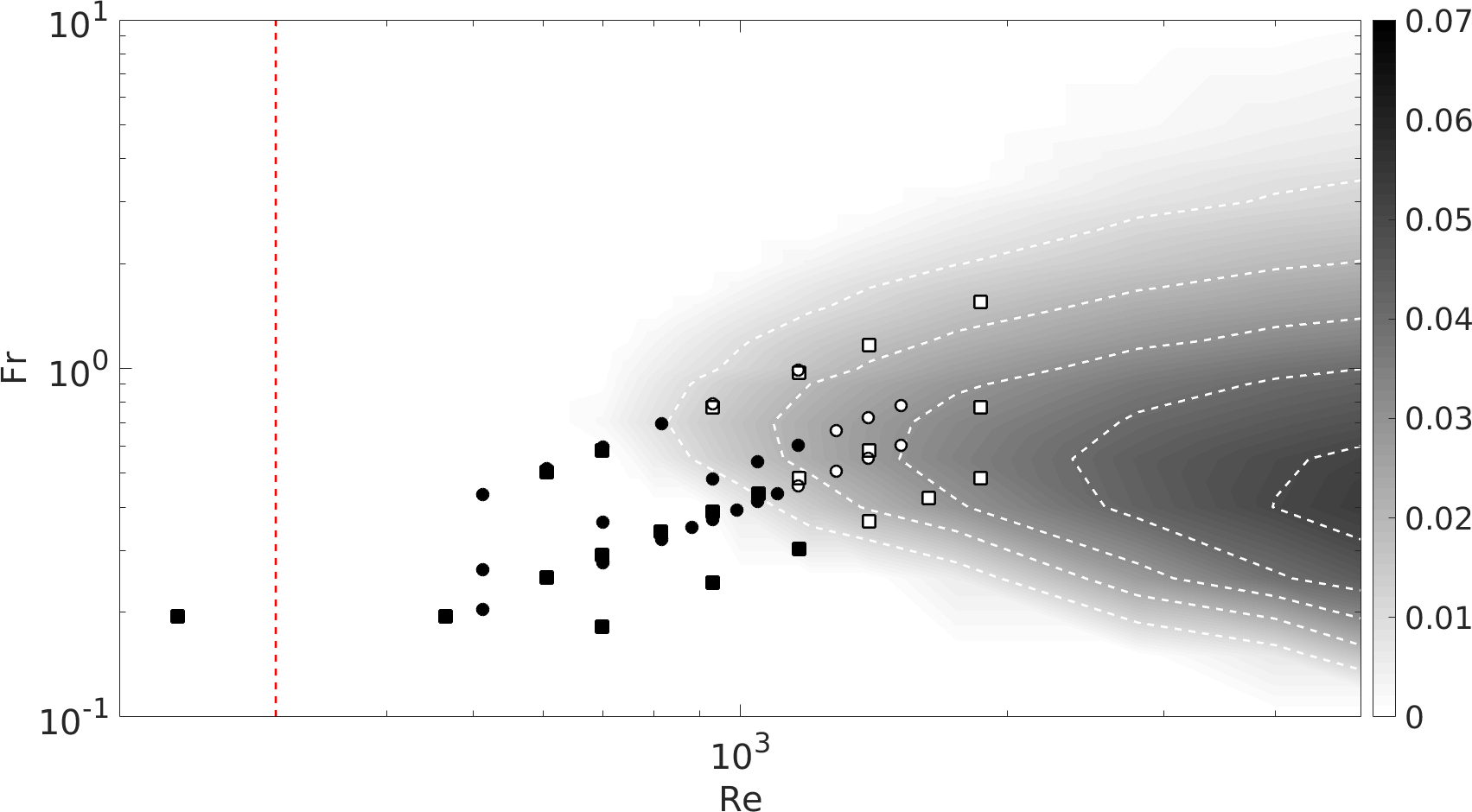}
\put (25,20) {L1,N1}
\put (29,16) {N3}
\put (23,12) {L2,N2}
\put (37, 12) {L3}
\end{overpic}
\caption{\label{fig:FrRe_LinExp}
Growth rate of the most unstable mode in the space $(Re,Fr)$
exactly as in figure \ref{fig:evalFrRe_all}.
Now we superpose the results of experiments (symbols) with
the same code as in figure \ref{fig:Ampl_all}.
One sees that transition from stable (closed symbols) to unstable (open symbols) cases happen close to the marginal
contour $Im(\omega)=0$.
The vertical dashed line corresponds to $Re=230$ at which 
the non stratified flow (i.e. pure water) becomes unstable. 
}
\end{figure}

Below we consider the stability of our experimental flow in the
$(Re,Fr)$ space which is the same for the linear stability analysis performed in section \ref{sec:stab_diagram}.
To this aim we need to define a common protocol to
assess the presence or not of the instability. 
One criterion may be the appearance of a vertical length.
Unfortunately the latter is a smooth process, for example
a vertical wavelength was often visible
at a shear $\sigma$ lower than what we assess to be the unstable case. 
Nonetheless the associated signal was generally weak and no growth process was observed.
The existence of the latter seems to be the most reliable
criterion, but demands longer experiments and
generally imposes to start from very small $\sigma$,
which imposes to slow down the flow after each experiment.
As a general rule we rather look at the saturated amplitude
of the order parameter $({\overline{u^2}})^{1/2}$ 
as a function of $\sigma$ and detect if an abrupt
change occurs, as it is clearly visible in figure \ref{fig:ExpRefCase} at large time.
In particular $({\overline{u^2}})^{1/2}$ is computed once the instability has saturated or alternatively after a time of the order of the viscous time $T_\nu$ after the actual value of the shear is imposed, for example to get rid of the base flow transient.
We choose the control parameter $\sigma$ (i.e. the imposed shear) as the most suitable one,  because it can be varied continuously, simply controlling the speed of the entraining motor.
As a drawback both $Re$ and $Fr$ are linear in $\sigma$, thus the stability diagram must be explored moving on tilted straight lines for which the ratio $Fr/Re=\nu/Nd^2$ is constant. 
Any change in the vertical stratification $N$ and gap width $d$
is considerably more laborious, which constrains
the exploration of the $(Re,Fr)$ space to a few different 
$Fr/Re=const$ lines.
In figure \ref{fig:Ampl_all} we report the value of 
$({\overline{u^2}})^{1/2}$ as a function of the control parameter $\sigma$
for a collection of $6$ series of experiments at different
$Fr/Re$ corresponding to different values of $N$ and $d$.
Experiments labelled with $L$ refer to large gap experiments ($D/d=\unit[2.4]{}$) while those labelled with $N$ 
refer to narrow gap experiments ($D/d=\unit[5.7]{}$).
Different experiments are performed at a different ratio $Fr/Re$
as it is visible in figure \ref{fig:FrRe_LinExp}. 
One remarks that in almost all the different series,
the quantity 
$({\overline{u^2}})^{1/2}$ abruptly increases  
when the imposed shear $\sigma_c$ crosses a threshold value, which corresponds to a
threshold point in the space $(Re_c,Fr_c)$.
This allows to assess that the experimental flow is stable (closed symbols) for
 $\sigma<\sigma_c$ and unstable (open symbols) for $\sigma\geq\sigma_c$.
An exception was constituted by the experiment $N2$ 
where the vertical pattern establishes but appears
less structured and less robust than in the other cases.
Also in the case $N3$ the transition stable-unstable
is anticipated by one position because a growth event was 
clearly visible at the corresponding value of $\sigma$.

\section{Discussion}\label{sec:discussion}
In figure \ref{fig:FrRe_LinExp} we superimpose the experimental results of figure \ref{fig:Ampl_all} on 
the stability diagram obtained with the linear stability analysis (figure \ref{fig:evalFrRe_all}).
Experimental points which belong to the same line correspond to 
a series of observation at fixed $Fr/Re$ ratio.
Closed symbols and open symbols correspond to stable and unstable cases respectively, as in figure \ref{fig:Ampl_all}.
One observes that the transition of 
the quantity $(\overline{u^2})^{1/2}$ 
(i.e. from close to open symbols) happens
close to the marginal contour where linear growing modes
appear according to linear stability analysis.
This strongly supports the claim that we experimentally observe
the signature of the instability predicted by the 
linear analysis.

\begin{figure}
\centering
\begin{overpic}[width=.6\linewidth]{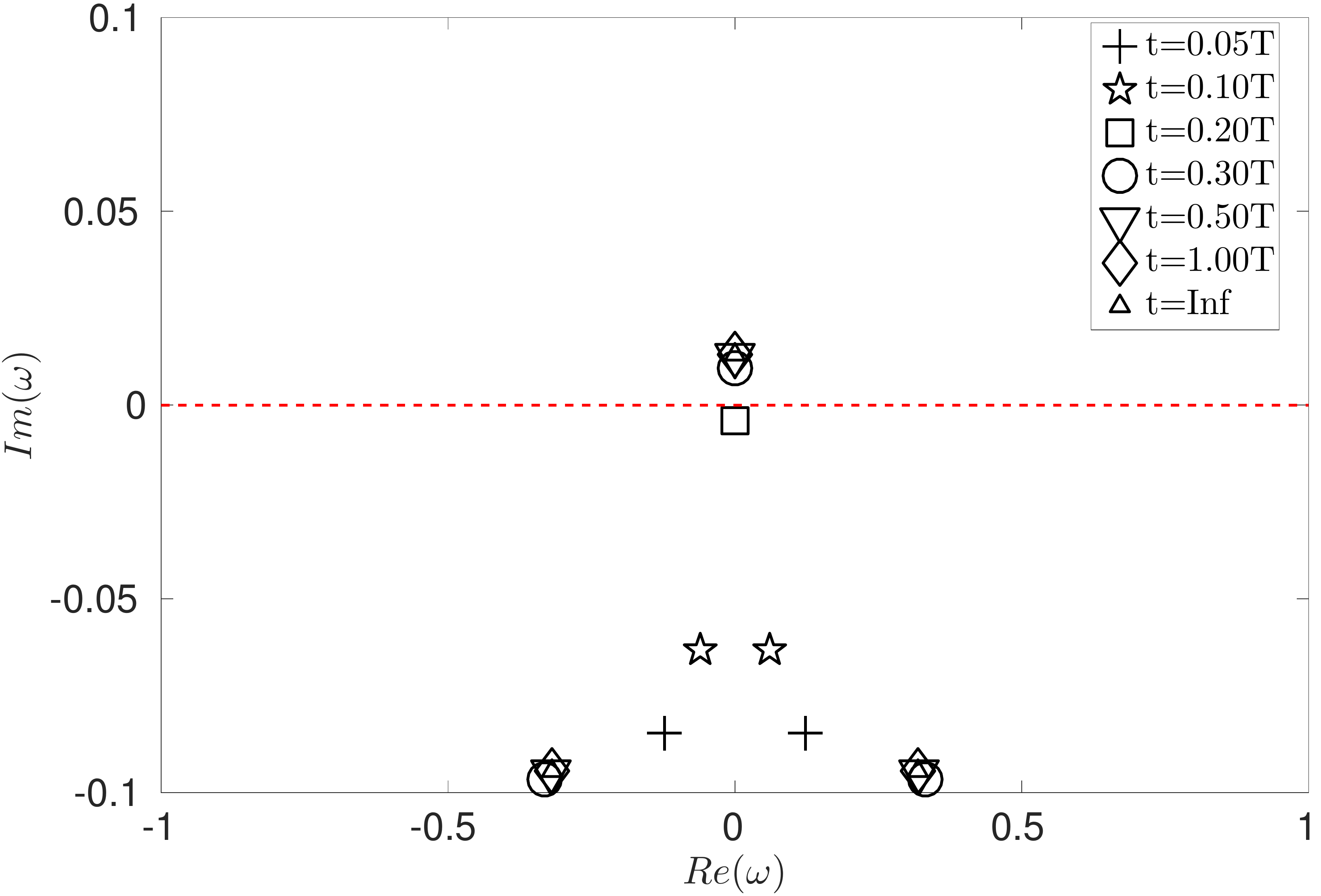}
\end{overpic}
\caption{\label{fig:eval_transient}
Solutions of the modified eigenvalue problem for a base flow given by expression (\ref{eq.cou_transient}).
Here we choose the same $Re$, $Fr$, and $U_i$ (i.e. the belt velocity at $t=0$) as the reference experiment.
The eigenvalues problem is solved for the combination $(k_x,k_z)$ which is the most unstable  according to linear analysis and at seven different transient times.
Different symbols correspond to different times.
The dashed line marks the limit for instability.}
\end{figure}

Below we compare the temporal behaviour of the observed instability
with the linear analysis.
Besides the growth rate, that precisely characterizes the instability onset, 
we want to discuss first what happens during the transient phase that 
necessarily comes with each experiment.
This constitutes a difference with the linear analysis where the base flow is always constant, and may affect the estimation of the observed growth rate. 
In other words, one may wonder at which time since the beginning of an experiment the instability is expected to grow.
The question becomes particularly relevant when considering that the expected growth rate is comparable with and even smaller than the viscous time.
Also we want to rule out the possibility that
the appearance of the unstable pattern is due
rather to the transient profile of our flow,
for example a non constant shear profile like
that considered by \cite{Chen2016}.
The temporal diagram of figure \ref{fig:ExpRefCase} shows that the exponential 
growth starts at $t\sim T_{\nu}/3$, which seems to be consistent with the description of the base flow 
given in section \ref{sec:base_flow}.
In order to give a more quantitative explanation we solved a modified eigenvalue
problem where the base flow is now given by the expression (\ref{eq.cou_transient})
for a collection of different times and with the same set of parameters as
the reference experiment presented in figure \ref{fig:ExpRefCase}.
The results are presented in figure \ref{fig:eval_transient}, where we report the eigenvalues of the most unstable mode focusing close to the transition region
$Im(\omega)=0$.
One remarks that no unstable eigenvalue is present for $t<0.2T_{\nu}$ while
one unstable mode appears for $t\geq0.3T_{\nu}$, thus confirming
that the base flow must be sufficiently established for the instability to develop.
This result was confirmed by specific DNS where
the initial condition is not the PC flow, but the 
flow is at rest and the shear profile is progressively
established through the no slip boundaries.
Also in this case the growth of perturbations is
delayed to the moment when the shear profile
has become almost constant.
We then conclude that what we observe is associated
with a constant shear plane Couette profile.
%

Besides instability threshold in the $Re$, $Fr$ 
space we also want to compare the shape of the mode
selected in our experimental device with the
unstable mode predicted by linear theory.
If we focus again on the reference case described
in figure \ref{fig:ExpRefCase}, we remark that
both $k_x$ and $k_z$ are larger than what is predicted
by the linear theory for the most unstable mode, and
the perturbation is oscillating in time
while linear theory predicts that instability appears
as a stationary mode.
One remarks that oscillations are quite regular and 
relatively slow, with a typical 
period $T=\unit[43\pm 3]{\sigma^{-1}}$, where
the uncertainty is taken as the width at
mid height of the peak in the average temporal
spectrum. 
We recall that according to linear analysis
oscillatory branches also exist (see figure \ref{fig:evalFR02_1_5Re1000})
which appear at higher $Re$ number, typically
$Re\gtrsim 2000$.
Interestingly the period associated to the first 
oscillating branch is always long, for example with the
$Fr$ of the reference experiment $N1$ and  
$Re=2500$ one has $T\sim\unit[42]{\sigma^{-1}}$.
Moreover this branch happens at larger wave numbers more
 compatible with the observed ones.
In this scenario what we look at may be either a single
propagative mode, for example of the type (b), as it
is visible in the lower part of the spatio-temporal diagram,
or a standing wave generated by two counterpropagating 
modes of type (b) as it is visible in the upper part of the 
same diagram.
These elements suggest that experimentally, 
the instability
is activated close to the absolute threshold 
(i.e. that where first stationary modes appear) predicted by linear analysis,
but a different non stationary mode is selected in the end.

In any case one should recall that our linear and experimental
problems are different, thus unstable modes are not expected to
share the same features.
The major difference between the theoretical system
and our experimental setup is the finite size
of the domain.
In principle, to mimic periodic boundary conditions imposed
in linear calculations, one wants to take the horizontal 
and vertical aspect ratio $D/d\gg1$ and $H/d\gg1$ 
while our best realization (i.e. narrow gap) of this hypothesis  was $D/d=5.7$ and $H/d=2.4$.
We observe that the impact of physical confinement is twofold. 
First from the point of view of modal analysis only the 
wavelengths which fit in the domain may have a chance to develop. 
This was confirmed by periodic DNS that do not show any instability whenever the box size does not fit the spatial shape of unstable modes.
We notice that the eigenvalue problem is solved assigning 
an arbitrary value of $k_x$ and $k_z$, thus when comparing to DNS and experiments one should retain that the ($k_x,k_z$) grid
of figure \ref{fig:evalFR1Re1000} is coarsed-grain,
especially at low $k_x$ and $k_z$.
Thus the ideal constraint for the aspect
ratio are $D/d\gg \lambda_x$ and $H/d\gg \lambda_z$, where $\lambda_x$ and $\lambda_z$ are the non dimensional
wavelengths of the unstable mode we want to observe.
A second problem appears in the stream-wise direction because the streamlines must turn and close up when getting close to the walls that close the domain in the stream-wise direction, as it
is clearly visible in the snapshot reported in figure \ref{fig:cou_profile_transient}.
We mentioned before that this feature does not modify significantly the
shape of the base flow in the bulk, but it may locally destabilize the flow (i.e. close to the corners) and successively affect the stability of the whole domain.
\subsection{Simulations in a finite domain}\label{sec:DNS_confined}
In order to closer investigate finite size effects, 
we performed new DNS where the 
computational domain is now closed 
in the stream-wise direction
by two solid walls with no-slip insulating boundary conditions
as sketched in figure \ref{fig:sh_device} (right),
while boundary conditions remain periodic in the vertical
direction.
Note that compared with the periodic case discussed in section \ref{sec:DNS} 
the mesh is further refined close to the two
additional stream-wise boundaries,
in order to properly solve for
the boundary layers.
In addition, the corners of the domain are now singular due to
the incompatibility between the velocity imposed at the side boundaries and the fixed stream-wise walls.
This is naturally smoothed by viscosity but is nevertheless
an inevitable source of vorticity.
In figure \ref{fig:exp_DNS} (top) we report the results of a 
DNS confined simulation which reproduces both the control 
parameters ($Fr$,$Re$) and the aspect ratio of the
reference experiment illustrated in figure \ref{fig:ExpRefCase}.
For a direct comparison we report again the results of the
reference experiment (bottom) already shown in figure
\ref{fig:ExpRefCase} with the only difference
that bulk oscillations are now filtered from the spatio temporal
diagram of $u$ at $x=y=0$ and the origin of time axis
is shifted forward to $t=T_{\nu}/3$ which is when 
we estimate that the 
PC flow is well established.
One observes a striking good agreement between 
our DNS results  and
experimental result
on both the spatial and temporal shapes of the selected mode.
Computing the temporal spectrum we find that the
temporal frequency as predicted by DNS is 
$T=\unit[50\pm 5]{\sigma^{-1}}$
which is compatible with the experimental one, while the
consistency of spatial wavelengths is evident because
in both cases an integer number of velocity maxima fit in 
the vertical mid plane $y=0$.
Finally we observe that as a whole the transition from 
the initial noise to the final non linear pattern takes
almost the same time in DNS and
the experiment.
As a summary, figure \ref{fig:exp_DNS} indicates that 
we correctly isolated the crucial factor which possibly
alters the selection of the unstable mode, 
that is the stream-wise confinement.
Incidentally we also report that additional DNS show
that the form of the late non-linear stage is quite sensitive
to initial and boundary conditions.
For example slightly varying the box dimensions or the amount of
initial noise the spatial shape of the selected
mode is different and travelling waves or 
standing waves patterns can be alternatively present. 
To better investigate the role of
stream-wise boundaries we perform 
additional DNS with a non stratified
($Fr=\infty$) PC flow in exactly the same confined geometry and same $Re$ number as the one
we just described (figure \ref{fig:exp_DNS}).
Results are reported in figure 
\ref{fig:nostrat_periodic}. 
We recall that $u$ perturbations do not grow when considering
periodic boundary conditions at $Fr=\infty$, as expected by the fact that the unstratified PC flow is linearly stable.
One observes that a strong 
vertical shear appears in a shorter time
compared to both the stratified 
DNS and experiment in figure \ref{fig:exp_DNS}
while the vertical length scale is larger.
A similar pattern is also observed in unstratified 
experiment as soon as $Re\gtrsim300$.
The vertical kinetic energy (black dashed line in figure
\ref{fig:GR_sim_all}) grows in an exponential way
suggesting that a linear instability may act at a first stage.
Conversely non linear mechanisms must act at larger time 
because the vertical kinetic energy rapidly saturates and
the spatio-temporal diagram of figure 
\ref{fig:nostrat_periodic} shows the presence of small scale features.
A detailed investigation of the nature of 
this instability of a confined
and unstratified PC flow is beyond the scope of the 
present work.
Formally we recognise the scenario 
of a locally perturbed PC flow 
already described in literature 
\citep[see][and citations inside]{Barkley1999}, 
where some of the (stable) modes of the ideal PC flow 
become unstable.
In particular the shape of the flow at the corners 
suggests that it may locally destabilise via 
centrifugal instability which will be studied in 
further investigations. 

We conclude that the presence of boundaries may destabilize the PC flow with or without the presence of stratification.
Now, the careful reader will agree that even if DNS fully 
justify differences between the linear analysis and the observed experimental pattern we are left with a cumbersome question
regarding the origin of the perturbation pattern observed
in the confined and stratified configuration:
does this pattern coincide with a boundary-induced 
modification of the linear instability, or rather with
the pure hydrodynamic boundary-induced instability modified by the stratification?
For example one may claim that the patterns observed in figure 
\ref{fig:exp_DNS} are a convoluted combination of the 
two showed in figure \ref{fig:DNS_GR_pattern} and \ref{fig:nostrat_periodic}. 
%
%
%
%
\def\s{.48}	
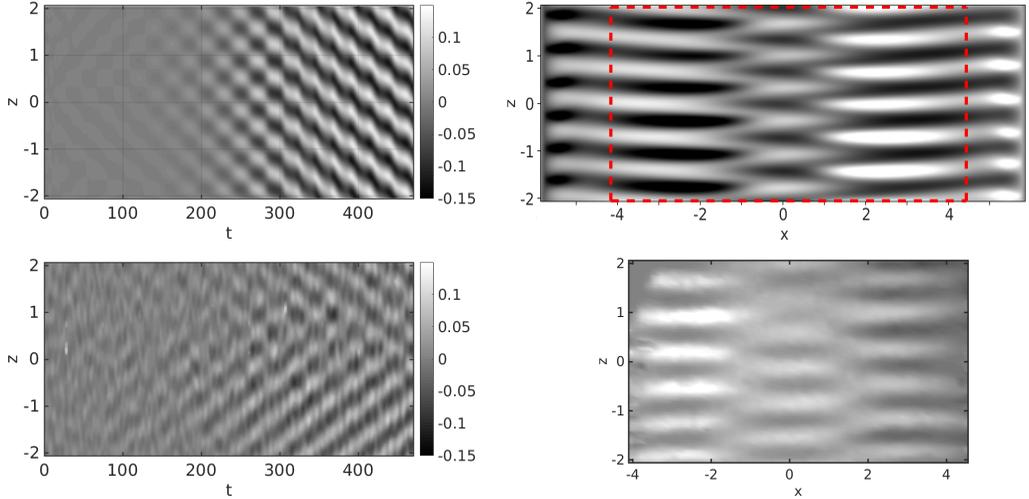
\begin{figure}
\begin{minipage}{0.48\linewidth}
\begin{overpic}[height=\s\linewidth]{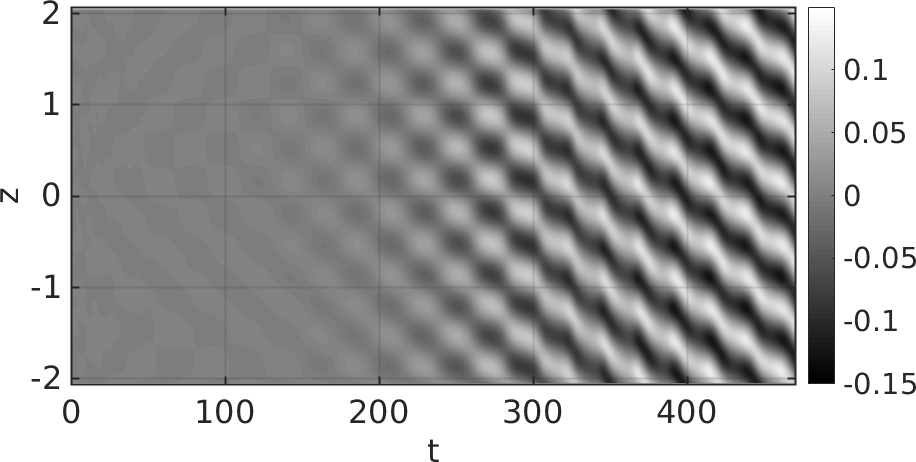}
\end{overpic}
\vspace{0.8em}

\noindent
\begin{overpic}[height=\s\linewidth]{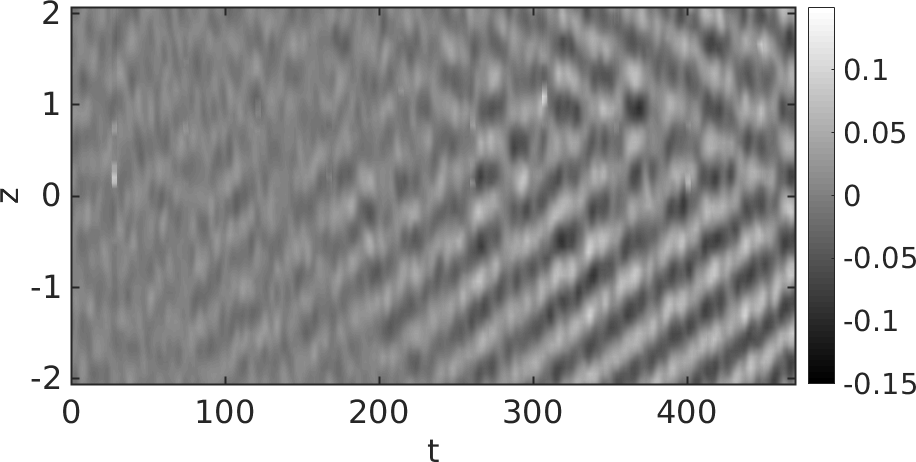}
\end{overpic}
\end{minipage}
\hfill
\begin{minipage}{0.48\linewidth}
\hspace{-1.6em}
\begin{overpic}[height=\s\linewidth]{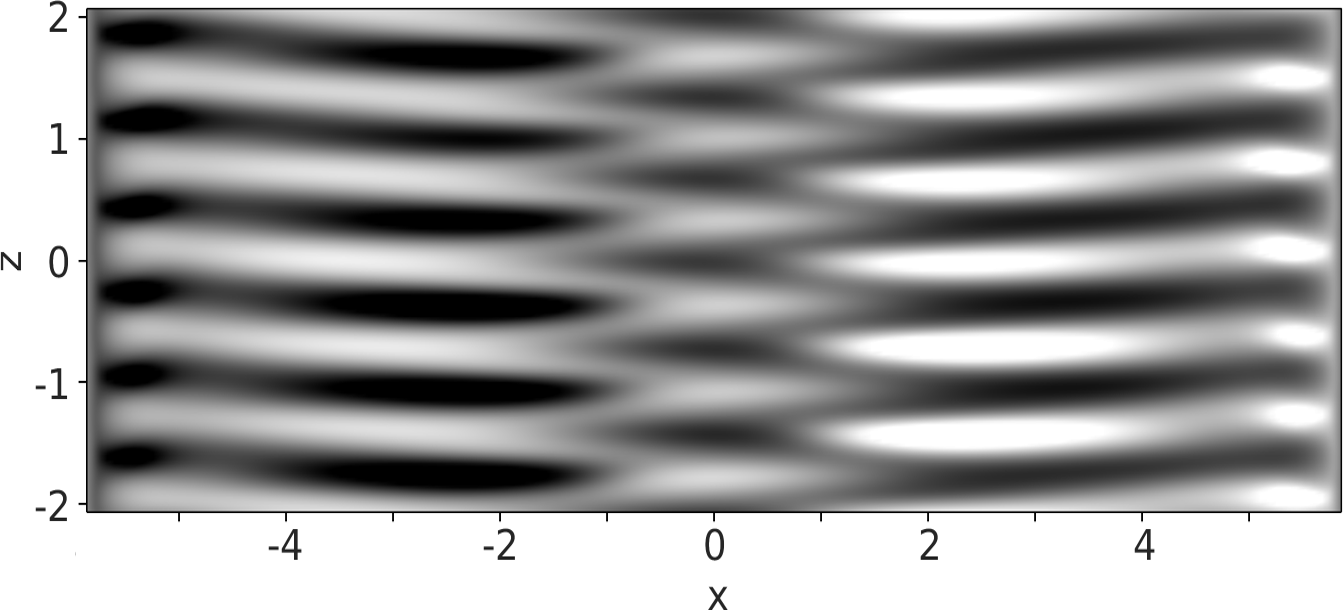}
\put (18,7) {
\begin{tikzpicture}
\def\LZ{2.55}
\def\LX{4.7}
\coordinate (C) at (0.1,0);
\draw [red,very thick,dashed] ($(C)$) -- ($(C)+(\LX,0)$);
\draw [red,very thick,dashed] ($(C)$) -- ($(C)+(0,\LZ)$);
\draw [red,very thick,dashed] ($(C)+(\LX,0)$) -- 
($(C)+(\LX,\LZ)$);
\draw [red,very thick,dashed] ($(C)+(0,\LZ)$) -- 
($(C)+(\LX,\LZ)$);
\end{tikzpicture}
}
\end{overpic}

\vspace{0.8em}

\noindent
\centering
\begin{overpic}[height=\s\linewidth]{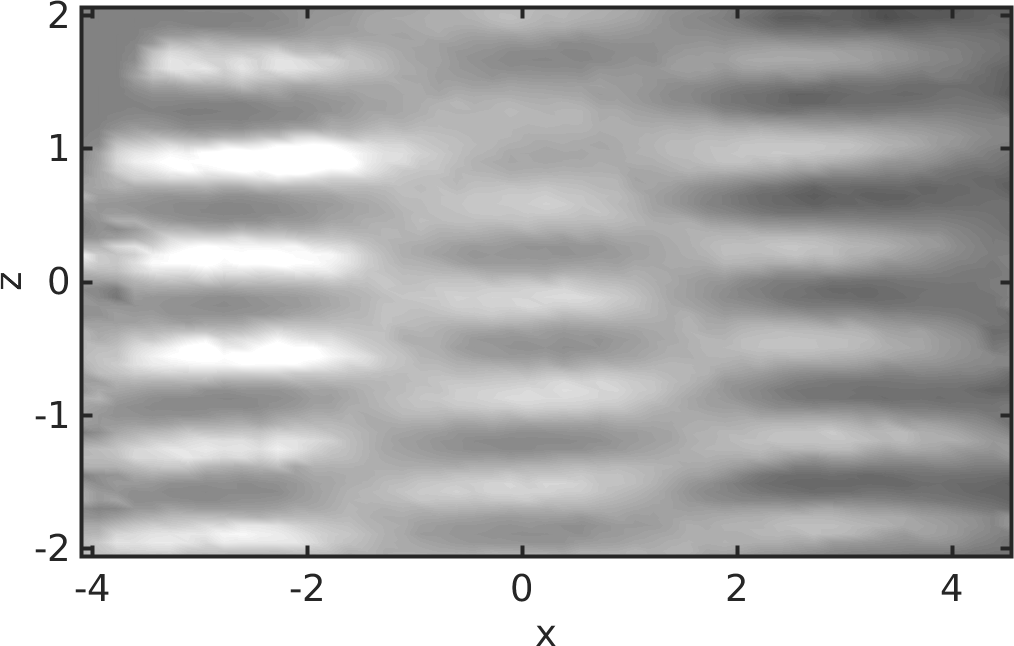}
\end{overpic}
\end{minipage}
\caption{\label{fig:exp_DNS}
Left: Spatio-temporal diagram of the perturbation $u$ at the center line $x=0$, $y=0$ for the reference case $Re=969$, $Fr=0.82$
for confined DNS (top) and the reference
 experiment (bottom).
Spurious bulk oscillations are filtered 
from the experimental data.
Right: perturbation $u$ in the plane $y=0$ once the 
flow has become unstable for confined DNS (top) and 
the experiment (bottom).
The red dashed rectangle indicates the area 
accessible to experimental measurements.}
\end{figure}
\begin{figure}
\centering
\begin{overpic}[width=.68\linewidth]{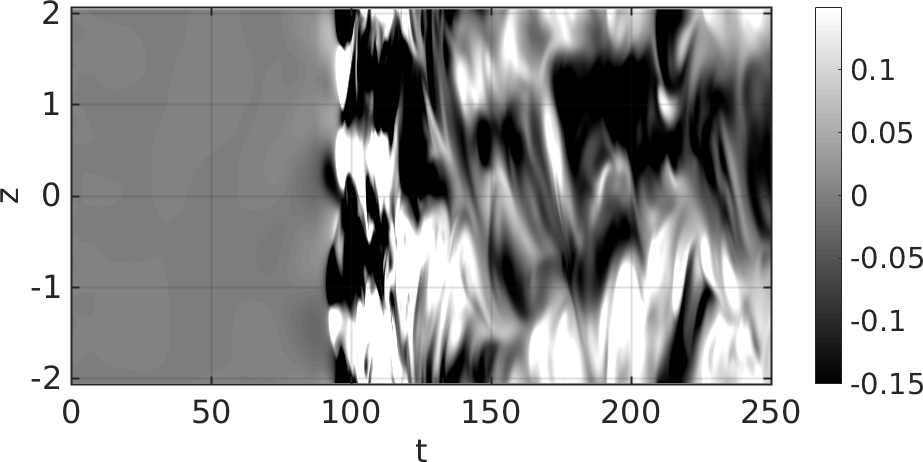}
\end{overpic}
\caption{\label{fig:nostrat_periodic}
Spatio temporal diagram of the perturbation $u$ at the center line $x=0$, $y=0$ from DNS.
Here $Re=966$, $Fr=\infty$ (i.e. unstratified) 
and the box geometry 
is the same as 
in the experiment.}
\end{figure}
The key to this answer resides in the same approach we followed 
with experiments, that is to detect if and when, our flow
abruptly changes when varying the control parameter $\sigma$.
We then consider further DNS which copy 
the parameters of another experiment of our reference 
series $N1$ where $Fr$ and $Re$ were 30\% smaller than the
unstable case described in figure \ref{fig:ExpRefCase}.
At the same time we repeat unstratified simulations 
at such a lower value of the $Re$ number.
We observe that the new unstratified case
is almost unchanged while the 
subcritical stratified case shows a dramatic change.
In this case perturbations are significant only close to the boundaries,
and no instability develops in the bulk.
In figure \ref{fig:GR_sim_all} we report the time
evolution of the vertical kinetic energy for all the stream-wise
confined DNS we have discussed above,
together with the one performed in a periodic domain.
One observes that in the unstratified case (dashed line), 
perturbations rapidly grow and saturate at the same value
independently of the $Re$ number.
If we add stratification (solid lines), perturbation grows
(black line) and saturates to the same value as periodic
simulation if $Re$ number is beyond the threshold predicted by
  the linear analysis. Conversely perturbations are damped 
(dashed line) when the $Re$ number is below the threshold.
\begin{figure}
\centering
\begin{overpic}[width=.88\linewidth]{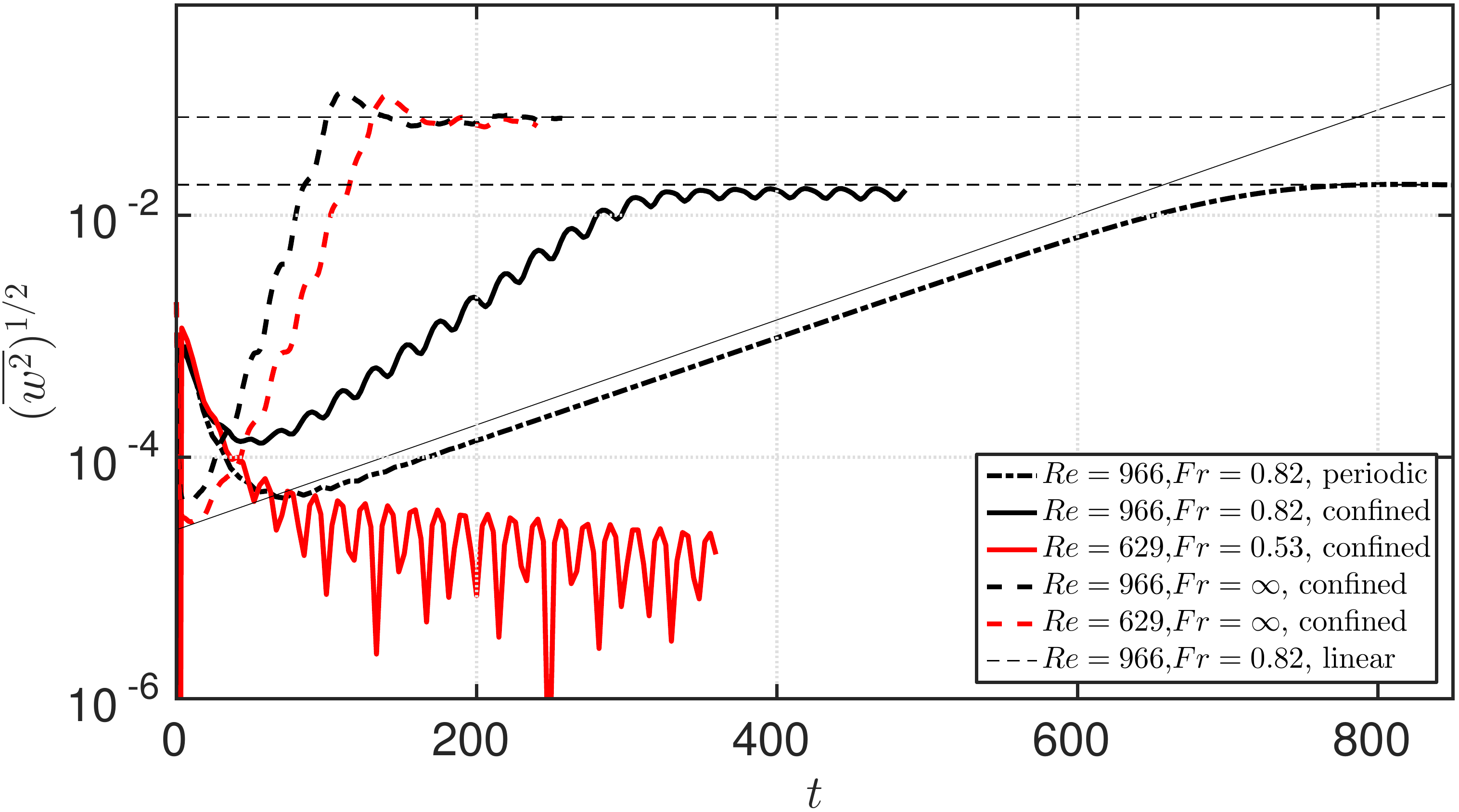}
\end{overpic}
\caption{\label{fig:GR_sim_all}
Vertical kinetic energy density $({\overline{{w}^2}})^{1/2}$ 
as a function of time, for five different DNS experiments.
Black lines correspond to supercritical simulations ($Re=966$)
performed at $Fr=0.82$ (solid line), $Fr=\infty$ (dashed line), and $Fr=0.82$ with periodic boundary conditions in the
stream-wide direction (dash-dotted line).
Red lines refer to subcritical simulations ($Re=629$) performed
at $Fr=0.53$ (solid line) and $Fr=\infty$ (dashed line).
The thin black line corresponds to the growth
of the most unstable mode at $Re=966$, $Fr=0.82$
as predicted by the linear theory.
Horizontal dashed lines highlight the saturation level
of supercritical unstratified and stratified DNS respectively. 
}
\end{figure}

We have now enough elements to
conclude that what we observe both in 
experiments and DNS 
(figure \ref{fig:exp_DNS}) is a signature
of the linear instability of a PC flow, 
vertically stratified in density.
Besides we observe that stream-wise boundaries are source of instability for a flow and likely affect
the features, or possibly just the selection, of the unstable
mode which shapes in the end the observed pattern.

Incidentally we report that additional DNS
were performed in a domain which is larger but still confined in the stream-wise direction, in order to explore when
finite size effects become negligible and suitably 
design a larger experiment.
Surprisingly we find that for a doubled size
domain, the growth rate decreases and almost matches
that of periodic simulations but the instability disappears
(we observe a pattern similar to the 
subcritical stratified case)
when further increasing the stream-wise domain (i.e. 4 times larger).
With respect to this trend 
the long computation time demanded to consider
even larger domains prevents us to 
be conclusive and further studies will be necessary.

The transition from the confined to the 
periodic case happens in a discontinuous way which needs to 
be further investigated.

At this stage we speculate that stream-wise boundaries 
may both introduce some forcing and inhibit the instability, 
perturbing the normal form and thus the resonance 
of the waves supported by the flow.
Horizontal aspect ratio possibly controls the mutual importance of these two effects in a non trivial,
non monotonous way, thus explaining the observed
scenario.

\section{Conclusions}\label{sec:conclusions}
We performed the linear stability analysis of the plane Couette Flow for a stably stratified fluid with a constant density gradient orthogonal to the shear.
The domain has rigid closed boundaries in the direction of the shear, and open periodic boundaries in both vertical and stream-wise directions.
Unstable stationary modes are found at strikingly moderate Reynolds number $Re\geq700$ and for a Froude number close to $1$ for non vanishing
horizontal and vertical wavenumbers with $k_x/k_z\sim0.2$.
We then explore the stability of the flow in the $(Re,Fr)$ space.
In the region we consider, the most unstable mode is always stationary and the growth rate remains relatively small, while the range of unstable $Fr$ numbers increases when increasing the $Re$ number.
Moreover the flow is unstable only to three dimensional perturbations,
i.e. only for $k_x$,$k_z\neq 0$. 
This result constitutes a fundamental difference with homogeneous shear flows for which the 
Squire theorem prescribes that the most unstable mode should be two-dimensional.
In the presence of stratification,
hyperbolic tangent profile
\citep{Deloncle2007} and Kolmogorov flow \citep{Lucas2017b} are also 
dominated by two-dimensional perturbations, while both
boundary layer profile \citep{Chen2016} and strato-rotational instability
\citep{Yavneh2001} similarly appear for three-dimensional perturbations.  
Curiously the study of \cite{Bakas2009b} included the same modal approach 
but no unstable case was reported.
Even more recently \cite{Chen2016} showed that a linear instability does occur in parallel flows free from inflection 
points, when a vertical stratification is added, but 
only the plane Poiseuille flow and the viscous boundary layer
were considered.
Remarkably our instability also appears close to $Fr=1$, confirming the necessary coexistence of shear and stratification.
The critical Reynolds number for the stratified plane Couette flow turns to be at least two times smaller than for the boundary layer and slightly larger than for the Poiseuille flow.

Looking at the most unstable mode of the linear problem,
vertical velocity and density perturbations develop close to 
the boundaries which suggests that a crucial role may be played by lateral boundaries.
Nonetheless, a comparable horizontal motion dominates in the mid vertical plane, and shows a vertically modulated pattern
which is reminiscent of the deep equatorial currents, and
staircase density layering in the Earth ocean \citep{Dunkerton1981,Dengler2002,dOrgeville2004}.

A mechanism was proposed to explain the onset of the instability
as the one suggested by \cite{Satomura1981a} adapted
to the case of internal gravity waves, instead of shallow
water waves.
In this picture internal gravity waves are trapped close to the boundaries and Doppler shifted, thus allowing two counter propagating waves to become stationary and mutually resonant.
The shape of the unstable region in the wave number space
and the appearance of discrete additional resonances are also
fully captured by the model, thus supporting its relevance.
An analogous mechanism was also invoked at the origin of
Strato-Rotational instability both in the plane Couette \citep{Kushner1998,Vanneste2007} and the Taylor-Couette \citep{Yavneh2001,Park2013} 
geometries.

The linear stability analysis was confirmed by DNS
which fairly reproduce the spatial pattern and the growth rate.
We report that no instability is observed 
when 
none of the unstable modes can properly fit in the domain.
This confirms that 
the instability sharply selects the spatial pattern of the perturbation. 

We analysed the experimental flow produced by a shearing
device immersed in a tank filled with salty water linearly stratified in density.
We report that when the $Fr$ number is close to $1$ and for $Re\geq 1000$
velocity perturbations are observed to grow in an exponential way.
Remarkably we observe that perturbations start to grow only when 
the plane Couette profile is almost completely established.
This was confirmed by ad-hoc versions of the linear problem
for a collection of transient profiles and 
by DNS which mimic the transient flow 
of experiments.
We conclude that the observed instability is crucially 
associated to the shape of the shear,
namely the plane Couette profile.

Then we explored the stability of the flow in the $(Re,Fr)$ 
space
varying the control parameter $\sigma$, that is equivalent to move along $Fr/Re=const$ lines, for a few different values of $Fr/Re$.
For each series of experiments we observe that an abrupt increase in the perturbation amplitude occurs, when $\sigma$ is bigger than a threshold value $\sigma_c$.
When adding experimental data to the stability diagram
predicted by linear theory we find that the threshold contour
indicated by experiments qualitatively matches the margin
of the linearly unstable region.
Also, close to the threshold, the velocity perturbation
shows a well organized pattern
and is almost horizontal, 
which 
is in agreement with the solution of the linear problem.
Nonetheless the unstable mode slowly oscillates in time
and appears at higher wave numbers than the 
most unstable (stationary) mode indicated by the 
linear analysis.
These two elements suggest that the 
mode selected in our experiment is not the most unstable of 
those predicted by linear analysis,
or that these latter are possibly not 
the same when considering the finite size
experimental apparatus.

In any case we  claim that the origin of the discrepancy 
relies on the critically low value of the horizontal aspect ratio of our experimental domain, which is necessarily bounded in the stream-wise direction.
The relevance of this hypothesis has been tested with complementary DNS where no slip rigid boundaries
are now implemented also in the stream-wise direction.
Remarkably, when copying the aspect ratio of our experiments
we minutely reproduce the perturbation pattern observed in 
experiments.
More generally DNS show that stream-wise confinement
affects the stability of the flow irrespectively of the $Fr$ 
number (i.e. also without stratification), which 
questions the link between the 
instability observed in experiment and 
that predicted by linear analysis. 
We performed then DNS of a subcritical stratified
experiment (i.e. $Re$ and $Fr$ below the critical value)
and show that the instability disappears.
We then acknowledge the unstable pattern observed in 
both experiments and DNS as a true
signature of the linear instability of a
plane Couette flow vertically stratified in density.

Future studies are planned to closer investigate 
which is the critical aspect ratio to recover
quantitatively the results of linear theory and 
periodic DNS.
To this aim new DNS will be performed in a 
larger domain, which will possibly indicate how to
correspondingly design a new set-up.

Quantitative measurements of the density field will be also performed in future experiments 
to quantify the density layering whose evidences
were already available in our highest Reynolds experiments in the form of 
regularly spaced optical distortion. 
Such measurements will possibly add cues to the comprehension 
of the diapycnal mixing in the presence of horizontal layering 
as recently studied with experiments \citep{Woods2010,Oglethorpe2013} 
and numerical simulations \citep{Lucas2017} in the case of the 
Taylor-Couette and Kolmogorov flows respectively.

\section*{Acknowledgments}

This work has been carried out thanks to the support of the A*MIDEX grant (ANR-11-IDEX-0001-02) funded by the French Government ``Investissements d'Avenir'' program.


\bibliographystyle{jfm}
\bibliography{biblioSHEAR}
\end{document}